\documentclass[traditabstract]{aa}
\usepackage{txfonts}
\usepackage{graphicx}
\usepackage{natbib}
\bibpunct{(}{)}{;}{a}{}{,}
\usepackage{amssymb}
\usepackage{longtable}

\begin{document}
\title{Manganese in dwarf spheroidal galaxies
\thanks{Based on
observations made with the  FLAMES-GIRAFFE multi-object spectrograph mounted on
the Kuyen  VLT telescope  at   ESO-Paranal   Observatory   (programs 
171.B-0588, 074.B-0415 and 076.B-0146)}}
\author{P. North\inst{1} \and G.~Cescutti\inst{2,1} \and P.~Jablonka\inst{1,3} \and V.~Hill\inst{4}
\and M.~Shetrone\inst{5} \and B.~Letarte\inst{6}
\and B.~Lemasle\inst{7}
\and K.A.~Venn\inst{8} 
\and G.~Battaglia\inst{9} 
\and E.~Tolstoy\inst{7} 
\and M.J.~Irwin\inst{10}
\and F.~Primas\inst{9} \and P.~Fran\c{c}ois\inst{3}}

\institute{Laboratoire d'astrophysique, 
           Ecole Polytechnique F\'ed\'erale de Lausanne (EPFL),
           Observatoire de Sauverny,
           CH-1290 Versoix, Switzerland
\and Leibniz-Institut f\"ur Astrophysik Potsdam (AIP), An der Sternwarte 16, D - 14482, Potsdam, Germany 
\and	   GEPI, Observatoire de Paris, CNRS, Universit\'e de Paris Diderot, F-92195
           Meudon, Cedex, France
\and       Laboratoire Lagrange, UMR7293, Universit\'e de Nice Sophia-Antipolis, CNRS,
           Observatoire de la C\^ote d'Azur, 06300 Nice, France
\and       McDonald Observatory, University of Texas, Fort Davis, TX 79734, USA
\and       South African Astronomical Observatory, P.O. Box 9, Observatory 7935,
           South Africa
\and       Kapteyn Astronomical Institute, University of Groningen, P.O. Box 800,
           9700AV Groningen, the Netherlands
\and       Dept. of Physics \& Astronomy, University of Victoria, 3800 Finerty
           Road, Victoria, BC V8P 1A1, Canada
\and       European Southern Observatory, Karl-Schwarzschild-str. 2, D-85748, Garching
           by M\"unchen, Germany
\and	   Institute of Astronomy, University of Cambridge, Madingley Road,
           Cambridge CB3 0HA, UK}

\date{Received 13 December 2011/ Accepted 1 March 2012}
\authorrunning{North et al.}
\titlerunning{Manganese in dwarf spheroidal galaxies}
\abstract{We provide manganese abundances (corrected for the effect of the
hyperfine structure) for a large number of stars in the
dwarf spheroidal galaxies Sculptor and Fornax, and for a smaller number in
the Carina and Sextans dSph galaxies.
Abundances had already been determined for a number of other
elements in these galaxies, including $\alpha$ and iron-peak ones, which
allowed us to build [Mn/Fe] and [Mn/$\alpha$] versus [Fe/H] diagrams.
The Mn abundances imply sub-solar [Mn/Fe] ratios for the stars in all four
galaxies examined. In Sculptor, [Mn/Fe] stays roughly constant between
[Fe/H]$\sim -1.8$ and $-1.4$ and decreases at higher iron abundance. In Fornax,
[Mn/Fe] does not vary in any significant way with [Fe/H].
The relation between [Mn/$\alpha$] and [Fe/H] for the dSph galaxies is clearly
systematically offset from that for the Milky Way, which reflects the different star formation histories
of the respective galaxies. The [Mn/$\alpha$] behavior can be interpreted
as a result of the metal-dependent Mn yields of type II and type Ia supernovae.
We also computed chemical evolution models for star formation
histories matching those determined empirically for Sculptor, Fornax, and Carina,
and for the Mn yields of SNe Ia, which were assumed to be either constant or variable with
metallicity. The observed [Mn/Fe] versus [Fe/H] relation in Sculptor, Fornax,
and Carina can be reproduced only by the chemical evolution models that include
a metallicity-dependent Mn yield from the SNe Ia.  }


\maketitle

\section{Introduction}

Manganese (Mn) is an iron-peak element that can be produced by both
type II and type Ia supernovae (SNe). Theoretical works indicate that
the  SNe II yields of Mn should increase with metallicity \citep{WW95},
which is
supported by observations such as the rise in [Mn/O] with [O/H] increasing
from $-0.5$ to $0.0$ \citep[e.g.,][]{FFB07}. Conversely, the
question of the metal dependence of the SNe Ia yields remains a
matter of debate.  \citet{SVT03} suggested that the SNe Ia yields of both
Cu and Mn increase with metallicity, and \citet{MWRS03}
brought additional arguments in favor of this hypothesis by comparing
the Mn abundances in the Milky Way bulge, the solar neighborhood,
and the Sagittarius dSph galaxy. These arguments in favor of a
metallicity-dependent Mn yield by SNe Ia were however challenged by
\cite{CGB04}, who judge that the observational results gathered so
far are too complex to allow a clear-cut conclusion to be drawn.
Nevertheless, \citet{CML08}, with their chemical
evolution model, and \citet{BBH08}, with their new method for
measuring the metallicity of Type Ia supernovae, independently found
additional evidence of the metal-dependence of the SNe Ia Mn yields,
which was also suggested by theoreticians such as \citet{OUN06}.

\citet{BBH08} suggest the following explanation of the phenomenon:
during the late evolution of the supernova (SN) Ia progenitor, the $^{14}$N
produced by the CNO cycle is converted into $^{18}$F (before being finally
transformed into $^{22}$Ne), which is transformed into $^{18}$O through $\beta^+$ decay.
This increases the number of neutrons in the stellar core, which is the
future white dwarf.  The neutron excess $\eta$ is proportional to the
metallicity $Z$ and is essentially preserved until the supernova
explosion.  Although this neutron excess leaves the production of the
most abundant species (e.g. Fe) unaffected, the formation of elements
with a larger number
of neutrons than protons is favored at high Z during the SN Ia explosion.
$^{55}$Mn, with its 25 protons and 30 neutrons, is the most abundant
of them; it is produced during incomplete Si burning (first as
$^{55}$Co, which then decays into $^{55}$Mn). When compared with the
abundance of an element insensitive to the neutron excess (especially
Cr, which is also built during incomplete Si burning), the resulting
Mn abundance can be expected to be an efficient tracer of the
progenitor metallicity.

To shed light on the production mechanisms of Mn, we clearly require to
investigate its abundances in a variety of galaxies, with different
star formation histories. To date the number of systems in which this
information is available is small and the number of stars is very
limited: besides the Milky Way, about two dozens of stars have been
analyzed in Sagittarius \citep{BHM2000, MWRS03,SBB07}, nine stars in
Sculptor \citep{SVT03,GSW05,TJH10}, and up to a maximum of five stars
per galaxy in Draco \citep{SCS01}, Sextans \citep{SVT03,TJH10}, Carina
\citep{SVT03}, Fornax \citep{SVT03,TJH10}, and LeoI \citep{SVT03}.

DART, the Dwarf galaxy Abundance and Radial-velocities Team, allows a
real step forward for Sculptor, Fornax, Carina, and Sextans.  This
ESO Large Program based on the FLAMES/GIRAFFE spectrograph at the
VLT can encompass stellar samples of up to 80 stars per galaxy with
optical spectra at relatively high resolution (R$\sim$20,000). The
abundances of most elements with measurable lines have already been
published, except for manganese: the equivalent widths of this element
are available, but the abundance determination is more
complicated. Since manganese has an odd atomic number ($Z=25$), it
has a significant hyperfine structure (hereafter HFS), which broadens the spectral
lines. This can lead to desaturation of the lines, which cannot be
neglected as soon as the equivalent widths exceed a few tens of
m\AA. Therefore, reliable abundances cannot be obtained by just using
the equivalent width and total oscillator strength of a given
line. All components of the hyperfine structure have to be taken into
account. This work provides Mn abundances (with HFS taken into
account) in the three Local Group dwarf spheroidal galaxies for an
unprecedentedly large number of stars. This constitutes by far the
largest sets of Mn abundances in any galaxy other than the Milky Way,
and the size of our sample is comparable to e.g. the samples of stars in
the thin and thick disks of our Galaxy considered by
\citet{FFB07}.

This paper is organized as follows. Section 2 introduces our sample of
stars. Section 3 describes how we derived Mn HFS-corrected abundances,
while Section 4 discusses the results.  Section 5 presents chemical
evolution models that reproduce the observations. Finally, Sect. 6
summarizes our results.

\section{Observational material and analysis}
\label{observations and sample}

In the following, we analyze five different samples. For four of them, 
the  FLAMES/GIRAFFE  HR10, HR13, and HR14 grisms were used, respectively, 
centered on $5488$, $6273$, and $6515$~\AA~  
\citep[see][]{ET06}.  The full abundance analysis papers of the DART
FLAMES/GIRAFFE in Sextans and Sculptor are being written up. Surveys of the
Fornax, Sculptor, and Sextans galaxies have already been been performed to
search for extremely metal-poor stars \citep{TJH10}. The results for the Mn
abundances of these stars, which were previously corrected for HFS, are incorporated
in the present work. The results of the analysis of all elemental abundances
besides Mn are published in \citet{LHT10} for Fornax and in \citet{LHT12}
for Carina.  In a companion work, \citet{VSI12} presented the chemical
composition of 23 elements in nine
bright Carina red giant branch stars observed with either the
FLAMES/UVES fibers or the Magellan/MIKE spectrograph. Their Mn abundances
were corrected for HFS structure and their sample complements ours. In summary, all
necessary data, such as equivalent widths and stellar parameters, were
available for the present analysis of manganese.

\subsection{Galaxy and stellar sample}

$\bullet$ The Fornax dSph galaxy was studied by \citet{LHT10}, who provided
and discussed the abundances of a large number of elements. There are 72 stars
with at least one measurable Mn line, 60 of which have three reliable Mn lines. 

\noindent $\bullet$ In Sculptor, 76 stars have at least one measurable Mn line,
50 of which have a reliable average Mn abundance based on three lines (Hill et al., in prep) . 

\noindent $\bullet$ Twenty-one stars constitute the stellar magnitude-limited sample
($I < 18$) in Sextans
(Jablonka et al, in prep.). However, only 5 stars have reliable Mn equivalent widths.

\noindent $\bullet$ In Carina, 17 stars have at least one Mn line \citep{LHT12},
but only 6 have detectable Mn\,\textsc{i} $\lambda 5407$ \AA\ and $\lambda
5420$ \AA\, lines, which were finally selected to compute the average
Mn abundance.

The detailed analysis of the Mn lines and the composition of the final sample of stars
is performed in Section \ref{final-lines}.

\subsection{Stellar atmosphere models and HFS corrections}

The abundance analysis was performed with two codes, \texttt{calrai} on the one
hand, and \texttt{moog} on the other, both used with the new
MARCS\footnote{{http://marcs.astro.uu.se/}} spherical models of
stellar atmospheres \citep{gustafsson03, gustafsson08}, under the LTE
approximation (for Sculptor, the \texttt{calrai} abundances were determined
using plane-parallel MARCS models). The computation of the radiative transfer was still done
in a plane-parallel geometry.

The stellar effective temperatures, gravities, and turbulence
velocities were adopted from the DART general analyses of each galaxy.
Temperatures and gravities were determined from photometric data for Fornax,
Sextans and Carina, and from spectroscopic data in the case of Sculptor.

In principle, equivalent widths were measurable for up to four Mn lines.
All of these lines belong to the wavelength range of the HR10 FLAMES/GIRAFFE setup.
One line, Mn\,\textsc{i} $\lambda 5432$, belongs to the multiplet No 1 and
is a resonance line, while the other three belong to the multiplet No
4. All four lines were significantly broadened by the hyperfine
structure.

The Mn hfs-corrected abundances were derived in two steps: 

$\rhd$ First, the uncorrected Mn abundances were computed with
\texttt{calrai}.  The code was initially developed by
\citet{spite67} \citep[see also the atomic part description in][]{cayrel91},
and has been continuously updated over the years.
\texttt{calrai} was used to analyze all DART data sets. The DART results
were partly summarized in \citet{tolstoy09}.  The homogeneity of
these analyses allows us to perform robust comparisons of the chemical patterns
for all metallicity ranges and between galaxies. 

%DEJA DIT PLUS HAUT
%\citet{LHT10} present the
%full analysis of the Fornax dSph giraffe sample, while \citet{TJH10}
%analyse extremely metal-poor stars with [Fe/H]\,$< -3.0$ in Sculptor,
%Fornax, and Sextans dSphs.

$\rhd$ Second, an HFS correction was computed 
with the August 2010 version of Chris Sneden's \texttt{moog}
code\footnote{http://www.as.utexas.edu/~chris/moog.html}.
On the one hand, for each line we computed the uncorrected Mn abundance (i.e.
neglecting the hyperfine structure), using the {\it abfind} driver
and the total $log(gf)$ value of the lines, taken from the Kurucz
file gfhy0600.100\footnote{available at http://kurucz.harvard.edu/linelist.html}.
The resulting abundances are very close to those given by \texttt{calrai}
(see the Appendix for a comparison between the \texttt{moog} and
\texttt{calrai} Mn abundances).
On the other hand, we computed the abundances {\it with} the hyperfine structure,
using the {\it blends} driver of \texttt{moog} and introducing all
hyperfine components listed in the above Kurucz file.

Finally, the HFS correction $\Delta_{hfs}$ for each line was defined as the difference
between the two above abundances.
The line parameters and hyperfine components are given in
Table~\ref{table:line_param}.

\begin{table*}
\caption{Parameters of the four Mn\,\textsc{i} lines used in this work and
of their hyperfine components, taken from Kurucz' database. The first line
gives the total $\log(gf)$ value of the line considered as single, while the
subsequent lines give the $\log(gf)$ values of the hyperfine components.}
\label{table:line_param}
\centering
\begin{tabular}{rrr|rrr|rrr|rrr}
\hline\hline
\multicolumn{3}{c|}{Mn\,\textsc{i}\,$\lambda 5407$\AA} &
\multicolumn{3}{c|}{Mn\,\textsc{i}\,$\lambda 5420$\AA} &
\multicolumn{3}{c|}{Mn\,\textsc{i}\,$\lambda 5432$\AA} &
\multicolumn{3}{c}{Mn\,\textsc{i}\,$\lambda 5516$\AA} \\
\hline
\multicolumn{1}{c}{$\lambda$}&$\chi_{exc}$ & $\log(gf)$ &
\multicolumn{1}{c}{$\lambda$}&$\chi_{exc}$ & $\log(gf)$ &
\multicolumn{1}{c}{$\lambda$}&$\chi_{exc}$ & $\log(gf)$ &
\multicolumn{1}{c}{$\lambda$}&$\chi_{exc}$ & $\log(gf)$  \\
\multicolumn{1}{c}{[\AA]}&\multicolumn{1}{c}{[eV]} & &
\multicolumn{1}{c}{[\AA]}&\multicolumn{1}{c}{[eV]} & &
\multicolumn{1}{c}{[\AA]}&\multicolumn{1}{c}{[eV]} & &
\multicolumn{1}{c}{[\AA]}&\multicolumn{1}{c}{[eV]} &   \\ \hline
5407.420&2.14&$-1.743$&5420.360&2.14&$-1.462$&5432.550&0.00&$-3.795$&5516.770&2.18&$-1.847$\\ \hline
5407.325&2.14&$-3.139$&5420.256&2.14&$-3.018$&5432.506&0.00&$-4.377$&5516.699&2.18&$-3.273$\\
5407.332&2.14&$-3.394$&5420.261&2.14&$-2.988$&5432.510&0.00&$-5.155$&5516.709&2.18&$-2.905$\\
5407.333&2.14&$-3.394$&5420.270&2.14&$-2.733$&5432.535&0.00&$-5.155$&5516.718&2.18&$-2.905$\\
5407.341&2.14&$-3.075$&5420.272&2.14&$-3.766$&5432.538&0.00&$-4.640$&5516.728&2.18&$-4.482$\\
5407.354&2.14&$-3.196$&5420.281&2.14&$-2.812$&5432.541&0.00&$-4.992$&5516.743&2.18&$-2.773$\\
5407.353&2.14&$-3.196$&5420.295&2.14&$-2.511$&5432.561&0.00&$-4.992$&5516.757&2.18&$-2.773$\\
5407.354&2.14&$-3.196$&5420.298&2.14&$-3.687$&5432.564&0.00&$-4.971$&5516.771&2.18&$-2.947$\\
5407.366&2.14&$-2.900$&5420.311&2.14&$-2.745$&5432.566&0.00&$-4.987$&5516.790&2.18&$-2.875$\\
5407.382&2.14&$-3.131$&5420.329&2.14&$-2.327$&5432.580&0.00&$-4.987$&5516.809&2.18&$-2.875$\\
5407.384&2.14&$-3.131$&5420.333&2.14&$-3.812$&5432.583&0.00&$-5.418$&5516.828&2.18&$-2.398$\\
5407.400&2.14&$-2.708$&5420.351&2.14&$-2.771$&5432.584&0.00&$-5.089$&	     &    &$	  $\\
5407.420&2.14&$-3.162$&5420.374&2.14&$-2.169$&5432.594&0.00&$-5.089$&	     &    &$ 	  $\\
5407.422&2.14&$-3.162$&5420.379&2.14&$-4.164$&5432.595&0.00&$-6.117$&	     &    &$	  $\\
5407.442&2.14&$-2.523$&5420.402&2.14&$-2.947$&5432.596&0.00&$-5.351$&	     &    &$	  $\\
5407.468&2.14&$-3.344$&5420.429&2.14&$-2.029$&5432.601&0.00&$-5.351$&	     &    &$	  $\\
5407.469&2.14&$-3.344$&        &    &$	    $&	      &	   &$	   $&	     &    &$	  $\\
5407.494&2.14&$-2.352$&        &    &$	    $&	      &	   &$	   $&	     &    &$	  $\\ \hline
\end{tabular}
\end{table*}

\begin{figure*}[t!]
\centering
\includegraphics[height=6cm]{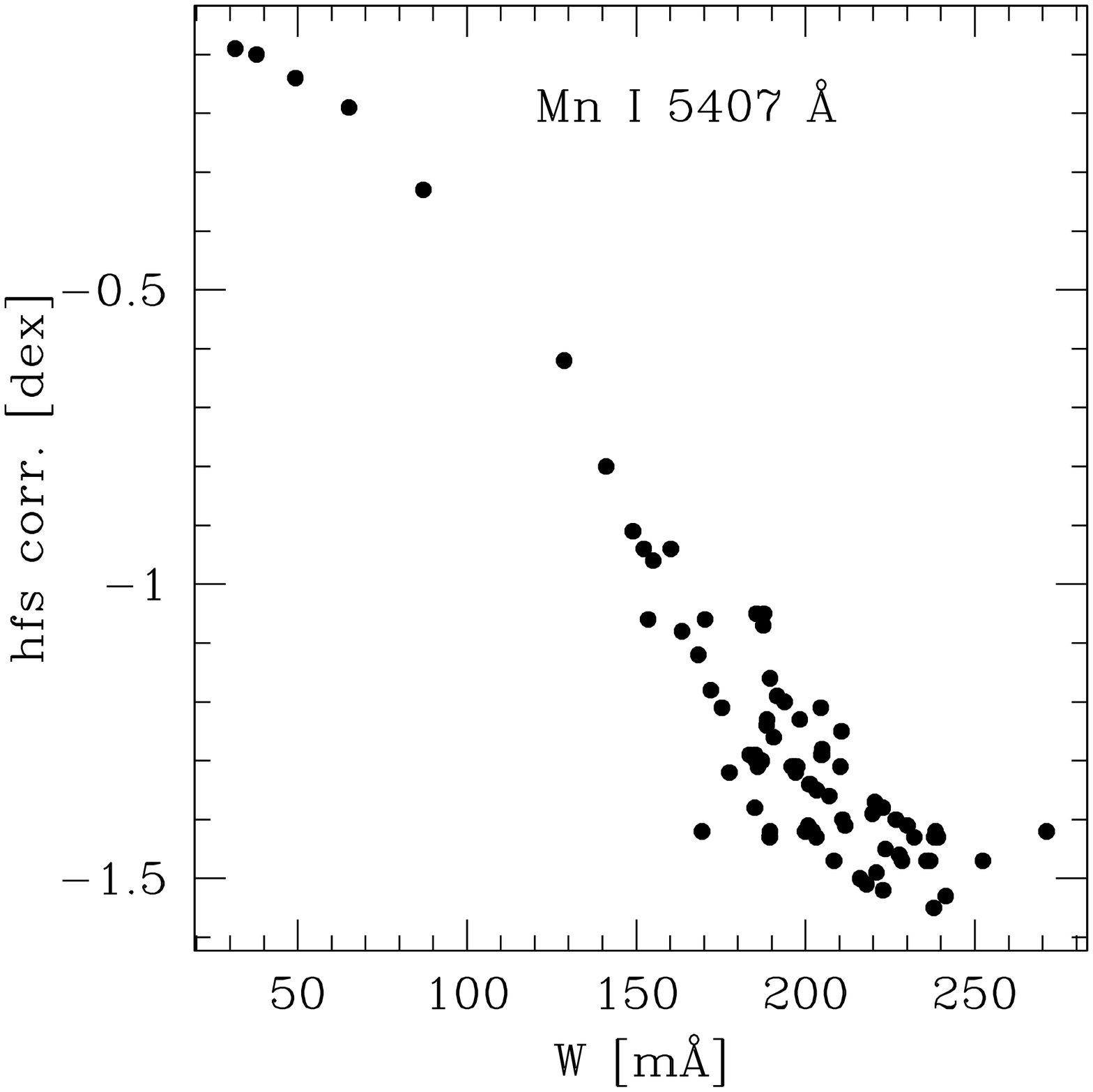}
\includegraphics[height=6cm]{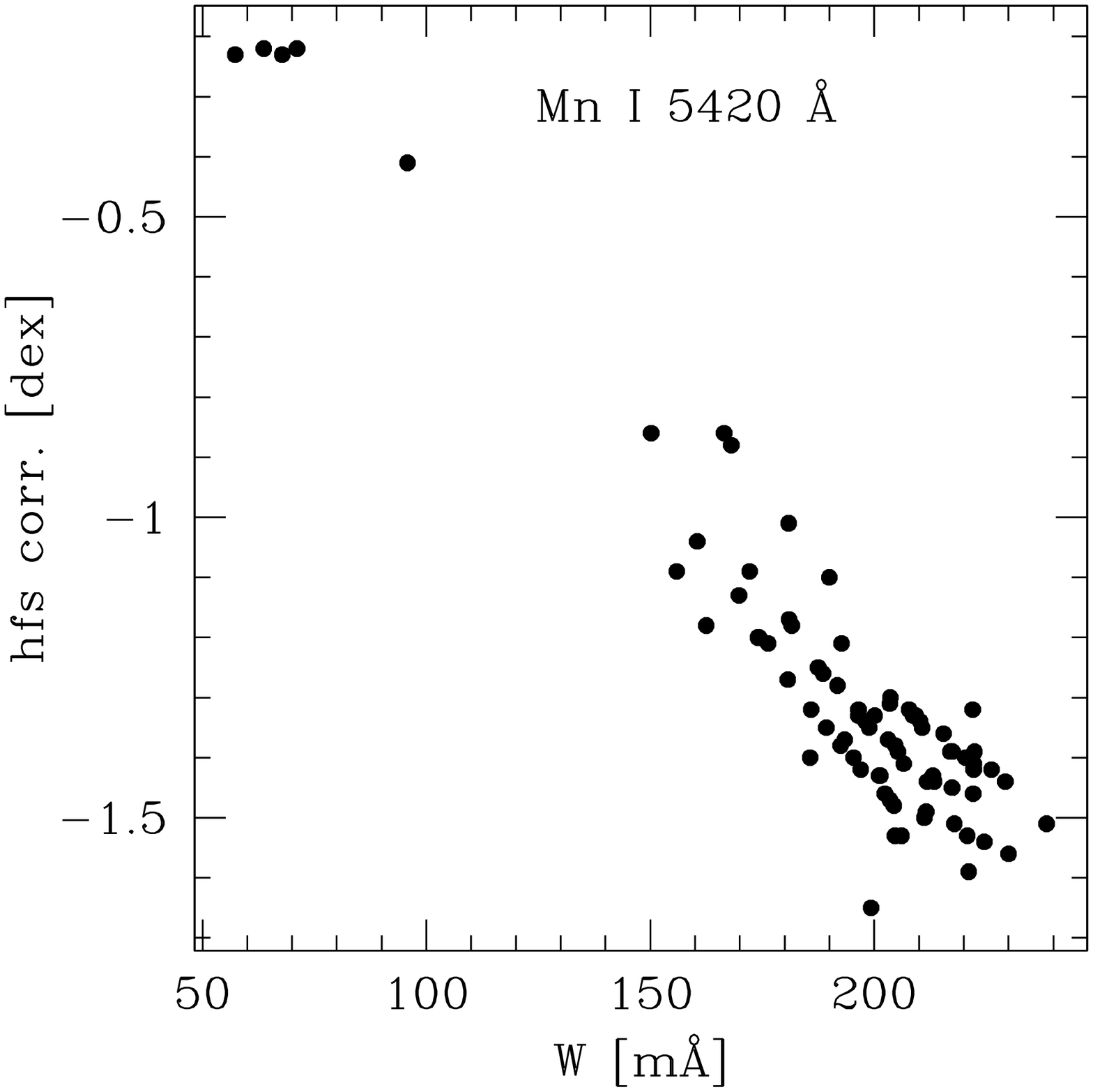}
\includegraphics[height=6cm]{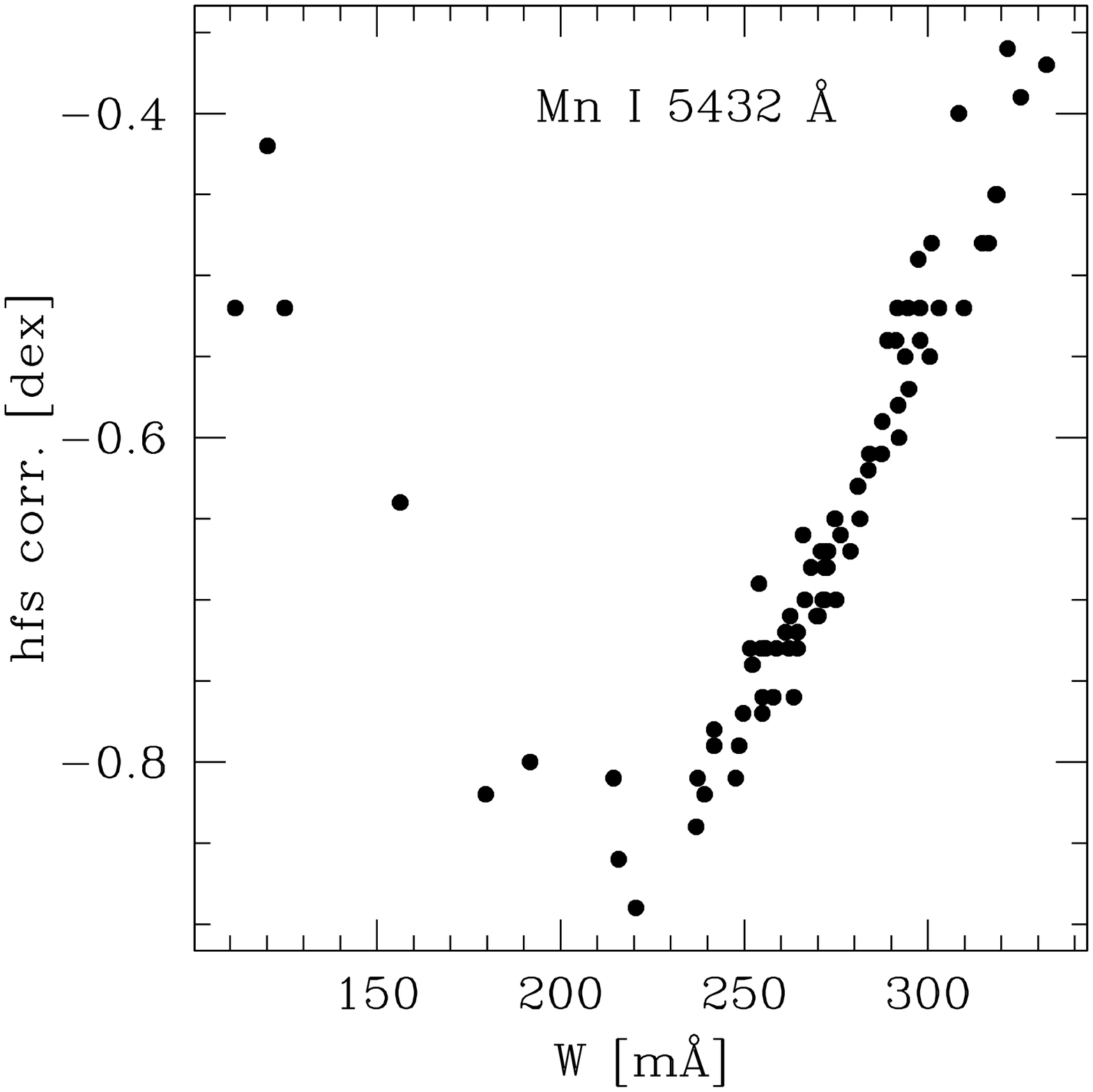}
\includegraphics[height=6cm]{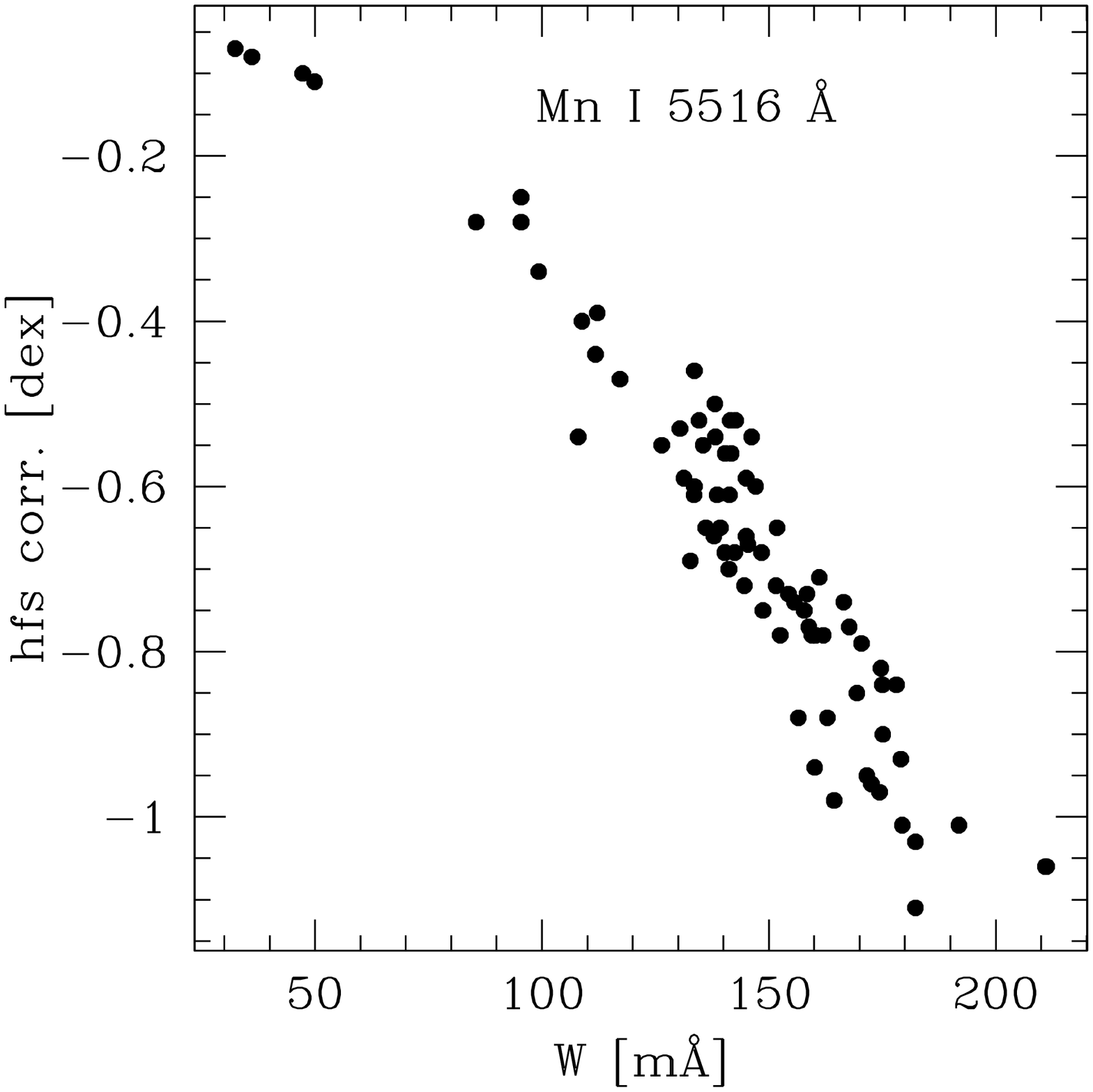}
\caption{Hyperfine structure correction (defined as the abundance with hfs
correction versus abundance without it) as a function of equivalent width for
the Mn\,\textsc{i} lines $\lambda5407$, $\lambda5420$, $\lambda5432$, and
$\lambda5516$~\AA~ for red giants in the Fornax dSph galaxy.
The $\lambda5432$ line was finally discarded (see text).}
\label{hfs_corr}
\end{figure*}

As an example,
the HFS corrections for the 72 stars in the Fornax dSph galaxy are shown in Fig.
\ref{hfs_corr} as a function of equivalent width, for the four available lines
(for 71 stars only in the case of the $\lambda5516$\AA\ line). The behavior of the
correction for the strongest line, Mn\,\textsc{i}\,$\lambda5432$\AA\ ,
is especially noteworthy: the correction becomes increasingly negative
as the equivalent width increases, then turns upward beyond
220~m\AA. This behavior reflects the curve of growth: the minimum
correction (or maximum of its absolute value) coincides with the
plateau of the curve of growth, while the desaturation effect of the
hyperfine structure become unimportant in the linear part on the one
hand, and on the strongly saturated part on the other. The
scatter in the HFS corrections at a given equivalent width is due to the
variety of stellar parameter values, especially for the micro-turbulent
velocities. In Sculptor, the behavior of the HFS correction is
similar, except that the rising branch (for the
Mn\,\textsc{i}\,$\lambda5432$\AA\ line) is much shorter, because of
the lower metallicity. In Sextans, the HFS corrections are never
larger than 0.35 dex, this maximum being reached for the
$\lambda5432$\AA\ line, which is the strongest. In Carina, the HFS
corrections are smaller than 0.3 dex for the $\lambda5407$\AA\ and
$\lambda5516$\AA\ lines, and smaller than 0.6 dex for the other two
lines.

We note that the amplitude of the HFS correction may reach 1.6~dex;  Fig.
\ref{hfs_corr} illustrates how inescapable this correction is.

\section{Mn final abundances}
\label{final-lines}
\subsection{Final line-by-line abundances}

The final Mn abundances were obtained by adding $\Delta_{hfs}$ to the
initial abundances derived with \texttt{calrai}. For Carina,
whose data was included later, we used only \texttt{moog}
to determine the Mn abundance, because the results of this code perfectly
match those of \texttt{calrai}, as shown in Fig.~\ref{fig:comp_ab_Fnx},
where we used the same spherical atmosphere models as for the abundance
determination of Fe and other elements.

\begin{figure*}[t!]
\centering
\includegraphics[height=8cm]{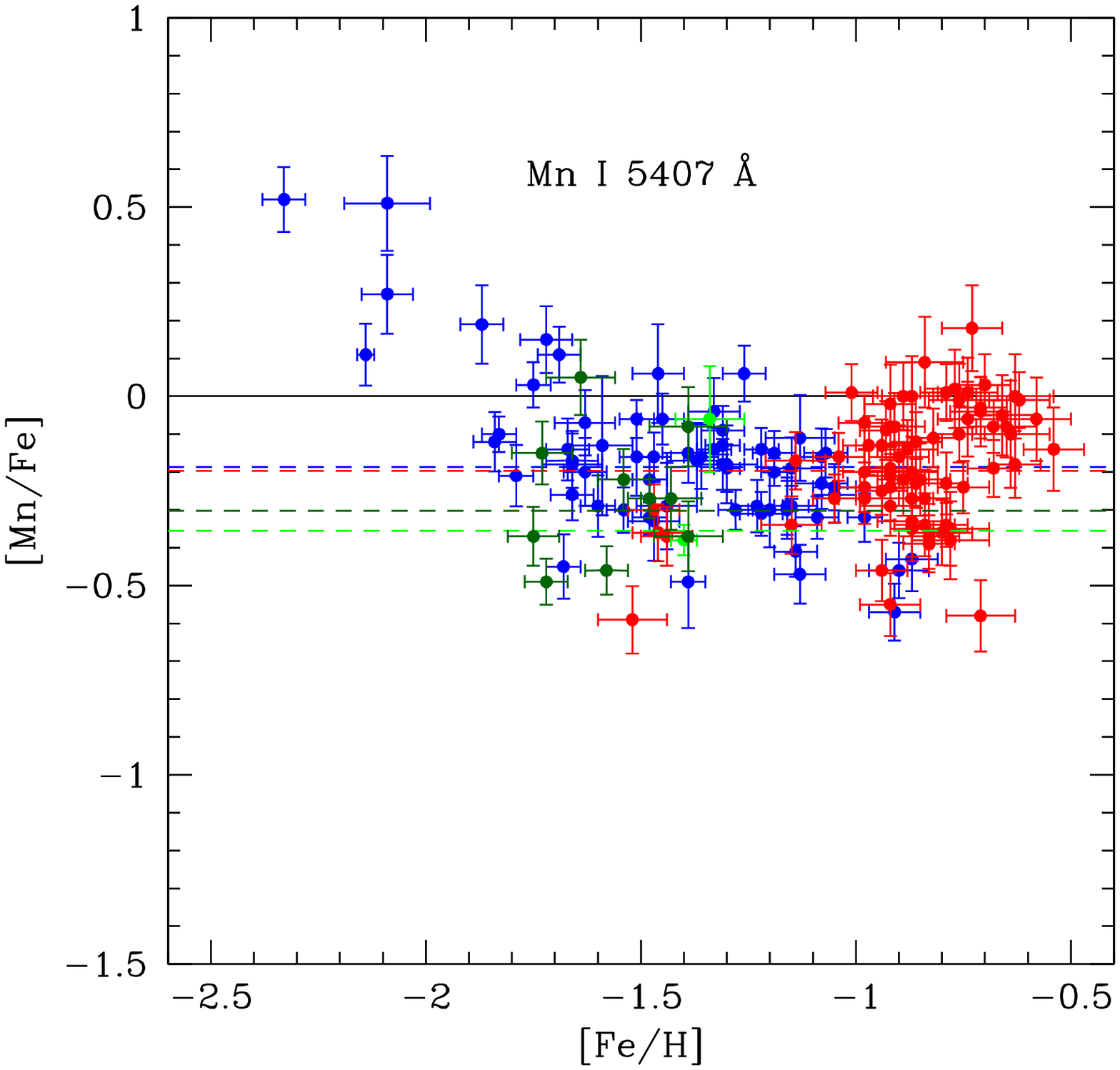}
\includegraphics[height=8cm]{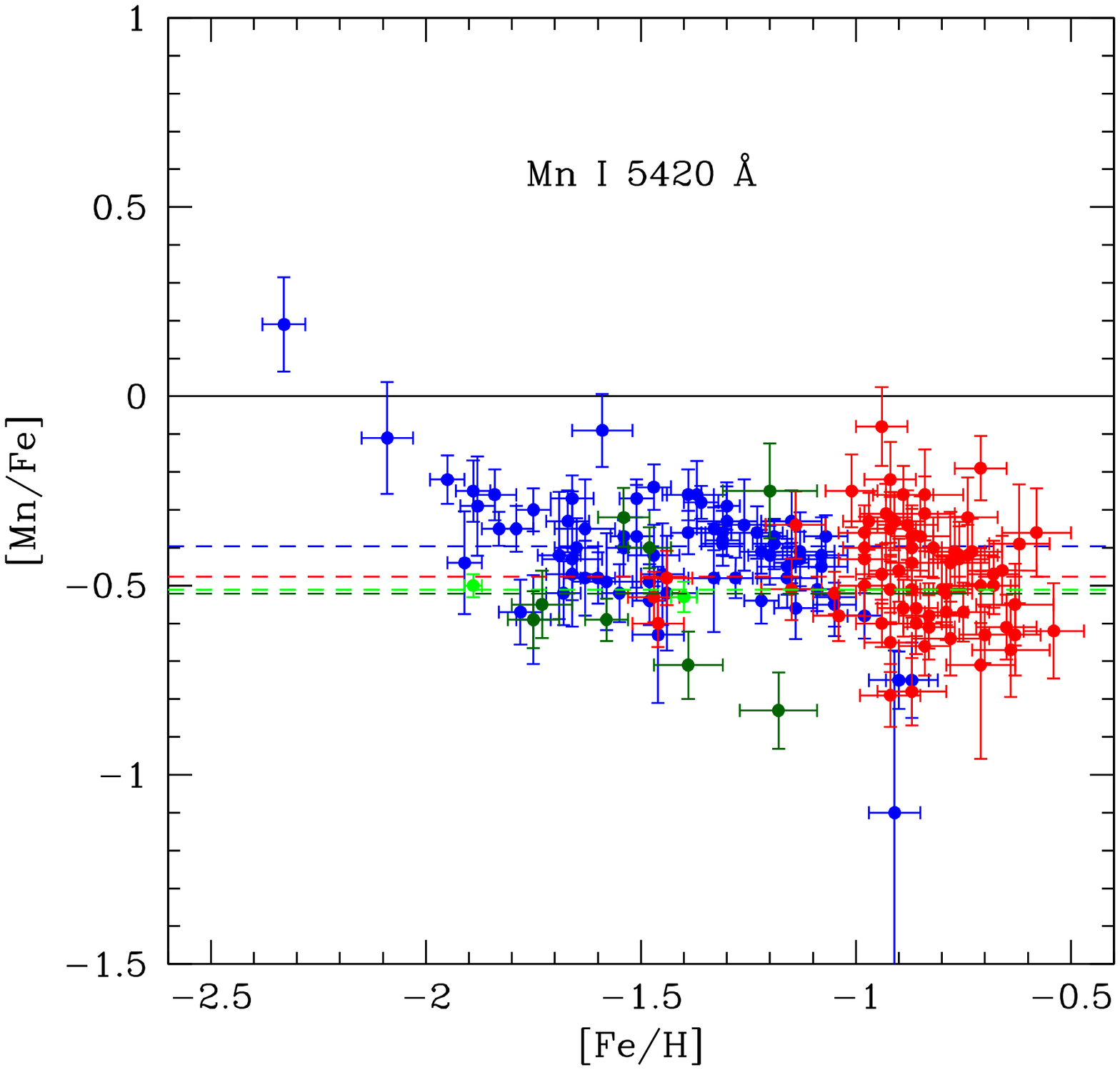}
\includegraphics[height=8cm]{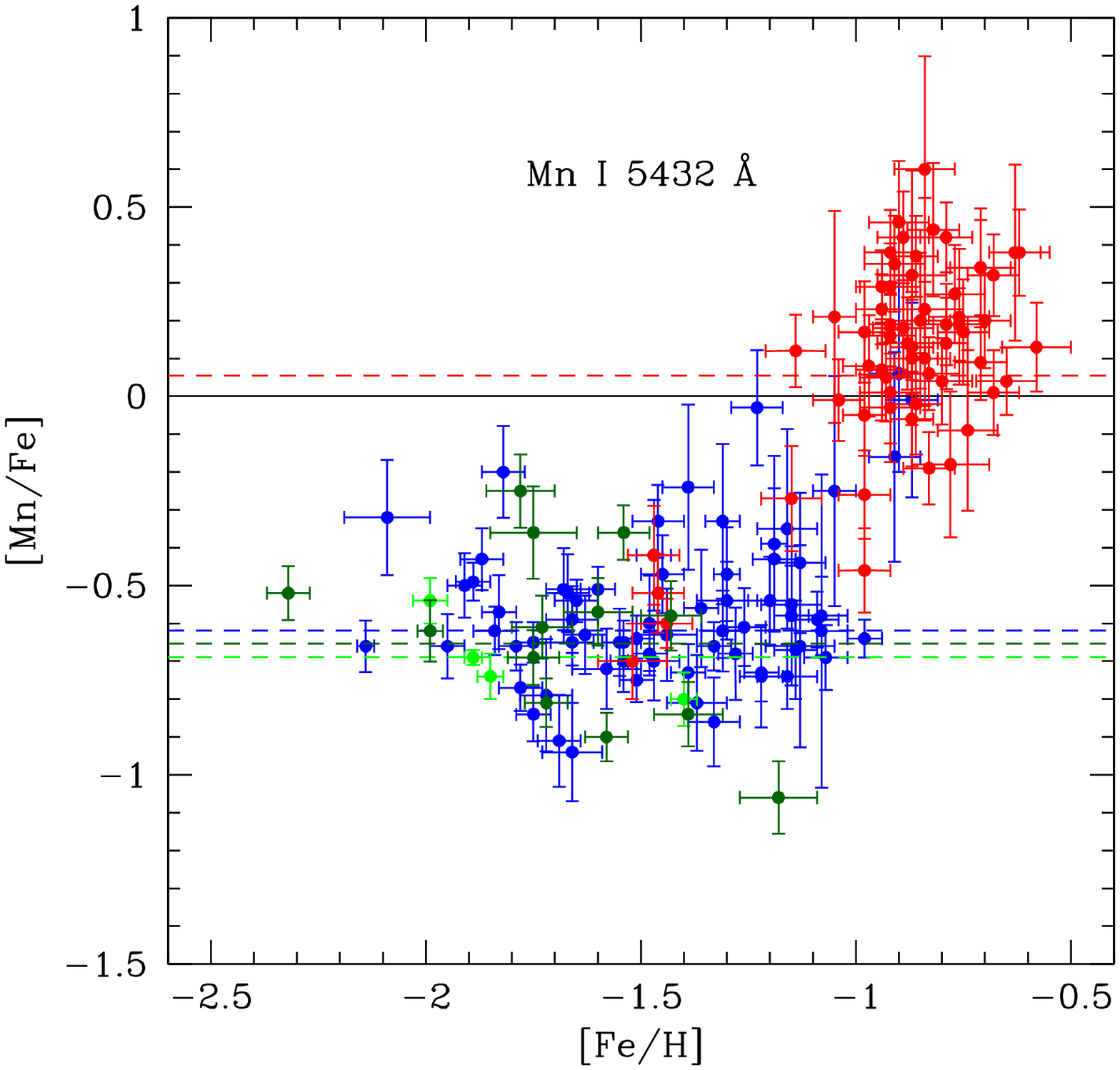}
\includegraphics[height=8cm]{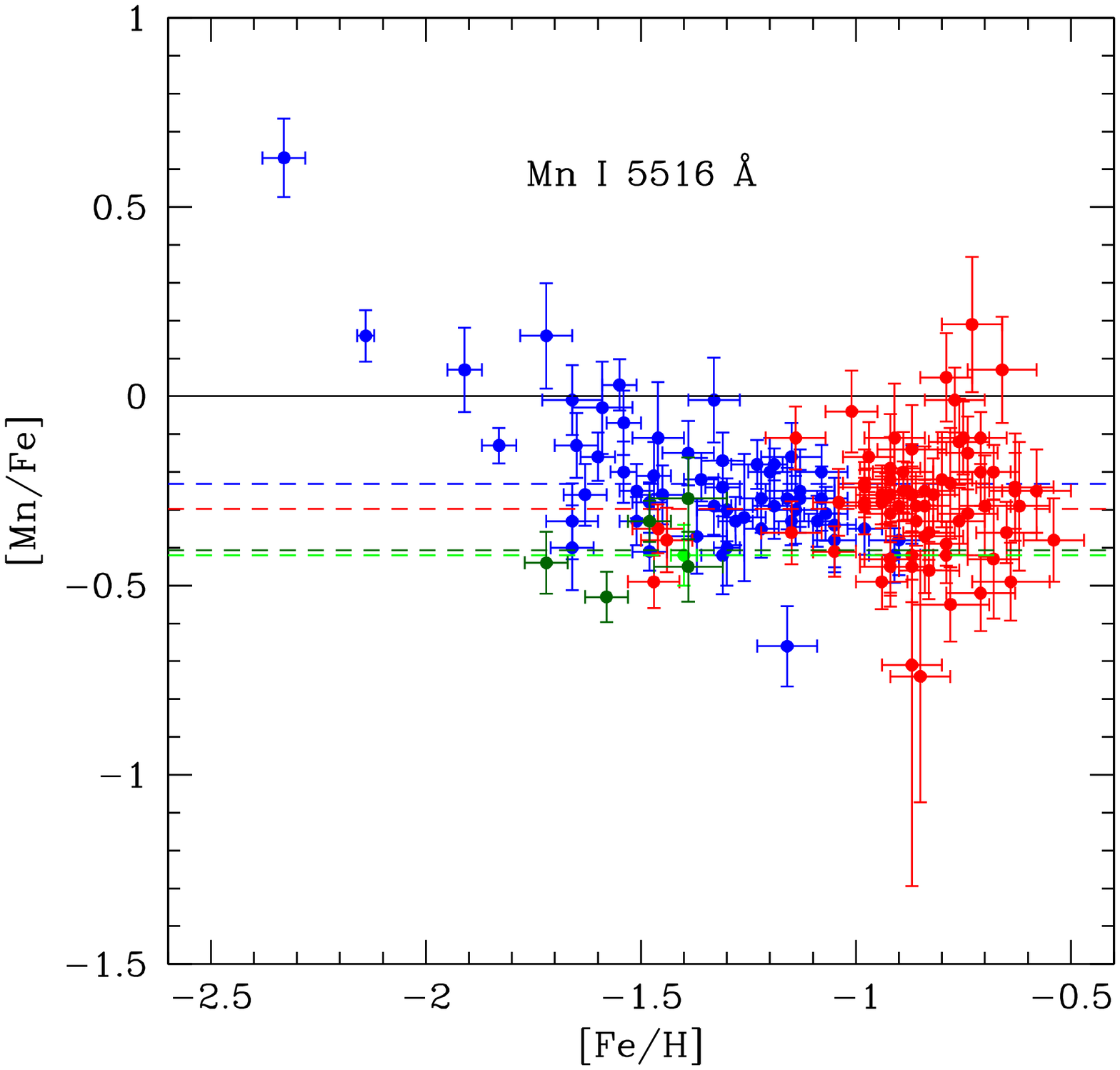}
\caption{Final Mn abundances for each of the four lines available for the stars
in the Sculptor (blue), Fornax (red), Sextans (green), and Carina (dark green)
dSph galaxies. The black
horizontal line indicates the zero value; the dashed lines are the weighted
averages for their respective galaxies. Note the strongly discrepant behavior of the
$\lambda 5432$ line.}
\label{MnFe_indiv}
\end{figure*}

Fig.~\ref{MnFe_indiv} displays [Mn/Fe] versus (vs) [Fe/H] for the Sculptor, Fornax,
Sextans, and Carina galaxies, for each of the four Mn\,\textsc{i} lines.
The three lines $\lambda 5407$\AA, $\lambda 5420$\AA, and $\lambda 5516$\AA\
follow very similar trends,
while the $\lambda 5432$\AA~ line behaves in a clearly different way.

While for the Sculptor dSph, the $\lambda 5432$\AA\ line
leads to differences in [Mn/Fe] of only a few tenths of a dex compared
to the other ones, for the Fornax dSph, both the mean
[Mn/Fe] level and the variation with [Fe/H] are affected. The
Giraffe sample of stars at the center of the Fornax dSph is indeed more
metal-rich than those at the center of the Sculptor dSph.
Therefore, the equivalent widths of the $\lambda 5432$\AA\ line are
larger in Fornax than in Sculptor and above 200~m\AA~ for most
stars. The $\lambda 5432$\AA\ line is the most sensitive to non-local
thermodynamic equilibrium (NLTE)
effects because of its low excitation potential. Furthermore, it is so
strong that its profile departs significantly from a Gaussian, thereby
severely biasing the equivalent width estimated by the
\texttt{daospec} code, which indeed assumes a Gaussian profile.  As a
consequence, we discarded the $\lambda 5432$\AA\ line in the computation
of the average Mn abundances.

The $\lambda 5407$\AA\ line in Fornax also behaves in a slightly different way,
with respect to the $\lambda 5420$\AA\ and $\lambda 5516$\AA\ lines.
As for the $\lambda 5432$\AA\ line, this is probably due to the
large equivalent widths of the most metal-rich stars of this galaxy,
which may greatly exceed $200$~m\AA . Therefore, we excluded all
lines with $EW>230$~m\AA\ (for the $\lambda 5407$\AA\ one, but also the
two others) from the Fornax sample when computing the average Mn abundances.
The safer and more stringent criterion of $EW>200$~m\AA\ would have left only
25 stars with an average Mn abundance based on three lines. Including stars
with $200 < EW < 230$~m\AA\ raises the average [Mn/Fe] ratio by no
more than $0.1$~dex, without biasing too substantially the distribution of stars
in terms of metallicity, hence this trade-off was deemed to be acceptable.

The line-by-line [Mn/H] and [Mn/Fe] abundances are listed in
Tables~\ref{table:abundFnxa} and \ref{table:abundFnxb}
for the Fornax dSph, in Tables~\ref{table:abundScla} and \ref{table:abundSclb}
for the Sculptor dSph, in Table~\ref{table:abundSex} for the Sextans
dSph, and in Table~\ref{table:abundCar} for the Carina dSph.

\subsection{Average abundances and compilation of the [Mn/Fe] vs [Fe/H] diagram}
To compute the final abundances, we used an average weighted by the
inverse variances of the abundances obtained from the individual lines;
these variances were propagated from the estimated errors in the corresponding
equivalent widths.

$\rhd$ In Fornax as well as
in Sculptor, the average abundances were computed from the three lines
$\lambda 5407$\AA, $\lambda 5420$\AA, and $\lambda 5516$\AA. Since
some stars lack one or more of these lines, or the equivalent
width of some of the lines is larger than $230$~\AA, only 60 stars are left
out of the initial 72 ones.
$\rhd$ in Sculptor, the average abundances could be computed from the same
three lines as in Fornax, for a final sample of 50 stars.
$\rhd$ In Sextans, keeping only those stars with three reliable lines would
have resulted in only one single object. Therefore, all 5 stars
\citep[in addition to the EMP stars of][]{TJH10} were included in the final
sample, even though the average abundances are based
on fewer than three lines in most cases. 
$\rhd$ In Carina, the initial sample of 17 stars shrinks to 6 objects
having at least the two Mn lines  $\lambda 5407$\AA, $\lambda 5420$\AA.
The average Mn abundances are based on these two lines.

The average Mn abundances are computed from three lines
in both Sculptor and Fornax but from only two lines in Carina, which might cause
a zero-point problem, when our results for the two galaxies are compared.
However, Fig.~\ref{MnFe_indiv} shows that the $\lambda 5516$\AA\ line, which
was not included in the average [Mn/Fe] values of Carina, yields [Mn/Fe]
ratios that are in-between those derived using the two other lines
(see e.g. the averages for Sculptor), such that neglecting the line does not change
the average values by more than a few hundredths of dex at most.
Another kind of zero-point problem does, however, arise between some published
values and those of this work because of the different solar abundances
adopted. We adopt $\log(N_\mathrm{Mn})+12=5.39$ and
$\log(N_\mathrm{Fe})+12=7.50$, \citet{SIS06} adopt $5.39$ and $7.52$
respectively, and \citet{VSI12} adopt $5.43$ and $7.50$. This
difference of a few hundredths of dex remains smaller than the uncertainties
and was therefore neglected.

\begin{figure*}[t!]
\centering
\includegraphics[width=13cm]{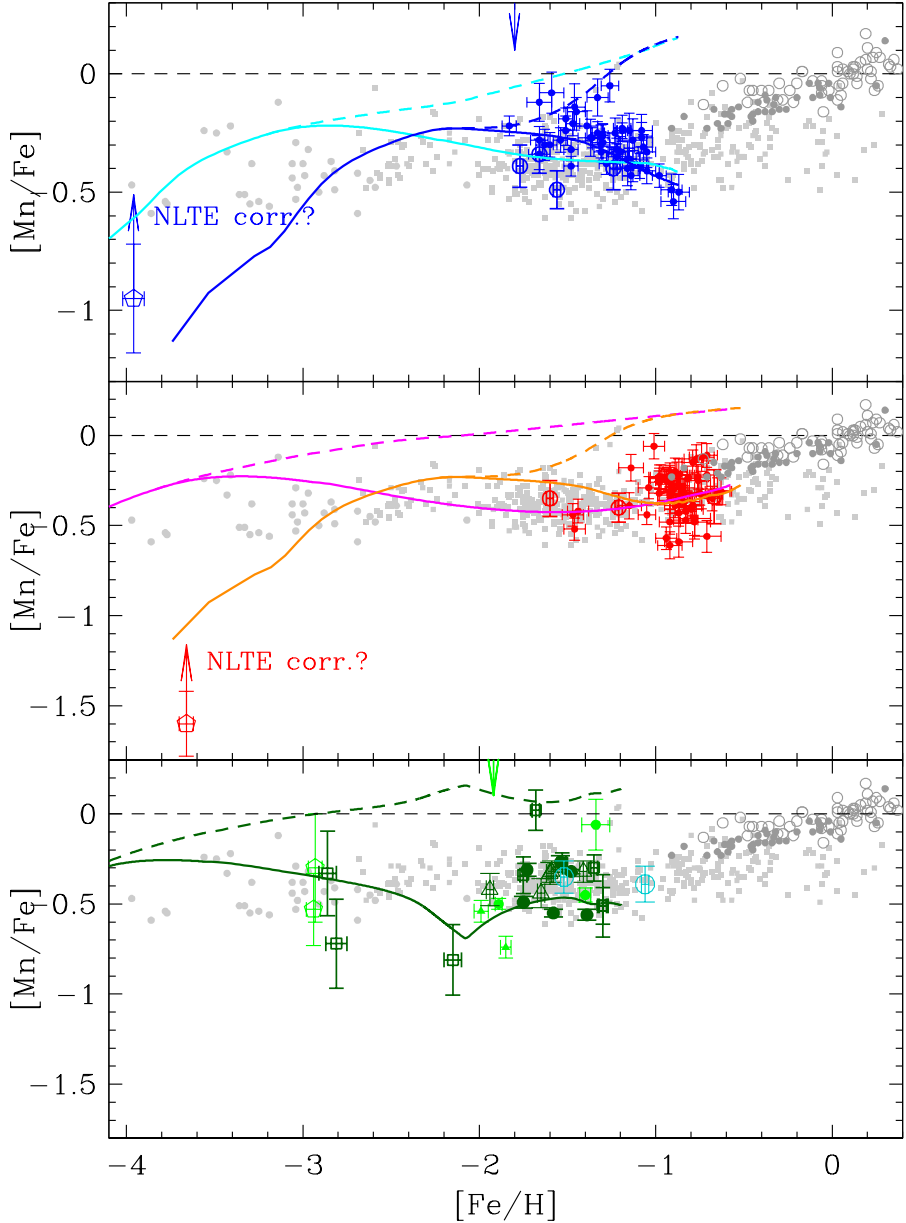}
\caption{The relation between [Mn/Fe] and [Fe/H]. {\bf Data:} Each
  dSph galaxy is shown with a different color: blue stands for
  Sculptor, red for Fornax, green for Sextans, dark green for Carina,
  and dark turquoise for Leo I.  The two green triangles identify the
  Sextans stars S05-010 and S08-038, whose Mn abundances are based on
  the $\lambda 5432$
  line only.  The colored filled circles identify the samples analyzed
  in this work; the open symbols stand for previous published
  analyzes. The four open pentagons at [Fe/H] $\lesssim -2.9$ are from
  \citet{TJH10} and are corrected for the HFS. The
  attached arrows gives an idea of the NLTE correction that they
  likely need, $+0.44$, as computed by \citet{BG08} for the halo giant
  HD 122563. The results of \citet{SVT03} for Sculptor, Fornax,
  and Leo I are displayed with open circles, and for Carina with
  open triangles. The dark green open squares are the Carina results of
  \citet{VSI12}.
  The dark gray open and filled circles
  represent the \citet{FFB07} Milky Way thin and thick disk stars,
  respectively. The \citet{SIS06} Milky Way cluster and field halo
  stars are shown with full gray squares; gray dots are very metal poor
  stars in the Milky Way halo from \citet{CDS04}.
  %, while diamonds are
  % halo stars from \citet{AHB05}.
  The vertical blue and green
  downward arrows indicate the "knee", where [Mg/Fe] starts to
  decrease with increasing [Fe/H], in the Sculptor and Sextans dSphs
  (DART). {\bf Models:} The chemical evolutions of the Sculptor (model
  A in cyan and model B in blue, see text), Fornax (model C in
  magenta and model D in orange, see text), and Carina (model E in
  dark green) dSphs are followed. The
  continuous lines show models with metallicity-dependent SNe Ia Mn
  yields as in \citet{CML08}.  The dashed lines follow the evolution
  of [Mn/Fe] for the same SFHs, but with
  metal-independent SNe Ia Mn yields.
  }
\label{MnFe}
\end{figure*}

All points corresponding to fewer than two Mn lines were ignored in
Fig.~\ref{MnFe}, except for a few stars in Sextans. Figure~\ref{MnFe}
is discussed further in Section~4.

\subsection{Discussion of possible NLTE effects}
Whilst we took the line HFS into account, our
abundances may still suffer from NLTE effects. Very few studies
address this problem for the manganese lines. \citet{BG07} examined
the solar atmosphere for a total of 39 lines belonging to ten
multiplets, and their line list includes the four Mn\,\textsc{i} lines
we use here. They showed that the NLTE correction (defined as $\Delta X
= \log\varepsilon^{NLTE}-\log\varepsilon^{LTE}$, where $\varepsilon$
is the ratio of the number densities of Mn to H) is at most on the
order of 0.1~dex in absolute value. The maximum correction, $\Delta X
=+0.11$, applies to the $\lambda 5432$\AA\ line, closely followed by the
other three ($+0.09$ for $\lambda 5420$\AA, and $+0.085$ for $\lambda
5407$\AA, and $\lambda 5516$\AA). Unfortunately, these corrections cannot
be applied directly to our case, because the surface gravities and
metallicities of our sample are very different from solar. \cite{BG08}
computed the NLTE abundances of 14 stars, all of which but one are
metal-deficient, down to [Fe/H]$\sim-2.5$. Unfortunately, all
but one are main sequence or subgiant stars, the only exception being a
giant star with $\log g\sim 1.5$ and [Fe/H]$=-2.51$. In this case, the
NLTE correction to [Mn/Fe] is on the order of $+0.44$.  However,
it is difficult to infer what the NLTE correction
should really be for our stars. \cite{BG08} consider 18 lines from
$\lambda 4018$\AA\ to $\lambda 6021$\AA\, but none of them coincide with the
lines used here. Nevertheless, the possibility remains, that our 
[Mn/Fe] values might increase by as much as 0.2~dex when corrected
for NLTE effects.

On the observational side, \citet{FFB07} argued that the excitation
balance is unaffected by departure from LTE in their sample, based
on the identical behavior of lines with different excitation
potentials, when plotting the abundance as a function of
$T_\mathrm{eff}$, $\log g$, and [Fe/H]. However, they did not exclude
possible departures from ionization balance. In addition, all their
stars are either on the main sequence or the subgiant branch, and
none have [Fe/H] $< -1$. Furthermore, we have only one line in common
with \citet{FFB07}, Mn\,\textsc{i} $\lambda 5432$, which, as argued
above, we chose to discard because it is probably the most sensitive
to NLTE effects and its equivalent width is biased owing to
its large strength. Therefore, the conclusion reached by \citet{FFB07}
cannot be generalized to our sample.

\citet{SIS06} determined Mn abundances for 200 stars in 19 globular
clusters and for a comparable number of field stars with similar
stellar parameters. They also neglected the NLTE effects, on the
grounds that they should be small when considering [Mn/Fe], which
involves two neutral species \citep{IKS01}. Interestingly, they found
an average constant value $<$[Mn/Fe]$>=-0.36$ for their halo field
stars, which is about 0.15~dex lower than the value found by
\citet{FFB07} for their most metal-poor stars
([Fe/H]$\approx-1$) in the thick disk. In the
range $-1 < $[Fe/H]$ < +0.4$, where [Mn/Fe] increases to values
above solar
for the thick disk stars of \citet{FFB07}, the [Mn/Fe] values of
\citet{SIS06} show the same trend but systematically lower by about 0.1~dex.
One possible explanation of this difference is a bias produced by
there being only one line, Mn\,\textsc{i} $\lambda 6013$\AA, in common to
\citet{SIS06}. Another explanation might be an NLTE correction that would be
0.15 dex larger for giants than for less evolved stars\footnote{Note
that \citet{FFB07} see no difference between dwarf and giant stars
in the [Mn/O] vs [O/H] diagram, which contradicts this explanation,
unless Fe alone is responsible for the difference.}, but this
remains to be confirmed on theoretical grounds.

\section{Discussion of the observational results}
\subsection{The [Mn/Fe] versus [Fe/H] diagram}
\subsubsection{Description}
The main results of the present study are summarized in
Fig.~\ref{MnFe}.  Our comparison sample is
composed of
the results of {\it i)} \citet{SIS06} for the Milky Way globular clusters and
field halo stars,   {\it ii)} \citet{CDS04} for field halo stars, 
and {\it iii)} \citet{FFB07} for the Milky Way thin
and thick disk stars. We also display the extremely metal-poor
(EMP) stars found by \citet{TJH10} in Sculptor, Fornax, and Sextans,
and the  nine stars of \citet{VSI12} in Carina.
Finally, we show the stars studied by \citet{SVT03} in the Sculptor, Fornax,
Sextans, Carina, and Leo I dSph galaxies.

\subsubsection{The extremely metal-poor (EMP) stars}
\citet{TJH10} noted that the manganese
abundances of their Sextans members, S11-04 and S24-72, were based on
only one line, Mn\,\textsc{i}\,$\lambda 4823.52$\AA\, which differs from the
lines we used. In spite of this difference, their [Mn/Fe] values are in
good agreement with those of other Sextans stars of higher metallicity. They
also agree with the values found in the Milky Way halo \citep{SIS06} and
our results in Fornax and Sculptor. The Mn abundance of the most extreme EMP
star, Scl07-50, was obtained from the three resonance lines of the triplet at
$\lambda\sim 4030$~\AA, while that of Fnx05-42 (which is only slightly less iron-poor
but has the lowest [Mn/Fe] ratio) was obtained from two lines of the same triplet.
These resonance lines are expected to be strongly affected by NLTE. Hence, we
drew an upward arrow at the position of the two most
iron-poor stars, with an amplitude of 0.44~dex matching the NLTE correction of
\citet{BG08} for the metal-poor giant HD 122563. These arrows, however, have to
be considered only as a qualitative indication, because the true NLTE correction
might be very different (possibly larger), with respect to the extremely low [Fe/H].

\subsubsection{[Mn/Fe] trends with [Fe/H] in Sculptor and Fornax: are  they real?}

In a broad sense, the [Mn/Fe] vs [Fe/H] relations for the three Local Group
dSphs of this study agree rather well with the trends found for the Galaxy by
both \citet{FFB07} in the thick disk and \citet{SIS06}
in the halo and globular clusters. This agreement can only be
considered qualitative, owing to the zero point issues raised earlier and
the different kind of stars considered (dwarfs instead of giants) in the case of
\citet{FFB07}.

A closer look reveals some interesting features. In Fornax (red dots
in Fig.~\ref{MnFe}), there seems to be a very slight correlation
between [Mn/Fe] and [Fe/H], but the trend is essentially due to the
small group of 4 stars near [Fe/H]$=-1.4$. Pearson's correlation
coefficient is only $0.17$ (for 60 stars), the more robust Spearman
correlation coefficient is $0.05$, and the Student-t
test is $0.38$. The relatively large difference between the two
correlation coefficients is due to the small group of 4 stars around
[Fe/H]$\sim -1.4$, which lie rather far away from the bulk of the data and
cause the large value of Pearson's coefficient. In conclusion, even though
future observations might confirm the trend suggested here in Fornax,
we can only tell for the time being that it is not statistically significant.
[Mn/Fe] might thus
be considered constant with [Fe/H], with an average value
$<[\mathrm{Mn/Fe}]>=-0.32\pm 0.02$. Any cosmic dispersion must be
smaller than about 0.09~dex, because the scatter in the
[Mn/Fe] values around the mean amounts to $\sim 0.12$~dex, while their
average error is $\sim 0.07$~dex. 

\begin{figure}[t!]
\centering
\includegraphics[width=8.5cm]{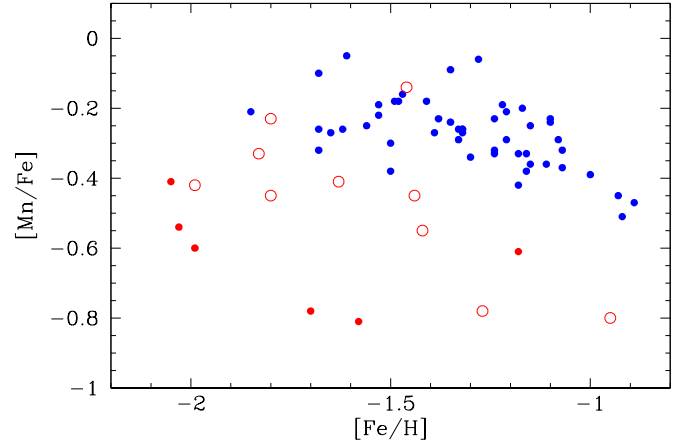}
\caption{Same as Fig.~\ref{MnFe}, but for Sculptor only (blue dots), showing a
plateau followed by a decreasing trend rather than a monotonic
decrease in [Mn/Fe]. Results for the globular cluster $\omega$ Centauri
are also shown as red circles \citep{CSB10} and red dots \citep{PMS11}.}
\label{MnFe_Scl}
\end{figure}

Conversely, the 50 Sculptor stars display a global negative trend
\[ \mathrm{[Mn/Fe](Scl)}
 = -0.299\, \mathrm{[Fe/H]} - 0.679 ~\mathrm{,}~~~\mathrm{rms} =0.085 ~\mathrm{dex~,}\]
where Pearson's correlation coefficient is $-0.569$, Spearman's
coefficient is $-0.546$, and the Student-t test is $-4.51$ (for 50 stars). This
is clearly significant because the null hypothesis has a probability well
below one percent. Zooming into Sculptor in Fig.~\ref{MnFe}, the
relation does not appear, however, to be a precisely monotically declining
line, but rather like a plateau followed by a decreasing linear
function (Fig.~\ref{MnFe_Scl}). If real, a ``knee'' appears 
between [Fe/H]=$-1.5$ and $-1.3$.

The contrasting behaviors of [Mn/Fe] in Fornax and Sculptor seem
difficult to explain entirely in terms of NLTE effects, primarily because the
average surface gravities are the same ($\sim 0.7$ dex) in both cases.
Moreover, while one would expect NLTE effects alone to produce in all
galaxies the same monotonic relation with metallicity as seen in the
Milky Way, the observed trends instead differ for each galaxy. An
alternative explanation could be a systematic error in the HFS
components (splitting, oscillator strengths), because on average,
the lower the [Fe/H], the smaller the HFS correction.

We conducted two different tests to explore the possibility of
erroneous HFS corrections:

First, we excluded from the sample all Sculptor stars with HFS
corrections larger than a given limit. Limiting the sample to the 45
stars with $|\Delta_{hfs}|\leq 1$ for the three lines at 5407\AA,
5420\AA, and 5516\AA\ still provides a Spearman rank correlation
coefficient of $-0.5$ and a t-test of $-3.75$, implying a probability
well below one percent that the correlation is random. Limiting ourselves
further to $|\Delta_{hfs}|\leq 0.3$ (24 stars), the correlation
remains, with a probability of random occurrence being well below five
percent.  This suggests that only a very large relative error in
$\Delta_{hfs}$, on the order of $50\%$, could account for the trend we see in
Sculptor, which spans almost $\sim 0.2$~dex in [Mn/Fe]. This seems
unlikely.

Second, instead of using the Kurucz line list, we extracted from Tables 1,
5, and 6 of \citet{VV03} the HFS components of the 5420\AA\ and 5432\AA\ lines,
which are based on the laboratory measurements of
\citet{BSW83}. We retained the uncorrected wavelength and oscillator
strength values (the ``$\lambda'$'' and ``$\log(gf)'$'' ones). We
computed the HFS correction for these data again for the seven
Sculptor stars for which the original HFS corrections ranged from
$-0.10$ to $-1.60$ for the 5420\AA\ line, and from $-0.34$ to $-1.07$
for the 5432\AA\ line.  For the 5420\AA\  line, we obtained the same
$\Delta_{hfs}$ values as for the Kurucz components within $0.02$~dex. 
For the 5432\AA\ line, $\Delta_{hfs}$ was recovered to be within
$0.01$~dex for six stars, and within $0.03$  for the last one.

Therefore, the HFS corrections appear to be very robust, especially as the
uncorrected $\log(gf)'$ values listed in the paper of \citet{VV03}
differ only slightly from Kurucz' ones (the total $\log(gf)$ value is
$-1.492$ instead of $-1.462$ for the 5420\AA\ line, and $-3.740$ instead
of $-3.795$ for the 5432\AA\ line).

In summary, the variation in [Mn/Fe] can probably be taken at face value and
 genuinely related to the nucleosynthesis of Mn.
The decreasing trend in [Mn/Fe] with increasing
[Fe/H] seen in Sculptor had been observed nowhere else, except
for giants and subgiants in the
globular cluster $\omega$ Centauri \citep{CSB10,PMS11}, where the
anti-correlation is even more pronounced (see Fig.~\ref{MnFe_Scl}).
\citet{RCM11} attempted to interpret these last sets of results, but
unsuccessfully, although they also found that a metallicity-dependent
yield of SNe Ia would be more realistic than a constant yield.

\subsection{Manganese and the $\alpha$ elements}

\begin{figure*}[t!]
\centering
\includegraphics[height=10cm]{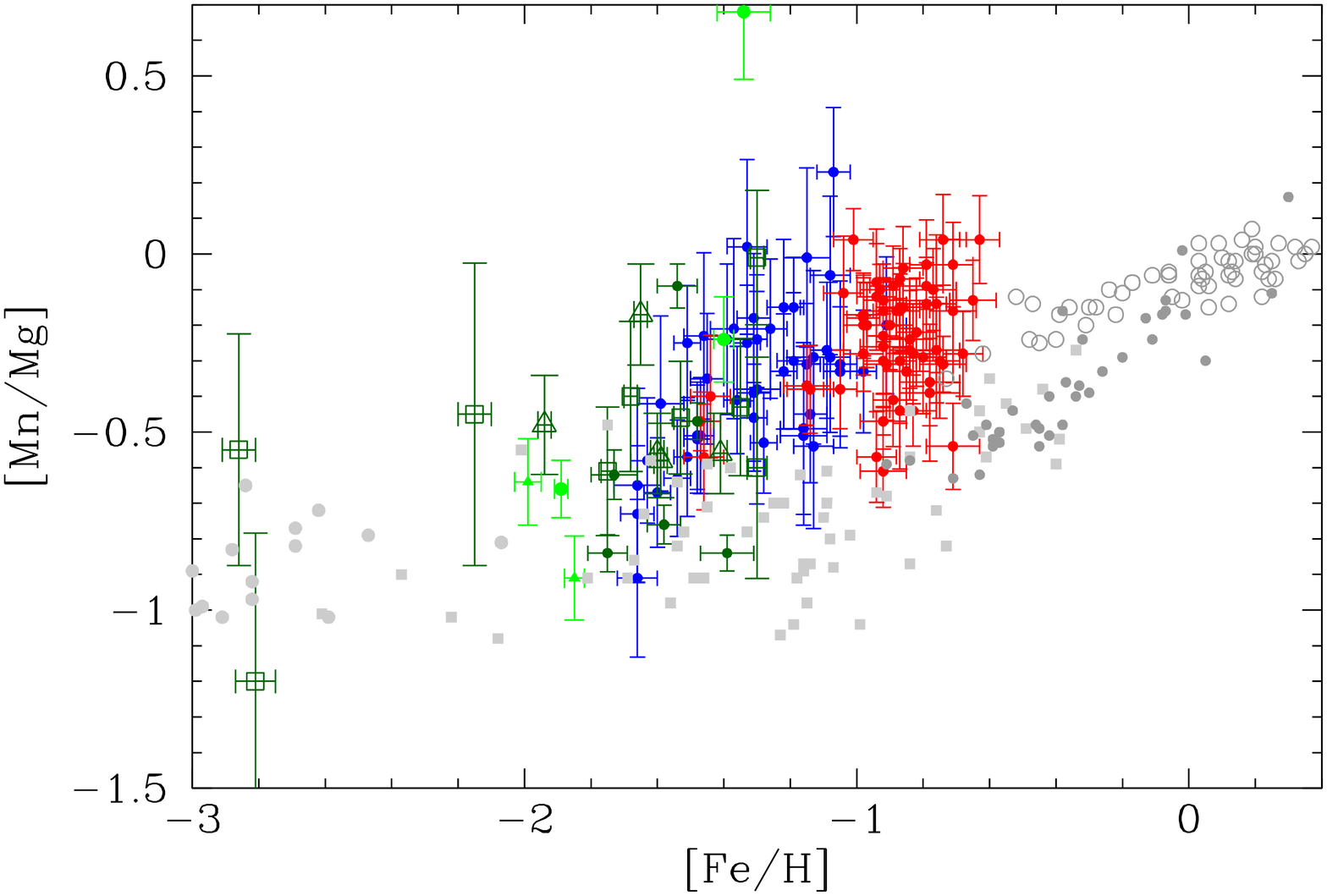}
\includegraphics[height=10cm]{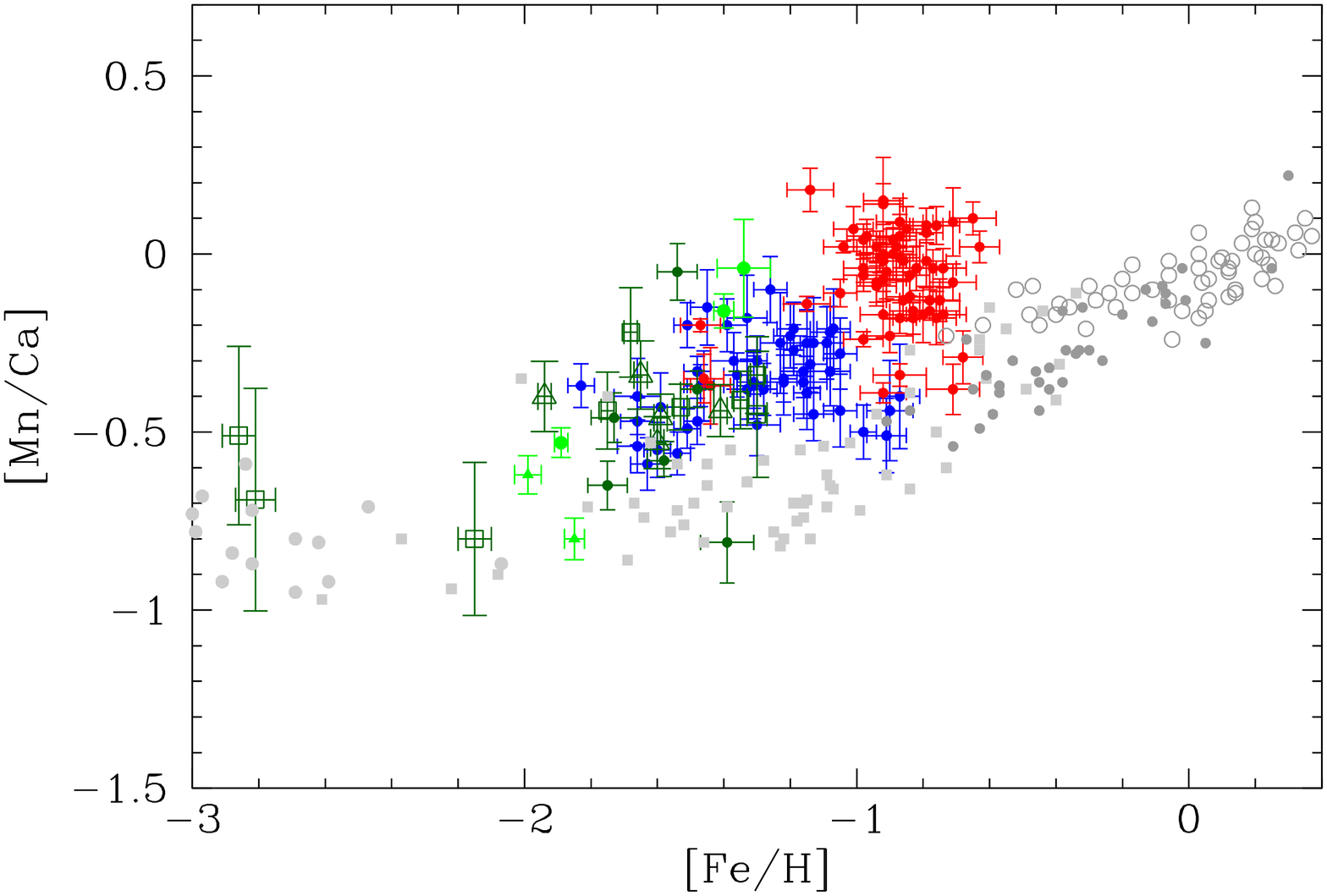}
\caption{[Mn/$\alpha$] versus [Fe/H] for each of the $\alpha$ elements Mg
and Ca, for the Sculptor (blue dots), Fornax (red dots), Carina
(dark green symbols), and Sextans (green symbols) galaxies. The key to the symbols
is the same as in Fig.~\ref{MnFe}. For comparison, we provide results for field Milky
Way stars, gathered from the literature:
  The open and full black dots are the thin and thick, respectively, disk dwarf
stars measured by \citet{BFL03}, \citet{BFL05}, and \citet{FFB07}. The full
grey symbols are from \citet{CDS04} (round dots), \citet{AHB05} (diamonds),
and \citet{GCC03} (squares).
}
\label{Mn_alpha}
\end{figure*}

Since the $\alpha$-elements are mostly produced in massive stars while
Mn can be produced by both SNe II and SNe Ia, the ratio of Mn to
some of the $\alpha$-elements may reveal at which point manganese is
produced by one or the other nucleosynthetic route.
Fig.~\ref{Mn_alpha} displays the cases of Mg and Ca, two
$\alpha$-elements with slightly different nucleosynthetic
origins: Mg is produced in a hydrostatic phase of the evolution of
massive stars, while Ca is instead produced during a type II
supernova explosion \citep{woosley02}.

Figure~\ref{Mn_alpha} clearly shows that the position in [Fe/H] of
the rising branch of [Mn/$\alpha$] depends on the galaxy star formation
history, similarly to the ``knee'' in [$\alpha$/Fe]. 
\citet{THT09} report  a decrease in [Mg/Fe] for Sculptor from $\sim+0.5$
to $\sim -0.2$~dex for [Fe/H] between $\sim -2.4$ and $\sim -1$, while
[Ca/Fe] decreases from $\sim+0.3$ to $\sim 0.0$.  Unfortunately, most
of the $-2.4<$[Fe/H]$<-1.8$ stars in our sample do not have reliable
equivalent widths for the Mn lines. Nevertheless, we have a dozen stars
between [Fe/H]$\sim-1.8$ and $\sim-1.4$ , where the slope of the
relation between [Mg/Fe] and [Fe/H] is strongly negative.  After an
initial increase in [$\alpha$/Mn] with metallicity [Fe/H], both [Mn/Mg] and
[Mn/Ca] are constant at [Fe/H]$> -1.4$.  The [Mn/Ca] mean level is
higher in Fornax than Sculptor stars as a consequence of the lower
[Ca/Fe] abundance ratio in Fornax. The dispersions around the mean are
similar for both galaxies. In Carina, the 20 stars (6 from
\citet{LHT12}, 9 from \citet{VSI12},
% after exclusion of 2 peculiar stars,
and 5 from \citet{SVT03}) lie close to the sequence defined by the
Sculptor stars. However, the star at [Fe/H]$=-1.4$ lies outside the
general trend defined by the sample of \citet{LHT12}, and the stars
of \citet{SVT03} do not show any trend. Our 5 Sextans stars define an
increasing trend similar to that of Sculptor and possibly steeper,
which needs confirmation by further observations.

In Sculptor, the variation in [Mn/Mg], from $\sim -0.8$ to $\sim -0.3$,
with [Fe/H], reflects i)
the plateau at [Mn/Fe] for [Fe/H] below $-1.4$ (see
Fig.~\ref{MnFe_Scl}), and ii) the decrease in [Mg/Fe] due to SNe Ia,
as can be most clearly seen at [Fe/H] $> -1.6$. The differential behavior of Mn
and the $\alpha$-elements can be attributed to their different
nucleosynthetic paths : Mn is produced ever more in increasingly metal-rich
core-collapse supernovae, and definitely more than in the metal-poor
type Ia supernovae \citep{MWRS03}. To further investigate the relative
roles of SNe II and SNe Ia, we introduce simple models of chemical
evolution in the next section.

\section{The nucleosynthesis of Mn}

We now discuss the chemical evolution of the three galaxies of our sample with
the largest number of stars, Sculptor, Fornax, and Carina, adopting a
differential approach in which we compare models with and without
metal-dependent SNe Ia Mn yields.

\subsection{Models of chemical evolution}

\begin{figure}[t!]
\includegraphics[height=7.5cm]{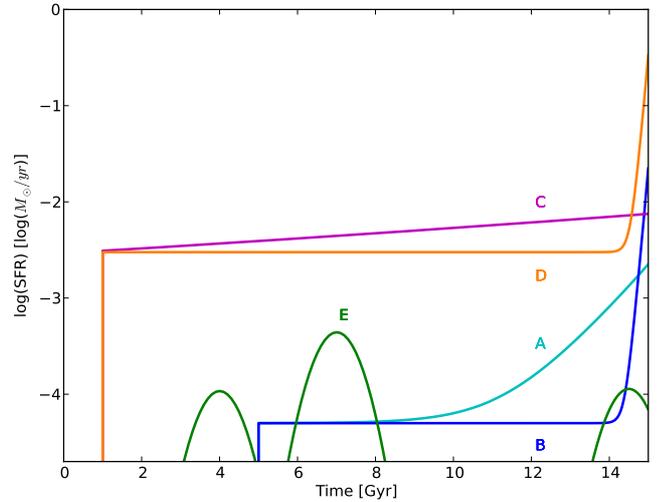}
\caption{The SF histories for the analyzed models.
In cyan model A and in blue model B, for Sculptor. In magenta model C
and in orange model D, for Fornax. In dark green, model E for Carina.}
\label{SFRhistory}
\end{figure}

Fig.~\ref{SFRhistory} presents the evolution of the star formation rate
(SFR) with time for our five different models.  They are based on the
observed star formation histories (SFHs) of \citet{dBTSO11} for
Sculptor, \citet{CdJ08} for Fornax, and \citet{RHBS03} for Carina.

Models A, C, and E are set up to follow the observations as closely as
possible, whereas  models B and D are extreme cases, with which
we intend to test the influence of the choice of SFH on the results.
These five models attempt to describe
the extremes of the possible SFH for these galaxies, the
true one lying somewhere within these boundaries. Their main characteristics
are summarized in Table~\ref{table:model_param}.

\begin{table*}
\caption{Parameters of our models. The models A, B, ... are further subdivided into
A1, A2, B1, B2 etc.: the X1 models have a SN Ia Mn yield $\propto(Z/Z_\odot)^{0.65}$,
while the X2 models have a constant SN Ia Mn yield (this subdivision is not shown
in the table).}
\label{table:model_param}
\centering
\begin{tabular}{lcccc}
\hline\hline
\multicolumn{1}{c}{Galaxy} &
\multicolumn{1}{c}{Model} &
\multicolumn{1}{c}{Star formation rate (SFR) [M$\odot$/yr]} &
\multicolumn{1}{c}{Final stellar}&
\multicolumn{1}{c}{Initial gas} \\
& & ($t$ is the look-back time expressed in Gyr)& mass [M$\odot$] & mass [M$\odot$] \\
\hline
Sculptor & A &$2.2\times 10^{-3}\exp\left(-\frac{15-t}{1}\right)+5\times10^{-5} ~~(t>5)$
&$\sim 1.5\times10^6$&$2\times10^7$\\
         &    &$ ~~~~~~~~~~~~~~~~~~~~~~~~~~~~~~~~~~~~~~~~~~~~0 ~~~~~~~~~(t <5)$& &  \\
         & B &$2.25\times 10^{-2}\exp\left(-\frac{15-t}{0.1}\right)+5\times10^{-5} ~~(t>5)$
&$\sim 1.5\times10^6$&$2\times10^7$\\
         &    &$ ~~~~~~~~~~~~~~~~~~~~~~~~~~~~~~~~~~~~~~~~~~~~0 ~~~~~~~~~(t <5)$& &  \\ \hline
Fornax   & C &$7.5\times 10^{-3}\exp\left(-\frac{15-t}{10}\right)+3\times10^{-3} ~~(t>1)$
&$\sim 4.5\times10^7$&$3\times10^8$\\
         &    &$ ~~~~~~~~~~~~~~~~~~~~~~~~~~~~~~~~~~~~~~~~~~~~0 ~~~~~~~~~(t <1)$& &  \\
         & D &$3.4\times 10^{-1}\exp\left(-\frac{15-t}{0.1}\right)+3\times10^{-3} ~~(t>1)$
&$\sim 4.5\times10^7$&$3\times10^8$ \\ \hline
 & & & & \\
Carina   & E  &$1.2\times 10^{-4}\exp\left(-\frac{(t-4)^2}{2(0.5)^2}\right)+
                4.8\times 10^{-4}\exp\left(-\frac{(t-7)^2}{2(0.5)^2}\right)+
		1.2\times 10^{-4}\exp\left(-\frac{(t-14.5)^2}{2(0.5)^2}\right)$
&$\sim 0.5\times10^6$&$13\times10^6$\\ \hline
\end{tabular}
\end{table*}

Models A and B refer to the Sculptor dSph. In model A, the SFR is a
decreasing exponential function on a timescale of 1 Gyr.  In model B, the SFR
is also a decreasing exponential function, although on a shorter
timescale of 100 Myr.  Both models have a low star formation rate
tail of $5\cdot10^{-5}$ M$_{\odot}$/yr, stopping 5 Gyr ago. They both
form a similar total mass of stars on the order of $\sim 1.5\cdot
10^{6}$ M$_{\odot}$, from a total initial mass of gas of $2\cdot
10^{7}$ M$_{\odot}$.

Models C and D refer to the Fornax dSph.  Model C assumes an
exponentially decreasing SFR on a long timescale of 10~Gyr, whereas
model D, with an exponentially decreasing SFR on a short timescale of
100~Myr, has an extended tail with a star formation rate of
$3\cdot10^{-3}$ M$_{\odot}$/yr. The evolution of the models was
stopped 1 Gyr ago and has an amplitude of star formation that is ten times
higher than for the Sculptor models. Both Fornax models form a total
mass of stars of $\sim 4.5\cdot 10^{7}$ M$_{\odot}$, from a total
initial mass of gas of $3\cdot 10^{8}$ M$_{\odot}$.

Model E refers to the Carina dSph. The star formation (SF) history is
modeled by three
Gaussian functions, with $\sigma=$500~Myr.  These three Gaussian
functions are centered at look-back times of 4, 7, and 14.5 Gyr, with
respective peak values of $\sim 10^{-4}$ M$_{\odot}$yr${-1}$, $\sim
4\cdot 10^{-4}$ M$_{\odot}$yr$^{-1}$, and $\sim 10^{-4}$
M$_{\odot}$yr$^{-1}$. The Carina model forms a total mass of stars of
$\sim 0.5\cdot 10^{6}$ M$_{\odot}$, out of a $13\cdot 10^{6}$
M$_{\odot}$ total initial mass of gas.

For all these models, we used the initial mass function (IMF) of \citet{K01},
in addition to the
chemical evolution parameters, such as stellar lifetimes and SNe Ia
scheme, of \citet{CMLFC07}.

For the SN Ia rate, which is a key component of our analysis,  
we underline that it was computed following \citet{MG86}, hence
expressed as:
\begin{equation}
R_{SNe Ia}=A\int\limits^{M_{BM}}_{M_{Bm}}\phi(M_{B})
\left[\int\limits^{0.5}_{\mu_{m}}f(\mu)\,\psi(t-\tau_{M_{2}})\,d\mu\right] dM_{B}\,\mathrm{,}
\end{equation}
where $\psi(t)$ is the SFR, $M_{2}$ is the mass of the secondary,
$M_{B}$ is the total mass of the binary system, $\mu=M_{2}/M_{B}$,
$\mu_{m}=max\left[M(t)_{2}/M_{B},(M_{B}-0.5M_{BM})/M_{B}\right]$, 
$M_{Bm}= 3 M_{\odot}$, and $M_{BM}= 16 M_{\odot}$. The IMF is represented
by $\phi(M_{B})$ and refers to the total mass of the binary system when
computing the SNe Ia rate, $f(\mu)$ is the distribution function
for the mass fraction of the secondary
\begin{equation}
f(\mu)=2^{1+\gamma}(1+\gamma)\mu^{\gamma}
\end{equation}
with $\gamma=2$ and $A$ is the fraction of systems in the appropriate mass range
that can give rise to SNe Ia events. This quantity is fixed to 0.05 by
reproducing the observed SNe Ia rate at
the present time \citep{CET99}.

\begin{figure}[t!]
\includegraphics[height=6.5cm]{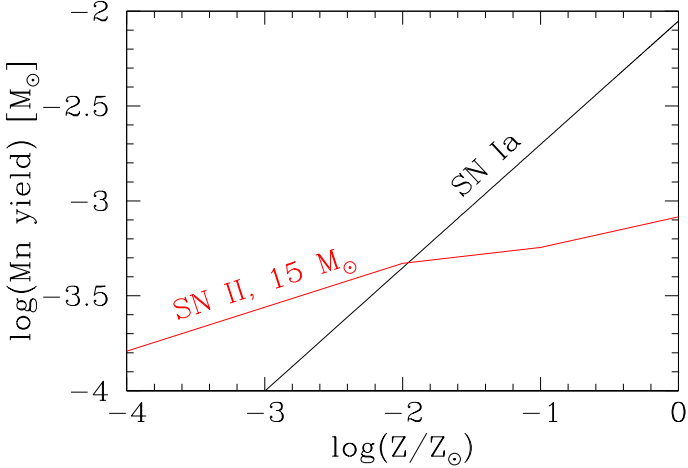}
\caption{Metallicity-dependent Mn yield of SNe Ia
according to the prescription of \citet{CML08} (black line),
and of SNe II with 15 M$_\odot$ progenitors (red line)
according to \citet{WW95}}
\label{yield}
\end{figure}

The metal-dependent yields of Fe and Mn
for SNe II are taken from
\citet{WW95}, with the difference that we halved the iron yields for
SNe II, as suggested by \citet{RKT10}. These yields are represented
by the red curve in Fig.~\ref{yield} for SNe II with a 15~M$_\odot$
progenitor, which is taken 
as representative of the majority of the core-collapse SNe.
We first implemented the hypothesis of \citet{CML08} that the metal dependence
of the Mn SNe Ia yields is y\,$\propto$\,(Z/Z$_{\odot}$)$^{0.65}$ (see the
black line in Fig.~\ref{yield}), which led to the five models
A1, B1, C1, D1, and E1. We then considered the \citet{IBNKU99} metal-independent
Mn yields for a solar metallicity, taking
the SNe Ia yields for iron from \citet{IBNKU99}. This
led to the five additional models A2, B2, C2, D2, and E2.

%Figure~\ref{SFRhistory}b shows the [Mg/Fe] ratio as a function of [Fe/H],
%resulting from the SFH adopted (Fig.~\ref{SFRhistory}). 

\subsection{Does the Mn yield depend on metallicity?}  
Figure~\ref{MnFe} unambiguously demonstrates that regardless of the galaxy and 
the assumed SFH, models for which there is no metallicity dependence for the Mn 
SNe Ia yields (dashed lines) predict a far too high [Mn/Fe].  In contrast,
all five models with a metal-dependence (solid lines) do pass through the
observed data points.   
\begin{figure}[t!]
\centering
\includegraphics[width=9.5cm]{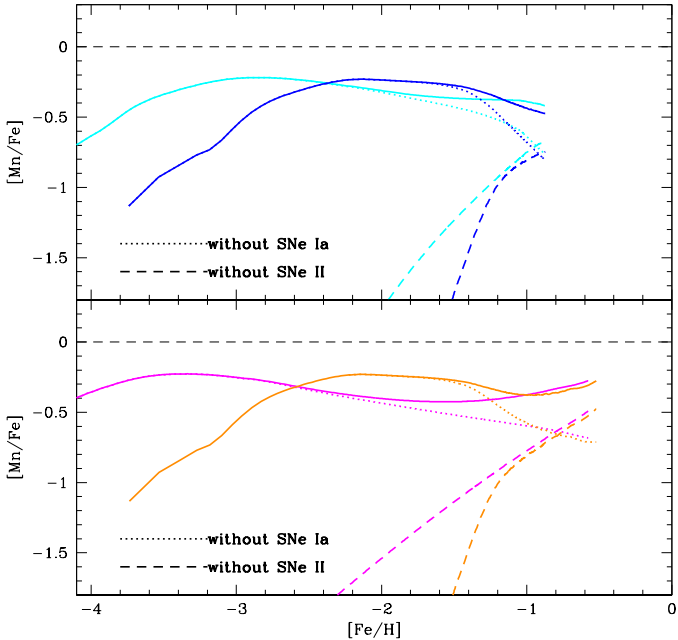}
\caption{Same as Fig.~\ref{MnFe} but for models only, showing the
respective contributions
of the SNe II and Ia to the Mn synthesis. Only the models A1 and B1
for Sculptor (continuous cyan and blue lines respectively), and C1
and D1 for Fornax (continuous magenta and orange lines) are shown. The
dotted lines show the evolution of [Mn/Fe] assuming no Mn is produced
by the SNe Ia, in contrast to Fe and all other elements. The dashed lines
indicate the same evolution, but assuming that no Mn is produced by the SNe II.}
\label{MnFe_models}
\end{figure}
 
Figure~\ref{MnFe_models} indicates the respective contributions of the
SNe II and SNe Ia to the Mn abundance. The plain lines show the
total contribution of both types of SNe, similarly to Fig.~\ref{MnFe}.
The dotted lines show the contribution to Mn of the SNe II {\sl only}.
In other words, the Mn yields of the SNe Ia are switched off, but not
the yield of Fe. Finally, the dashed curves show how [Mn/Fe] evolves
when only the SNe Ia contribute to the Mn abundance.

Below [Fe/H]$\sim -2.5$, [Mn/Fe] is set by the SNe II up to an average
level of $-0.2$ dex. The SNe II Mn ejecta are also metal-dependent and
increase with [Fe/H]. At [Fe/H]$\sim -2.5$, the exploding SNe Ia
produce [Mn/Fe]$_\mathrm{SNe Ia}$ $<$ [Mn/Fe]$_\mathrm{SNe II}$. This
situation holds up to [Fe/H]$\sim -1$, when our nucleosynthesis
predicts for the progenitors of SNe Ia that [Mn/Fe]$_\mathrm{SNe Ia}$ $\sim$
[Mn/Fe]$_\mathrm{SNe II}$. This explains the decreasing trend for
[Mn/Fe] vs [Fe/H] in the metallicity range $-2.5<$[Fe/H]$<-1$ for
Sculptor and Fornax.

The Sculptor model A1, which has the closest SFH to the observations,
results in a shallower decline of [Mn/Fe] with [Fe/H] than observed.
This may well indicate that the form of the assumed metal-dependence
of the SNe Ia yields is not fully correct. We did not try any
fine-tuning at this stage, since our relatively simple models
imply clearly enough that metal-poor SNe
Ia should produce less Mn than metal-rich ones.
  
Unfortunately, the data for Fornax only span a narrow range of [Fe/H],
making it more difficult to check the suitability of the models. The 
model C1 shows an increase in [Mn/Fe] above [Fe/H]$\sim -1$ after 
an initial decrease as in the case of Sculptor.  While the origin 
of the first drop is the same as in the case of Sculptor, the increase for 
[Fe/H]$>-1$ is due to metal-rich SNe Ia progenitors, which are a consequence 
of the longer SFH.  
 
As in Fornax, the data in Carina span a narrow range of metallicities,
except for the three stars with [Fe/H] $< 2.0$ published by \citet{VSI12}.
We see that the model E1 reproduces
the observations quite satisfactorily.
It is interesting to see the consequence of the  bursty Carina SFH. [Mn/Fe]
 decreases  between the first
and the second star formation peaks ($-3.5<$[Fe/H] $<-2.2$), because of
the very low production of Mn by metal-poor SNe Ia.  
The second decrease in [Mn/Fe] (-1.6$<$[Fe/H] $<$-1.4),
at the end of the intermediate age peak of star formation,
is shallower owing to the higher
metallicity of the SNe Ia at that time.

\subsection{Comparison with Lanfranchi's models}

The chemical evolution models adopted here are very similar to
the ones computed by Lanfranchi et al. \citeyearpar{LM03,LMC08}, although there are
two major differences : we do not consider
galactic winds and our SFHs are quite different.
While we adopt  SFHs initially derived from color-magnitude diagrams, Lanfranchi
et al. adjust their SF efficiency until the observations are reproduced.

Galactic winds or any other dynamical effects such as tidal and
ram pressure stripping must have removed the gas in these
dSphs, because none is detected. Moreover, as
shown by Lanfranchi et al., the galactic winds can influence
the chemical evolution at the end of
the evolution of these galaxies, if one considers differential
winds, i.e., that different elements can be expelled
with different  wind efficiencies. Nevertheless, to keep our models as simple as possible,
galactic winds were not an option in our analysis. This does not affect our conclusions.
Indeed the evidence that SN Ia Mn yields depend on metallicity does not arise
from the latest stages of the galaxy chemical evolution, when winds
would play a role, but much earlier.  Moreover, given that Lanfranchi's wind efficiency
is essentially the same for Fe and Mn, [Mn/Fe] is definitely not expected to change.

\section{Conclusion}

On the basis of the three Mn\,\textsc{I}\, lines at $\lambda
5407$, $5420$, and $5516$~\AA, we have derived the stellar
abundances of manganese in three dSph galaxies, Sculptor (50 stars),
Fornax (60 stars), and Carina (6 stars); Mn abundances in a fourth
dSph galaxy, Sextans (5 stars), was based on only one to three Mn
lines. These Mn abundances are corrected for HFSs, the correction   
reaching $1.6$~dex for strong lines (EW$\sim 200$~m\AA).

Our analysis of the relation between the [Mn/Fe] and [Mn/$\alpha$] abundance
ratios and [Fe/H] has highlighted the following features :

$\bullet$ The Mn abundances lead to sub-solar [Mn/Fe] ratios for all
stars in all four of the studied galaxies, as expected from their
low metallicity.

$\bullet$ The variation in [Mn/Fe] with [Fe/H] in Sculptor has
two phases : a plateau at [Fe/H] $< -1.4$, followed by a $\sim
0.3$ dex decrease at higher metallicity.  This decreasing trend of
[Mn/Fe] with [Fe/H] had only been observed previously in the globular
cluster $\omega$ Centauri. In Fornax, there is a marginal suggestion
of an increasing trend, but without any statistical significance.

$\bullet$ Our datasets in four different galaxies, and their
comparison with the case of the Milky Way clearly demonstrates that
the evolution of [Mn/$\alpha$] as a function of [Fe/H] depends on the
galaxy SFH. The variation in [Mn/$\alpha$] can be interpreted
in terms of the balance between the metal-dependent yields of type
II and type Ia supernovae.

$\bullet$ Three simple chemical evolution models for Sculptor,
Fornax, and Carina have been developed. The impacts of the
type II and type Ia Mn
yields, with and without any metal-dependence, have been investigated.
They unambiguously demonstrate that the reproduction of the
observations requires SNe Ia metal-dependent yields. The successive
increase and decrease in [Mn/Fe] as a function of [Fe/H], as well as
the amplitude of these variations, are the result of the increasing
SNe II Mn yields with [Fe/H], combined with initially low SNe Ia
yields that subsequently augment with metallicity.

\begin{acknowledgements}

  PJ, PN, and GC gratefully acknowledge support from the Swiss National Science Foundation.
  The work greatly benefited from the International Space Science Institute (ISSI)
  in Bern, thanks to the funding of the team ``Defining the full life-cycle
  of dwarf galaxy evolution: the Local Universe as a template''.

\end{acknowledgements}

\bibliographystyle{aa}
\bibliography{spectro}
\listofobjects

\appendix
\section{Comparison between MOOG and CALRAI abundances}
The abundances of Mn that were uncorrected for the HFS were computed
with both codes \texttt{calrai} and \texttt{moog} for the same atmosphere
models. Therefore, it is possible to compare the results and check the
consistency between the two codes.

\begin{figure*}[t!]
\centering
\includegraphics[height=6cm]{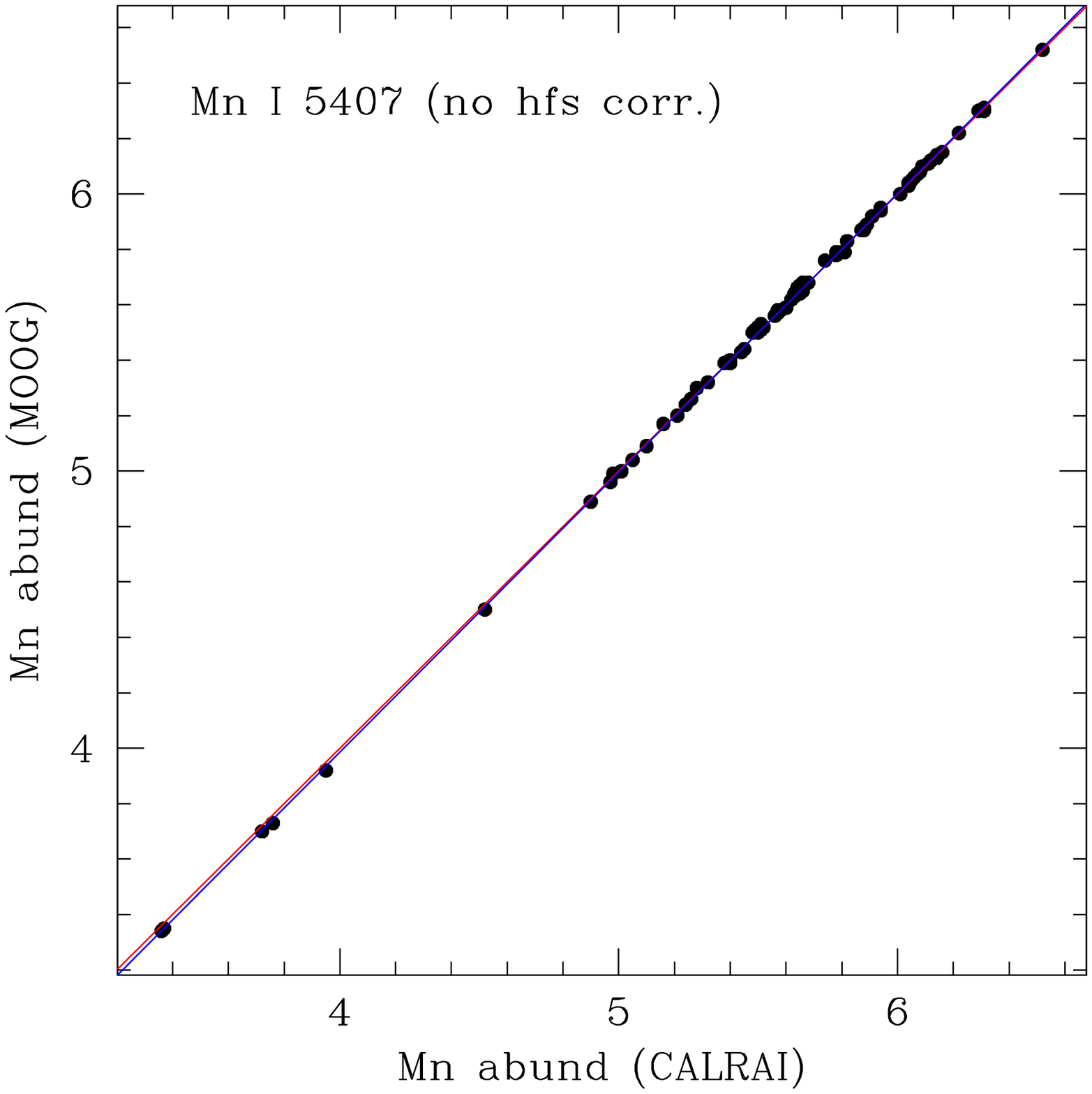}
\includegraphics[height=6cm]{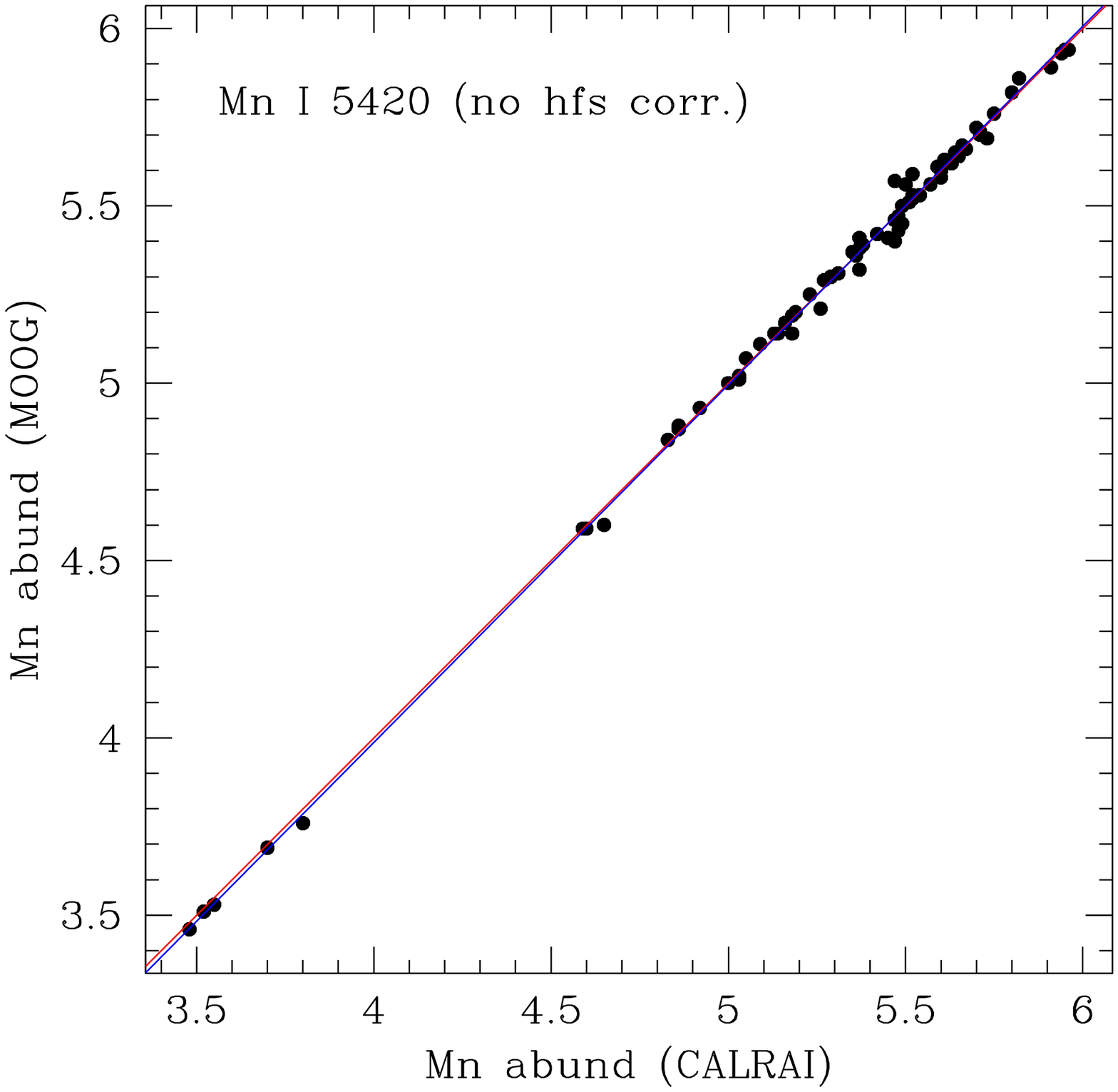}
\includegraphics[height=6cm]{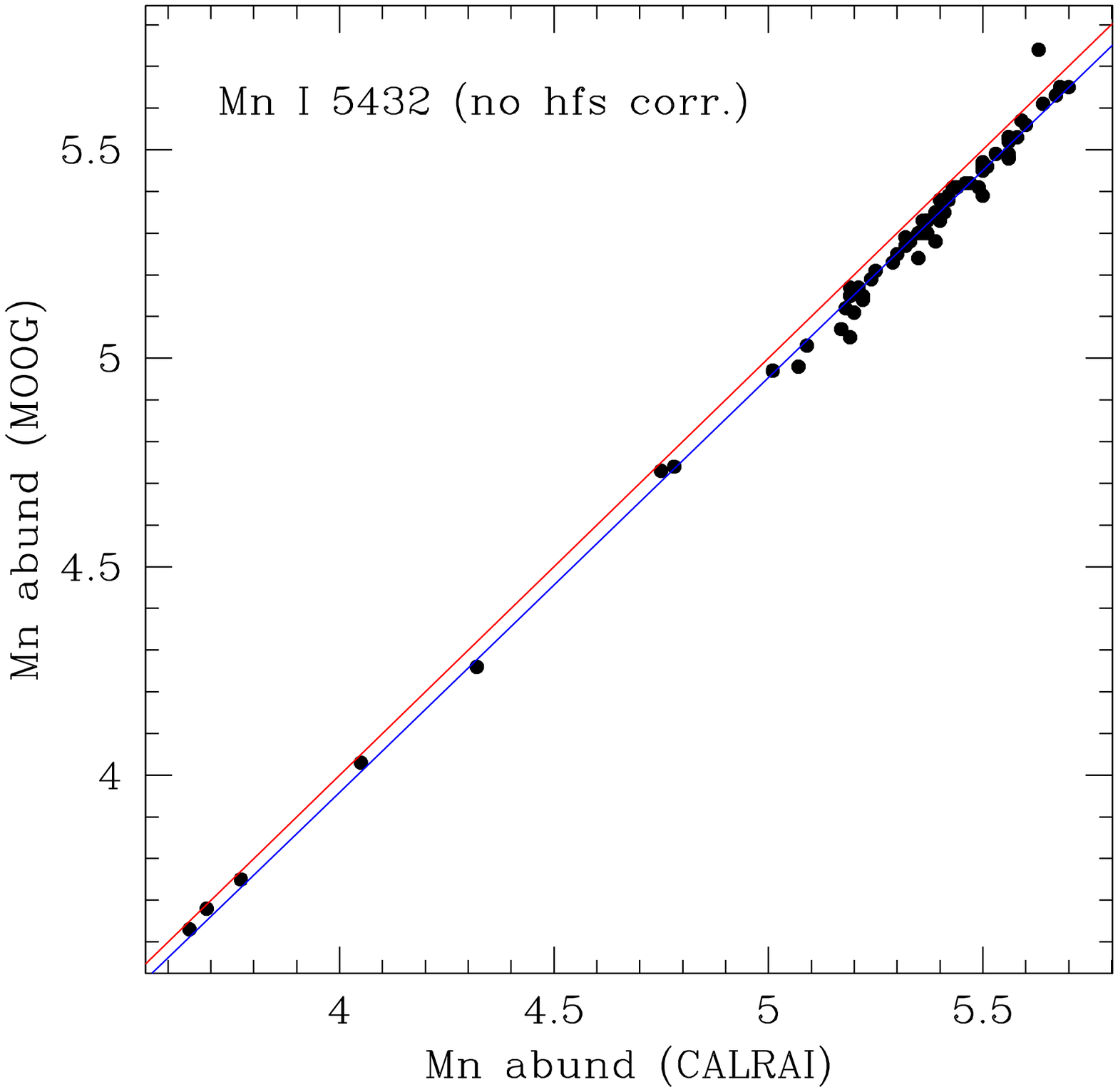}
\includegraphics[height=6cm]{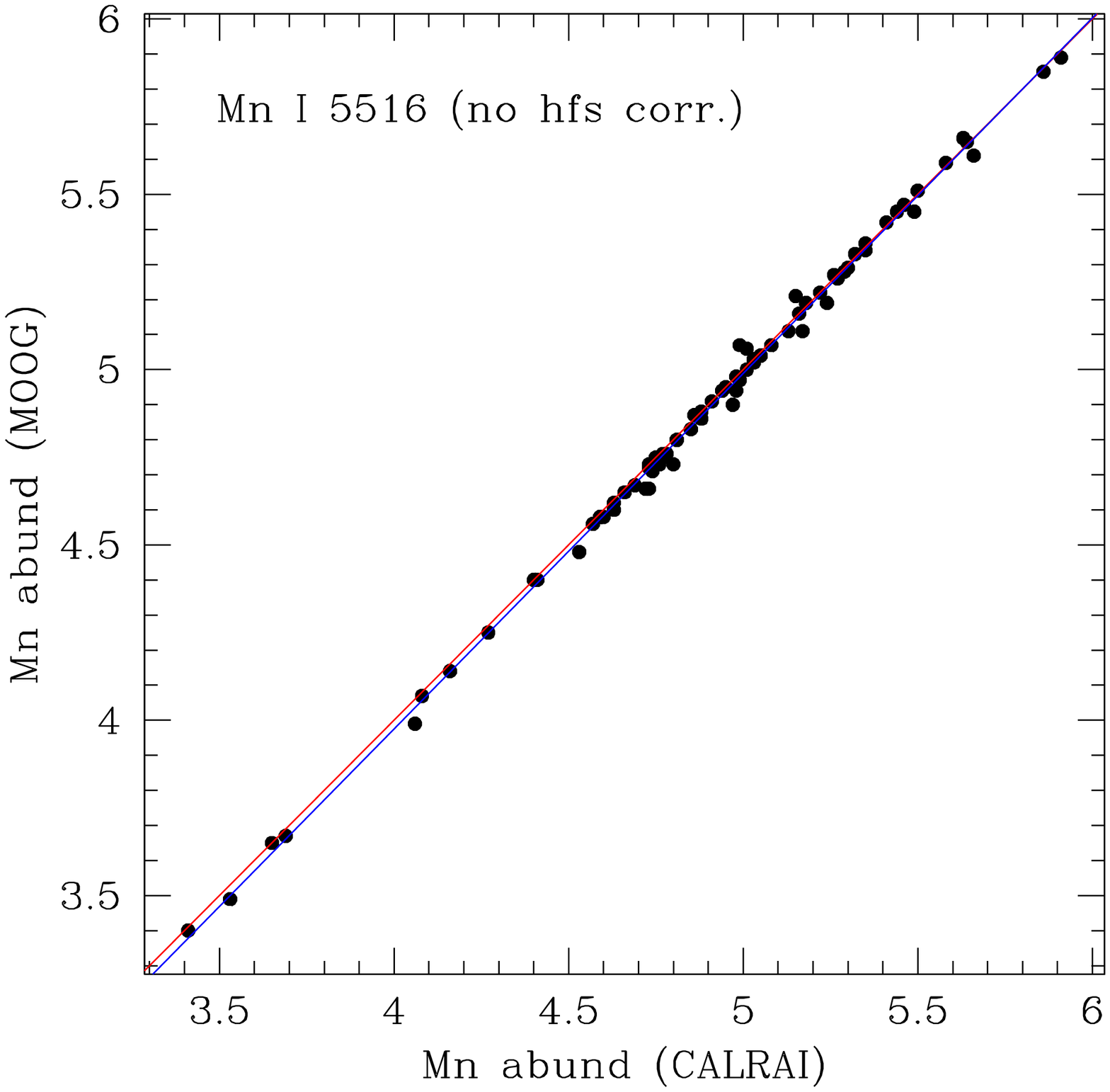}
\caption{Comparison between the Mn abundances (not corrected for hfs) obtained
for the Fornax dSph galaxy using the \texttt{moog} code, and the ones obtained using
the \texttt{calrai} code, for each of the 4 lines Mn\,\textsc{i}\,$\lambda 5407$,
$\lambda 5420$, $\lambda 5432$, and $\lambda 5516$. In both cases, the abundances
were determined using spherical atmosphere models.}
\label{fig:comp_ab_Fnx}
\end{figure*}

For Fornax, the raw (i.e. uncorrected for HFS) Mn abundances given by
the two codes prove to be perfectly consistent (Fig.~\ref{fig:comp_ab_Fnx}).

For Sculptor, however, there is a systematic shift of about 0.1 to 0.2~dex,
in the sense that the \texttt{moog} abundances are lower than the
\texttt{calrai} ones for all four lines. The slopes are very close to 1, but
tend to be slightly above unity.
\begin{figure*}[t!]
\centering
\includegraphics[height=6cm]{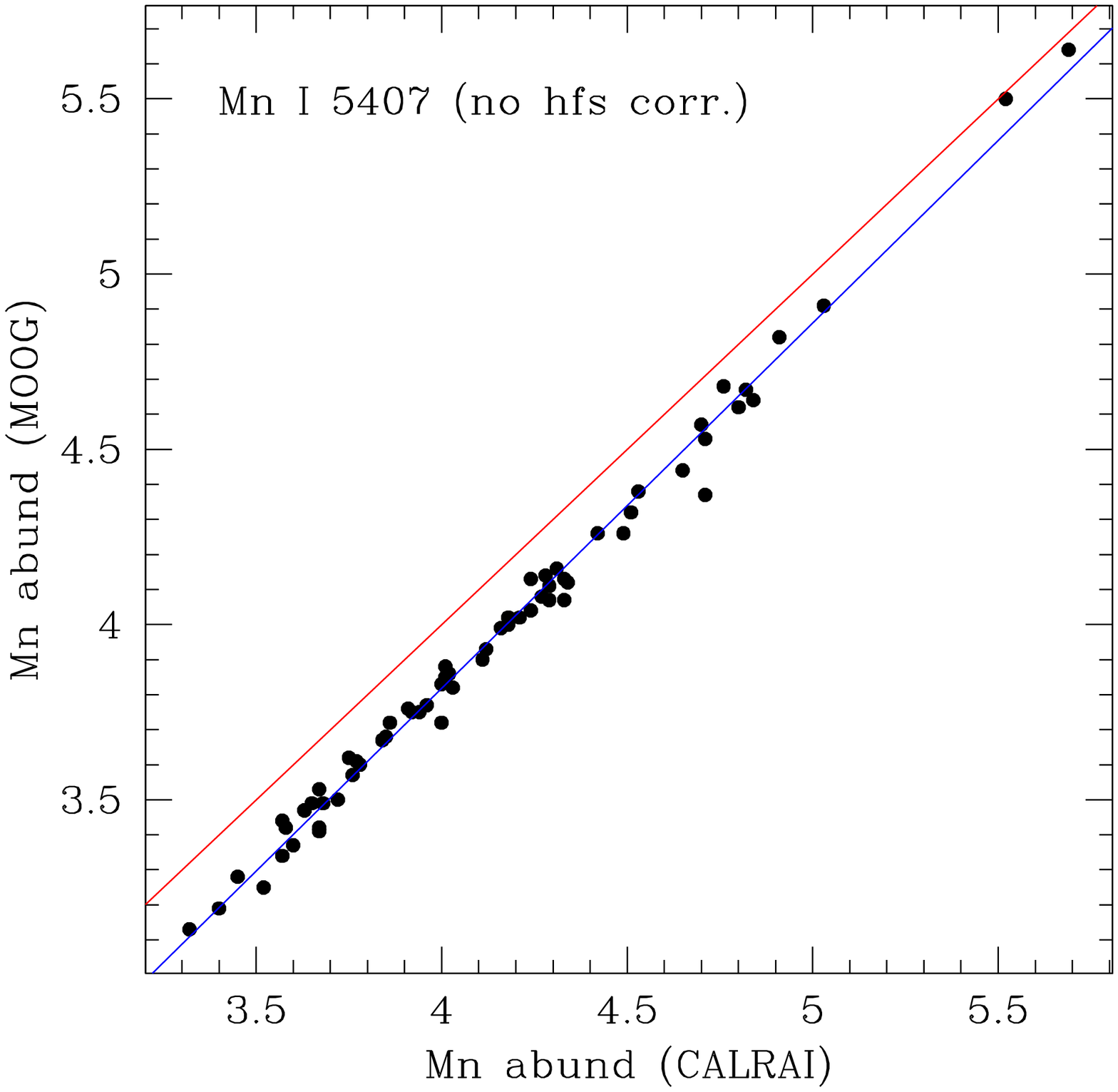}
\includegraphics[height=6cm]{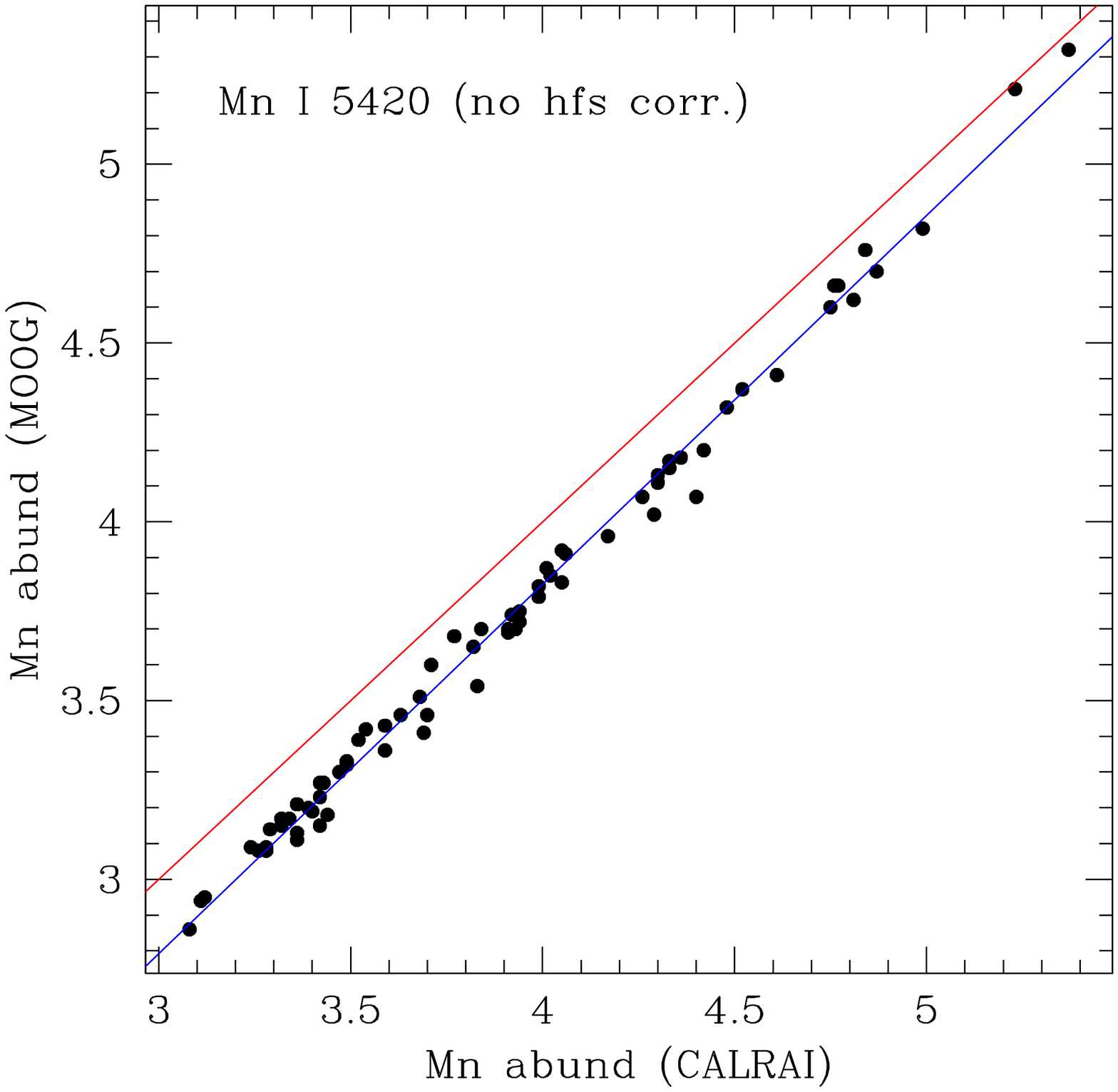}
\includegraphics[height=6cm]{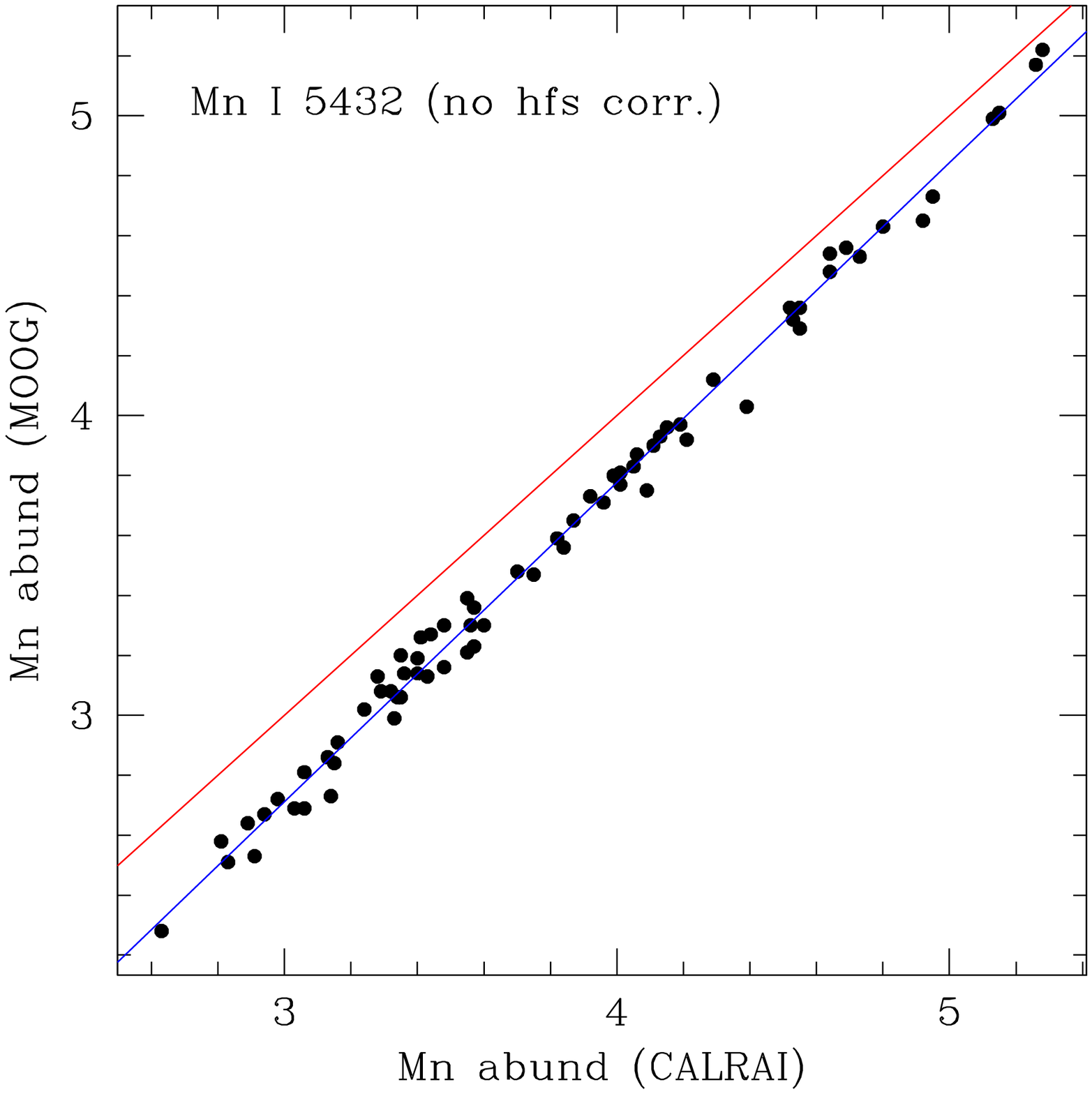}
\includegraphics[height=6cm]{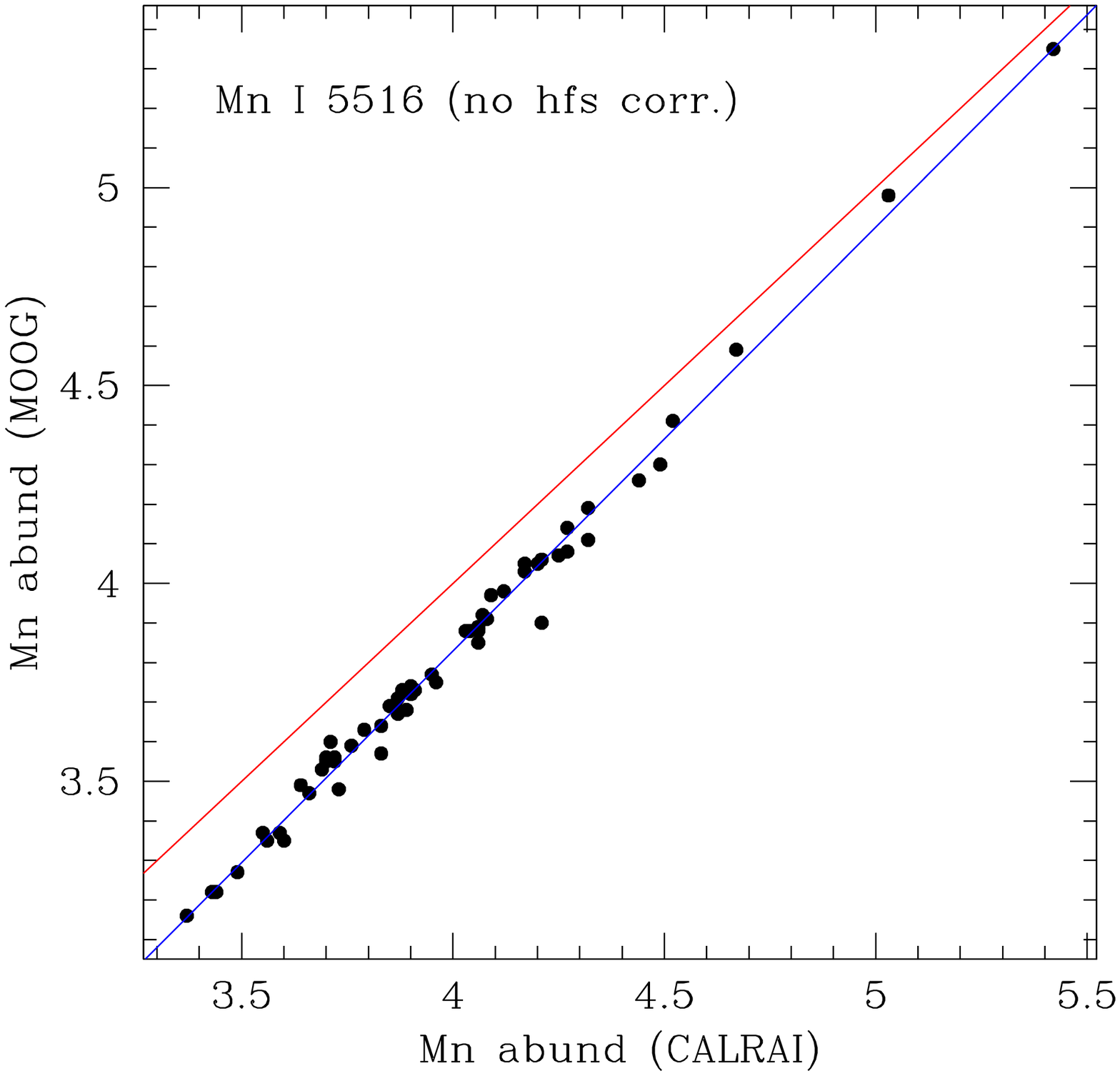}
\caption{Same as Fig.~\ref{fig:comp_ab_Fnx}, but for the Sculptor dSph galaxy.
Here the \texttt{calrai} abundances were determined using plane-parallel
atmosphere models, while the \texttt{moog} abundances are based on spherical
models.}
\label{fig:comp_ab_Scl}
\end{figure*}

The reason why the systematic zero-point shift is much larger in Sculptor
than Fornax lies in the atmosphere models used. While spherical models
were used in connection with the \texttt{moog} spectral synthesis code for
both galaxies, plane-parallel models were used in connection with the
\texttt{calrai} code in the case of Sculptor, leading to the overestimated
abundances seen in Fig.~\ref{fig:comp_ab_Scl}.

%\section{Long Tables}
\Online
\begin{table*}
\caption{Equivalent widths, HFS corrections (labeled ``HFS'' on top of the respective
columns), and Mn abundance for the Mn\,\textsc{i} $\lambda 5407$ and $\lambda 5420$
lines observed for the stars of the Fornax dSph galaxy.}
\label{table:abundFnxa}
\centering
\scriptsize
\begin{tabular}{l|rrrr|rrrr}
    &\multicolumn{4}{c|}{$\lambda 5407$}&\multicolumn{4}{c}{$\lambda 5420$} \\
Star& EW & HFS & [Mn/H] & [Mn/Fe]& EW & HFS  & [Mn/H] & [Mn/Fe] \\ \hline

BL038&$210.9\pm 5.6$&$-1.40$&$-0.99$&$-0.08\pm0.09$&$217.4\pm 3.4$&$-1.45$&$-1.24$&$-0.33\pm0.08$\\
BL045&$152.2\pm 5.6$&$-0.94$&$-1.32$&$-0.27\pm0.06$&$160.5\pm 3.9$&$-1.04$&$-1.57$&$-0.52\pm0.06$\\
BL065&$ 49.3\pm 4.4$&$-0.14$&$-1.81$&$-0.37\pm0.08$&$ 67.8\pm 4.3$&$-0.23$&$-1.92$&$-0.48\pm0.07$\\
BL076&$190.6\pm 4.9$&$-1.26$&$-1.09$&$-0.23\pm0.06$&$188.6\pm 6.8$&$-1.26$&$-1.42$&$-0.56\pm0.07$\\
BL077&$188.5\pm 3.4$&$-1.24$&$-1.13$&$-0.34\pm0.06$&$198.1\pm 6.4$&$-1.34$&$-1.36$&$-0.57\pm0.07$\\
BL081&$218.0\pm 6.3$&$-1.51$&$-0.81$&$-0.18\pm0.09$&$203.5\pm 9.7$&$-1.47$&$-1.26$&$-0.63\pm0.11$\\
BL084&$185.9\pm 6.4$&$-1.31$&$-1.22$&$-0.35\pm0.07$&$202.4\pm 5.8$&$-1.46$&$-1.38$&$-0.51\pm0.07$\\
BL091&$155.0\pm 4.5$&$-0.96$&$-1.25$&$-0.27\pm0.07$&$176.3\pm 5.7$&$-1.21$&$-1.41$&$-0.43\pm0.07$\\
BL092&$153.5\pm 6.3$&$-1.06$&$-1.47$&$-0.55\pm0.08$&$162.5\pm 5.6$&$-1.18$&$-1.71$&$-0.79\pm0.08$\\
BL096&$172.0\pm 6.6$&$-1.18$&$-1.29$&$-0.58\pm0.09$&$192.5\pm27.9$&$-1.38$&$-1.42$&$-0.71\pm0.25$\\
BL097&$189.5\pm 3.0$&$-1.16$&$-1.11$&$-0.19\pm0.06$&$209.3\pm 5.0$&$-1.33$&$-1.24$&$-0.32\pm0.07$\\
BL100&$149.0\pm 7.9$&$-0.91$&$-1.40$&$-0.46\pm0.08$&$174.0\pm 2.6$&$-1.20$&$-1.54$&$-0.60\pm0.06$\\
BL104&$201.4\pm 4.3$&$-1.34$&$-1.05$&$-0.07\pm0.07$&$187.5\pm 6.7$&$-1.25$&$-1.48$&$-0.50\pm0.08$\\
BL113&$203.3\pm 5.3$&$-1.35$&$-0.86$&$-0.10\pm0.08$&$200.1\pm 7.8$&$-1.33$&$-1.18$&$-0.42\pm0.09$\\
BL115&$ 87.1\pm 4.1$&$-0.33$&$-1.77$&$-0.30\pm0.07$&$ 95.8\pm 4.6$&$-0.41$&$-2.00$&$-0.53\pm0.07$\\
BL123&$191.6\pm 4.3$&$-1.19$&$-1.18$&$-0.20\pm0.06$&$203.6\pm 7.0$&$-1.30$&$-1.38$&$-0.40\pm0.08$\\
BL125&$199.9\pm 4.3$&$-1.42$&$-0.99$&$-0.24\pm0.07$&$197.0\pm 6.2$&$-1.42$&$-1.32$&$-0.57\pm0.08$\\
BL132&$220.9\pm 8.6$&$-1.49$&$-0.94$&$-0.02\pm0.10$&$220.8\pm 5.7$&$-1.53$&$-1.27$&$-0.35\pm0.09$\\
BL138&$219.8\pm 5.6$&$-1.39$&$-1.00$&$ 0.01\pm0.08$&$222.4\pm 7.8$&$-1.39$&$-1.26$&$-0.25\pm0.10$\\
BL140&$185.4\pm 5.9$&$-1.30$&$-1.20$&$-0.37\pm0.09$&$195.4\pm 5.7$&$-1.40$&$-1.41$&$-0.58\pm0.09$\\
BL141&$185.1\pm 5.2$&$-1.29$&$-1.11$&$-0.27\pm0.08$&$174.3\pm 6.6$&$-1.20$&$-1.50$&$-0.66\pm0.08$\\
BL146&$198.3\pm 6.4$&$-1.23$&$-1.02$&$-0.09\pm0.08$&$207.8\pm 5.5$&$-1.32$&$-1.24$&$-0.31\pm0.08$\\
BL149&$170.3\pm 4.7$&$-1.06$&$-1.21$&$-0.29\pm0.08$&$181.0\pm 8.1$&$-1.17$&$-1.43$&$-0.51\pm0.09$\\
BL150&$211.7\pm 5.1$&$-1.41$&$-0.93$&$-0.11\pm0.08$&$213.1\pm 5.8$&$-1.43$&$-1.22$&$-0.40\pm0.08$\\
BL151&$200.8\pm 3.5$&$-1.41$&$-1.02$&$-0.14\pm0.08$&$211.6\pm 6.4$&$-1.49$&$-1.22$&$-0.34\pm0.09$\\
BL155&$226.7\pm 7.0$&$-1.40$&$-0.75$&$-0.04\pm0.09$&$210.3\pm 7.9$&$-1.34$&$-1.21$&$-0.50\pm0.10$\\
BL156&$128.7\pm 4.1$&$-0.62$&$-1.49$&$-0.34\pm0.08$&$150.2\pm 6.4$&$-0.86$&$-1.66$&$-0.51\pm0.08$\\
BL158&$189.5\pm 6.9$&$-1.42$&$-1.07$&$-0.22\pm0.10$&$206.1\pm 8.5$&$-1.53$&$-1.22$&$-0.37\pm0.11$\\
BL160&$196.6\pm 4.8$&$-1.31$&$-1.07$&$-0.20\pm0.08$&$203.1\pm 6.7$&$-1.37$&$-1.31$&$-0.44\pm0.09$\\
BL163&$222.8\pm 7.1$&$-1.38$&$-0.73$&$ 0.01\pm0.09$&$221.2\pm 8.4$&$-1.40$&$-1.06$&$-0.32\pm0.11$\\
BL166&$210.3\pm 6.9$&$-1.31$&$-0.89$&$ 0.00\pm0.09$&$215.5\pm 3.5$&$-1.36$&$-1.15$&$-0.26\pm0.08$\\
BL168&$187.1\pm 5.4$&$-1.30$&$-1.06$&$-0.16\pm0.08$&$189.3\pm 5.1$&$-1.35$&$-1.36$&$-0.46\pm0.08$\\
BL171&$204.5\pm 6.9$&$-1.21$&$-1.00$&$-0.08\pm0.08$&$222.0\pm 8.3$&$-1.32$&$-1.14$&$-0.22\pm0.10$\\
BL173&$220.5\pm 8.4$&$-1.37$&$-0.87$&$ 0.00\pm0.11$&$210.7\pm 8.1$&$-1.35$&$-1.27$&$-0.40\pm0.10$\\
BL185&$222.9\pm 8.1$&$-1.52$&$-0.75$&$ 0.02\pm0.10$&$211.2\pm 6.0$&$-1.50$&$-1.18$&$-0.41\pm0.09$\\
BL190&$204.9\pm 7.5$&$-1.28$&$-1.02$&$-0.23\pm0.08$&$208.7\pm 7.8$&$-1.33$&$-1.30$&$-0.51\pm0.09$\\
BL195&$141.1\pm 3.9$&$-0.80$&$-1.22$&$-0.24\pm0.07$&$169.8\pm 5.8$&$-1.13$&$-1.34$&$-0.36\pm0.07$\\
BL196&$187.5\pm 2.9$&$-1.07$&$-1.20$&$-0.16\pm0.06$&$168.1\pm 4.3$&$-0.88$&$-1.62$&$-0.58\pm0.07$\\
BL197&$201.1\pm 8.5$&$-1.34$&$-1.11$&$-0.22\pm0.10$&$198.9\pm 5.5$&$-1.35$&$-1.45$&$-0.56\pm0.08$\\
BL203&$183.5\pm 5.2$&$-1.29$&$-1.18$&$-0.34\pm0.08$&$217.9\pm 7.0$&$-1.51$&$-1.15$&$-0.31\pm0.10$\\
BL205&$207.0\pm 5.7$&$-1.36$&$-0.74$&$-0.03\pm0.08$&$222.1\pm 6.0$&$-1.46$&$-0.90$&$-0.19\pm0.09$\\
BL208&$202.1\pm 4.7$&$-1.42$&$-0.87$&$-0.19\pm0.08$&$201.1\pm 5.0$&$-1.43$&$-1.18$&$-0.50\pm0.08$\\
BL210&$223.6\pm 7.4$&$-1.45$&$-0.77$&$-0.01\pm0.10$&$213.4\pm 5.0$&$-1.44$&$-1.19$&$-0.43\pm0.09$\\
BL211&$237.9\pm 3.5$&$-1.55$&$-0.63$&$-0.01\pm0.07$&$230.0\pm12.5$&$-1.56$&$-1.01$&$-0.39\pm0.16$\\
BL213&$197.1\pm 3.7$&$-1.32$&$-1.07$&$-0.13\pm0.07$&$191.8\pm 6.0$&$-1.28$&$-1.41$&$-0.47\pm0.08$\\
BL216&$228.5\pm 4.1$&$-1.47$&$-0.78$&$ 0.01\pm0.08$&$206.6\pm 6.8$&$-1.41$&$-1.31$&$-0.52\pm0.09$\\
BL218&$241.4\pm 6.9$&$-1.53$&$-0.63$&$ 0.00\pm0.11$&$221.1\pm 5.4$&$-1.59$&$-1.18$&$-0.55\pm0.11$\\
BL221&$175.3\pm 5.8$&$-1.21$&$-1.22$&$-0.39\pm0.08$&$185.9\pm 7.2$&$-1.32$&$-1.44$&$-0.61\pm0.09$\\
BL227&$177.5\pm 8.6$&$-1.32$&$-1.20$&$-0.33\pm0.11$&$155.9\pm 6.2$&$-1.09$&$-1.65$&$-0.78\pm0.09$\\
BL228&$193.8\pm 5.9$&$-1.20$&$-1.14$&$-0.27\pm0.07$&$217.5\pm 6.5$&$-1.39$&$-1.24$&$-0.37\pm0.08$\\
BL229&$230.1\pm 3.3$&$-1.41$&$-0.76$&$-0.08\pm0.07$&$222.2\pm 6.8$&$-1.41$&$-1.15$&$-0.47\pm0.09$\\
BL233&$227.7\pm 3.9$&$-1.46$&$-0.73$&$-0.08\pm0.08$&$204.7\pm 5.3$&$-1.38$&$-1.26$&$-0.61\pm0.09$\\
BL239&$168.3\pm 2.9$&$-1.12$&$-1.19$&$-0.25\pm0.06$&$224.6\pm 8.0$&$-1.54$&$-1.02$&$-0.08\pm0.10$\\
BL242&$163.5\pm 4.2$&$-1.08$&$-1.31$&$-0.17\pm0.08$&$180.7\pm 6.7$&$-1.27$&$-1.48$&$-0.34\pm0.09$\\
BL247&$232.1\pm 8.7$&$-1.43$&$-0.75$&$ 0.09\pm0.12$&$226.2\pm 8.0$&$-1.42$&$-1.10$&$-0.26\pm0.12$\\
BL250&$236.8\pm11.2$&$-1.47$&$-0.74$&$-0.10\pm0.14$&$211.8\pm 9.2$&$-1.44$&$-1.31$&$-0.67\pm0.12$\\
BL253&$238.4\pm 6.5$&$-1.42$&$-0.67$&$ 0.03\pm0.08$&$203.5\pm 5.7$&$-1.31$&$-1.33$&$-0.63\pm0.08$\\
BL257&$238.0\pm 7.5$&$-1.43$&$-0.68$&$-0.14\pm0.11$&$220.4\pm10.9$&$-1.40$&$-1.16$&$-0.62\pm0.13$\\
BL258&$235.7\pm 6.7$&$-1.47$&$-0.64$&$-0.06\pm0.11$&$238.5\pm 7.8$&$-1.51$&$-0.94$&$-0.36\pm0.12$\\
BL260&$210.6\pm 5.9$&$-1.25$&$-0.98$&$-0.12\pm0.08$&$190.0\pm 7.3$&$-1.10$&$-1.46$&$-0.60\pm0.08$\\
BL261&$185.0\pm 4.4$&$-1.38$&$-1.13$&$-0.35\pm0.10$&$185.7\pm 5.0$&$-1.40$&$-1.42$&$-0.64\pm0.10$\\
BL266&$ 65.1\pm 5.1$&$-0.19$&$-1.82$&$-0.36\pm0.08$&$ 71.1\pm 2.1$&$-0.22$&$-2.06$&$-0.60\pm0.06$\\
BL267&$208.4\pm 7.1$&$-1.47$&$-0.80$&$-0.06\pm0.10$&$201.4\pm 4.4$&$-1.43$&$-1.16$&$-0.42\pm0.08$\\
BL269&$189.4\pm 5.6$&$-1.43$&$-1.16$&$-0.36\pm0.09$&$204.6\pm 4.8$&$-1.53$&$-1.31$&$-0.51\pm0.09$\\
BL278&$252.4\pm 8.0$&$-1.47$&$-0.55$&$ 0.18\pm0.11$&$222.2\pm 5.4$&$-1.42$&$-1.14$&$-0.41\pm0.09$\\
BL279&$ 31.5\pm 2.7$&$-0.09$&$-2.11$&$-0.59\pm0.09$&$            $&$	 $&$	 $&$	    	$\\
BL295&$239.1\pm 8.1$&$-1.43$&$-0.71$&$-0.05\pm0.11$&$229.3\pm 4.8$&$-1.44$&$-1.12$&$-0.46\pm0.09$\\
BL300&$195.9\pm 4.9$&$-1.31$&$-1.13$&$-0.21\pm0.08$&$196.5\pm 6.6$&$-1.33$&$-1.43$&$-0.51\pm0.09$\\
BL304&$204.6\pm 5.5$&$-1.29$&$-1.10$&$-0.13\pm0.08$&$217.0\pm 4.7$&$-1.39$&$-1.30$&$-0.33\pm0.08$\\
BL315&$169.4\pm 6.5$&$-1.42$&$-1.16$&$-0.38\pm0.10$&$199.3\pm10.1$&$-1.65$&$-1.22$&$-0.44\pm0.14$\\
BL323&$204.9\pm 7.1$&$-1.29$&$-1.16$&$-0.24\pm0.09$&$192.7\pm 6.6$&$-1.21$&$-1.57$&$-0.65\pm0.08$\\
\hline
\end{tabular}
\end{table*}

\begin{table*}
\caption{Equivalent widths, HFS corrections (labeled ``HFS'' on top of the respective
columns), and Mn abundance for the Mn\,\textsc{i} $\lambda 5432$ and $\lambda 5516$
lines observed for the stars of the Fornax dSph galaxy.}
\label{table:abundFnxb}
\centering
\scriptsize
\begin{tabular}{l|rrrr|rrrr}
    &\multicolumn{4}{c|}{$\lambda 5432$}&\multicolumn{4}{c}{$\lambda 5516$} \\
Star& EW & HFS & [Mn/H] & [Mn/Fe]& EW & HFS  & [Mn/H] & [Mn/Fe] \\ \hline

BL038&$287.6\pm 7.9$&$-0.59$&$-0.56$&$ 0.35\pm0.13$&$169.4\pm12.8$&$-0.85$&$-1.02$&$-0.11\pm0.14$\\
BL045&$247.7\pm19.3$&$-0.81$&$-0.84$&$ 0.21\pm0.28$&$ 99.3\pm 5.9$&$-0.34$&$-1.46$&$-0.41\pm0.07$\\
BL065&$120.2 \pm2.3$&$-0.42$&$-2.04$&$-0.60\pm0.07$&$ 36.1\pm 3.9$&$-0.08$&$-1.82$&$-0.38\pm0.09$\\
BL076&$280.9\pm 6.9$&$-0.63$&$-0.49$&$ 0.37\pm0.11$&$145.4\pm 6.9$&$-0.67$&$-1.15$&$-0.29\pm0.08$\\
BL077&$272.6\pm 8.6$&$-0.68$&$-0.65$&$ 0.14\pm0.12$&$145.0\pm 3.6$&$-0.66$&$-1.18$&$-0.39\pm0.06$\\
BL081&$291.3\pm18.2$&$-0.54$&$-0.25$&$ 0.38\pm0.23$&$171.6\pm 7.7$&$-0.95$&$-0.88$&$-0.25\pm0.10$\\
BL084&$252.2\pm 7.0$&$-0.74$&$-0.93$&$-0.06\pm0.13$&$137.9\pm 5.1$&$-0.66$&$-1.29$&$-0.42\pm0.06$\\
BL091&$215.8\pm 6.6$&$-0.86$&$-1.24$&$-0.26\pm0.12$&$117.2\pm 6.0$&$-0.47$&$-1.27$&$-0.29\pm0.08$\\
BL092&$251.6\pm 8.7$&$-0.73$&$-0.91$&$ 0.01\pm0.14$&$132.7\pm 6.2$&$-0.69$&$-1.35$&$-0.43\pm0.09$\\
BL096&$289.0\pm10.7$&$-0.54$&$-0.37$&$ 0.34\pm0.16$&$141.2\pm 6.9$&$-0.70$&$-1.23$&$-0.52\pm0.10$\\
BL097&$271.4\pm 6.6$&$-0.70$&$-0.73$&$ 0.19\pm0.11$&$141.7\pm 6.0$&$-0.56$&$-1.18$&$-0.26\pm0.07$\\
BL100&$249.7\pm 9.3$&$-0.77$&$-0.87$&$ 0.07\pm0.13$&$111.8\pm 5.3$&$-0.44$&$-1.43$&$-0.49\pm0.07$\\
BL104&$264.5\pm 9.7$&$-0.72$&$-0.81$&$ 0.17\pm0.13$&$138.6\pm 7.3$&$-0.61$&$-1.26$&$-0.28\pm0.09$\\
BL113&$263.5\pm 4.8$&$-0.76$&$-0.57$&$ 0.19\pm0.10$&$135.5\pm 7.8$&$-0.55$&$-1.09$&$-0.33\pm0.09$\\
BL115&$179.6\pm 8.8$&$-0.82$&$-1.89$&$-0.42\pm0.13$&$ 47.3\pm 2.8$&$-0.10$&$-1.96$&$-0.49\pm0.07$\\
BL123&$254.6\pm 6.0$&$-0.73$&$-1.03$&$-0.05\pm0.11$&$147.1\pm 9.1$&$-0.60$&$-1.21$&$-0.23\pm0.09$\\
BL125&$266.5\pm10.1$&$-0.70$&$-0.58$&$ 0.17\pm0.14$&$174.4\pm 7.0$&$-0.97$&$-0.86$&$-0.11\pm0.10$\\
BL132&$291.7\pm 7.6$&$-0.52$&$-0.54$&$ 0.38\pm0.11$&$162.9\pm 9.8$&$-0.88$&$-1.14$&$-0.22\pm0.11$\\
BL138&  	    &$     $&$     $& 	      	   &$175.0\pm 9.6$&$-0.84$&$-1.05$&$-0.04\pm0.11$\\
BL140&$262.5\pm 5.5$&$-0.71$&$-0.77$&$ 0.06\pm0.10$&$136.1\pm 3.5$&$-0.65$&$-1.29$&$-0.46\pm0.08$\\
BL141&$255.0\pm 8.7$&$-0.76$&$-0.74$&$ 0.10\pm0.13$&$144.6\pm 3.9$&$-0.72$&$-1.13$&$-0.29\pm0.07$\\
BL146&$255.6\pm 8.1$&$-0.73$&$-0.88$&$ 0.05\pm0.12$&$138.2\pm 4.0$&$-0.54$&$-1.20$&$-0.27\pm0.07$\\
BL149&$241.8\pm 9.3$&$-0.78$&$-0.95$&$-0.03\pm0.14$&$112.2\pm10.1$&$-0.39$&$-1.37$&$-0.45\pm0.11$\\
BL150&$293.8\pm12.9$&$-0.55$&$-0.38$&$ 0.44\pm0.18$&$155.6\pm 8.2$&$-0.74$&$-1.08$&$-0.26\pm0.10$\\
BL151&$261.2\pm 8.0$&$-0.72$&$-0.74$&$ 0.14\pm0.12$&$148.7\pm 3.6$&$-0.75$&$-1.13$&$-0.25\pm0.08$\\
BL155&$294.8\pm 8.2$&$-0.57$&$-0.37$&$ 0.34\pm0.13$&$170.4\pm 6.2$&$-0.79$&$-0.91$&$-0.20\pm0.09$\\
BL156&$214.4\pm 8.6$&$-0.81$&$-1.42$&$-0.27\pm0.14$&$ 95.4\pm 5.9$&$-0.28$&$-1.51$&$-0.36\pm0.08$\\
BL158&$255.9\pm11.1$&$-0.73$&$-0.65$&$ 0.20\pm0.18$&$ 85.5\pm37.5$&$-0.28$&$-1.59$&$-0.74\pm0.33$\\
BL160&$264.5\pm12.8$&$-0.73$&$-0.77$&$ 0.10\pm0.18$&$151.6\pm 4.8$&$-0.72$&$-1.13$&$-0.26\pm0.08$\\
BL163&  	    &$     $&$     $& 	      	   &$166.5\pm 7.5$&$-0.74$&$-0.89$&$-0.15\pm0.10$\\
BL166&$269.7\pm 7.4$&$-0.71$&$-0.71$&$ 0.18\pm0.13$&$151.8\pm 9.3$&$-0.65$&$-1.09$&$-0.20\pm0.11$\\
BL168&$274.6\pm10.2$&$-0.65$&$-0.44$&$ 0.46\pm0.16$&$133.5\pm 6.1$&$-0.61$&$-1.19$&$-0.29\pm0.09$\\
BL171&  	    &$     $&$     $& 	      	   &$138.1\pm 5.4$&$-0.50$&$-1.23$&$-0.31\pm0.08$\\
BL173&  	    &$     $&$     $& 	      	   &$ 95.4\pm64.5$&$-0.25$&$-1.58$&$-0.71\pm0.58$\\
BL185&$272.8\pm 7.5$&$-0.67$&$-0.50$&$ 0.27\pm0.13$&$182.3\pm 4.6$&$-1.03$&$-0.78$&$-0.01\pm0.09$\\
BL190&$287.4\pm 6.9$&$-0.61$&$-0.60$&$ 0.19\pm0.11$&$145.0\pm 6.4$&$-0.59$&$-1.21$&$-0.42\pm0.07$\\
BL195&$191.7\pm 6.9$&$-0.80$&$-1.44$&$-0.46\pm0.11$&$108.8\pm 4.2$&$-0.40$&$-1.22$&$-0.24\pm0.07$\\
BL196&$258.7\pm 7.5$&$-0.73$&$-1.05$&$-0.01\pm0.11$&$133.6\pm 8.2$&$-0.46$&$-1.32$&$-0.28\pm0.09$\\
BL197&$297.8\pm 8.8$&$-0.52$&$-0.47$&$ 0.42\pm0.12$&$159.5\pm 5.8$&$-0.78$&$-1.14$&$-0.25\pm0.08$\\
BL203&$297.4\pm24.2$&$-0.49$&$-0.24$&$ 0.60\pm0.30$&$152.5\pm10.3$&$-0.78$&$-1.09$&$-0.25\pm0.12$\\
BL205&$248.6\pm 6.6$&$-0.79$&$-0.62$&$ 0.09\pm0.10$&$158.4\pm 3.5$&$-0.73$&$-0.82$&$-0.11\pm0.07$\\
BL208&$272.1\pm 6.0$&$-0.67$&$-0.36$&$ 0.32\pm0.11$&$133.6\pm17.8$&$-0.60$&$-1.11$&$-0.43\pm0.16$\\
BL210&$276.2\pm12.0$&$-0.66$&$-0.55$&$ 0.21\pm0.18$&$175.1\pm 8.8$&$-0.90$&$-0.88$&$-0.12\pm0.11$\\
BL211&$294.5\pm 7.4$&$-0.52$&$-0.24$&$ 0.38\pm0.11$&$172.6\pm14.9$&$-0.96$&$-0.91$&$-0.29\pm0.17$\\
BL213&$270.2\pm 4.8$&$-0.71$&$-0.71$&$ 0.23\pm0.09$&$141.3\pm 5.7$&$-0.61$&$-1.20$&$-0.26\pm0.08$\\
BL216&$297.9\pm 5.3$&$-0.54$&$-0.37$&$ 0.42\pm0.09$&$191.9\pm 9.9$&$-1.01$&$-0.74$&$ 0.05\pm0.12$\\
BL218&  	    &$     $&$     $& 	      	   &$182.3\pm 9.2$&$-1.11$&$-0.87$&$-0.24\pm0.14$\\
BL221&$237.3\pm 4.6$&$-0.81$&$-1.02$&$-0.19\pm0.10$&$140.3\pm 4.8$&$-0.68$&$-1.19$&$-0.36\pm0.07$\\
BL227&$266.0\pm18.0$&$-0.66$&$-0.55$&$ 0.32\pm0.28$&$160.1\pm 8.1$&$-0.94$&$-1.01$&$-0.14\pm0.12$\\
BL228&$278.9\pm 8.7$&$-0.67$&$-0.74$&$ 0.13\pm0.16$&$134.6\pm10.3$&$-0.52$&$-1.32$&$-0.45\pm0.09$\\
BL229&$274.8\pm 8.2$&$-0.65$&$-0.67$&$ 0.01\pm0.11$&$178.1\pm 3.8$&$-0.84$&$-0.88$&$-0.20\pm0.07$\\
BL233&$271.8\pm 5.6$&$-0.68$&$-0.61$&$ 0.04\pm0.09$&$160.3\pm 4.3$&$-0.78$&$-1.01$&$-0.36\pm0.08$\\
BL239&$257.9\pm 5.0$&$-0.76$&$-0.65$&$ 0.29\pm0.10$&$126.4\pm 4.0$&$-0.55$&$-1.22$&$-0.28\pm0.07$\\
BL242&$239.2\pm 5.2$&$-0.82$&$-1.02$&$ 0.12\pm0.10$&$131.3\pm 5.9$&$-0.59$&$-1.25$&$-0.11\pm0.08$\\
BL247&$281.5\pm21.8$&$-0.65$&$-0.61$&$ 0.23\pm0.29$&$140.4\pm14.7$&$-0.56$&$-1.21$&$-0.37\pm0.15$\\
BL250&  	    &$     $&$     $& 	      	   &$157.8\pm 5.8$&$-0.75$&$-1.13$&$-0.49\pm0.10$\\
BL253&$292.1\pm 8.2$&$-0.60$&$-0.50$&$ 0.20\pm0.13$&$167.7\pm12.9$&$-0.77$&$-0.99$&$-0.29\pm0.13$\\
BL257&  	    &$     $&$     $& 	      	   &$174.7\pm 8.9$&$-0.82$&$-0.92$&$-0.38\pm0.11$\\
BL258&$283.8\pm 7.1$&$-0.62$&$-0.45$&$ 0.13\pm0.12$&$179.1\pm 6.8$&$-0.93$&$-0.83$&$-0.25\pm0.11$\\
BL260&$268.2\pm 9.9$&$-0.68$&$-0.88$&$-0.02\pm0.13$&$146.2\pm 3.1$&$-0.54$&$-1.19$&$-0.33\pm0.07$\\
BL261&  	    &$     $&$     $& 	      	   &$156.5\pm12.0$&$-0.88$&$-1.01$&$-0.23\pm0.15$\\
BL266&$156.3\pm 7.8$&$-0.64$&$-1.98$&$-0.52\pm0.10$&$ 49.9\pm 2.4$&$-0.11$&$-1.81$&$-0.35\pm0.07$\\
BL267&$236.9\pm14.1$&$-0.84$&$-0.83$&$-0.09\pm0.21$&$139.3\pm10.3$&$-0.65$&$-1.05$&$-0.31\pm0.11$\\
BL269&$254.0\pm 7.1$&$-0.69$&$-0.76$&$ 0.04\pm0.11$&$164.4\pm 9.9$&$-0.98$&$-1.02$&$-0.22\pm0.13$\\
BL278&  	    &$     $&$     $& 	      	   &$210.9\pm14.3$&$-1.06$&$-0.54$&$ 0.19\pm0.18$\\
BL279&$111.4\pm 5.7$&$-0.52$&$-2.22$&$-0.70\pm0.10$&		  &$     $&$	 $&	    	 \\
BL295&  	    &$     $&$     $& 	      	   &$211.2\pm10.8$&$-1.06$&$-0.59$&$ 0.07\pm0.14$\\
BL300&$281.0\pm 6.6$&$-0.63$&$-0.63$&$ 0.29\pm0.11$&$158.8\pm13.4$&$-0.77$&$-1.11$&$-0.19\pm0.14$\\
BL304&$270.9\pm 9.6$&$-0.67$&$-0.89$&$ 0.08\pm0.13$&$161.1\pm 7.9$&$-0.71$&$-1.13$&$-0.16\pm0.09$\\
BL315&$220.5\pm 9.7$&$-0.89$&$-0.96$&$-0.18\pm0.19$&$108.0\pm 4.6$&$-0.54$&$-1.33$&$-0.55\pm0.10$\\
BL323&$291.9\pm 6.9$&$-0.58$&$-0.76$&$ 0.16\pm0.11$&$144.2\pm 9.1$&$-0.59$&$-1.35$&$-0.43\pm0.10$\\
\hline
\end{tabular}
\end{table*}

\begin{table*}
\caption{Equivalent widths, HFS corrections (labeled ``HFS'' on top of the respective
columns), and Mn abundance for the Mn\,\textsc{i} $\lambda 5407$ and $\lambda 5420$
lines observed for the stars of the Sculptor dSph galaxy.}
\label{table:abundScla}
\centering
\scriptsize
\begin{tabular}{l|rrrr|rrrr}
    &\multicolumn{4}{c|}{$\lambda 5407$}&\multicolumn{4}{c}{$\lambda 5420$} \\
Star& EW & HFS & [Mn/H] & [Mn/Fe]& EW & HFS  & [Mn/H] & [Mn/Fe] \\ \hline

ET009&$ 52.0\pm 3.7$&$-0.11$&$-1.83$&$-0.17\pm0.07$&$ 50.0\pm 3.3$&$-0.10$&$-2.13$&$-0.47\pm0.07$\\
ET013&$ 35.0\pm 3.9$&$-0.12$&$-1.84$&$-0.18\pm0.09$&$ 38.0\pm12.1$&$-0.14$&$-2.09$&$-0.43\pm0.18$\\
ET024&$150.0\pm 5.5$&$-0.84$&$-1.52$&$-0.29\pm0.07$&$183.0\pm 4.9$&$-1.19$&$-1.59$&$-0.36\pm0.07$\\
ET026&$ 	   $&$     $&$     $&$	      	  $&$ 28.0\pm 4.2$&$-0.08$&$-2.35$&$-0.57\pm0.09$\\
ET027&$ 57.0\pm 2.2$&$-0.13$&$-1.80$&$-0.32\pm0.05$&$ 69.0\pm 2.6$&$-0.17$&$-1.97$&$-0.49\pm0.05$\\
ET028&$135.0\pm 3.8$&$-0.80$&$-1.39$&$-0.20\pm0.04$&$155.0\pm 4.7$&$-1.04$&$-1.56$&$-0.37\pm0.04$\\
ET031&$ 	   $&$     $&$     $&$	      	  $&$ 39.0\pm 4.8$&$-0.08$&$-2.05$&$-0.40\pm0.08$\\
ET033&$ 	   $&$     $&$     $&$	      	  $&$ 24.0\pm 5.5$&$-0.06$&$-2.34$&$-0.59\pm0.12$\\
ET039&$ 32.0\pm 4.7$&$-0.13$&$-1.58$&$ 0.51\pm0.13$&$		 $&$	 $&$	 $&$	    	$\\
ET043&$ 74.0\pm 4.0$&$-0.32$&$-1.53$&$-0.31\pm0.06$&$ 97.0\pm 4.0$&$-0.54$&$-1.63$&$-0.41\pm0.06$\\
ET048&$ 	   $&$     $&$     $&$	      	  $&$ 17.0\pm 4.6$&$-0.04$&$-2.17$&$-0.29\pm0.13$\\
ET051&$165.0\pm 2.6$&$-1.49$&$-1.36$&$-0.46\pm0.07$&$165.0\pm 4.1$&$-1.49$&$-1.65$&$-0.75\pm0.08$\\
ET054&$ 26.0\pm 3.8$&$-0.06$&$-2.00$&$-0.21\pm0.08$&$ 35.0\pm 3.1$&$-0.09$&$-2.14$&$-0.35\pm0.06$\\
ET057&$109.0\pm 3.2$&$-0.58$&$-1.44$&$-0.13\pm0.05$&$114.0\pm 1.6$&$-0.63$&$-1.69$&$-0.38\pm0.04$\\
ET059&$ 65.0\pm 3.7$&$-0.18$&$-1.57$&$-0.06\pm0.05$&$ 73.0\pm 4.1$&$-0.22$&$-1.78$&$-0.27\pm0.05$\\
ET060&$ 	   $&$     $&$     $&$	      	  $&$ 60.0\pm 2.9$&$-0.21$&$-1.91$&$-0.37\pm0.04$\\
ET063&$111.0\pm 4.6$&$-0.71$&$-1.45$&$-0.29\pm0.08$&$134.0\pm 3.5$&$-1.03$&$-1.61$&$-0.45\pm0.07$\\
ET064&$ 85.0\pm10.0$&$-0.34$&$-1.52$&$-0.16\pm0.09$&$110.0\pm 2.9$&$-0.58$&$-1.64$&$-0.28\pm0.04$\\
ET066&$ 68.0\pm 4.2$&$-0.22$&$-1.58$&$-0.30\pm0.05$&$ 80.0\pm 4.4$&$-0.30$&$-1.76$&$-0.48\pm0.05$\\
ET067&$ 40.0\pm 5.7$&$-0.12$&$-1.83$&$-0.20\pm0.09$&$ 40.0\pm 6.7$&$-0.11$&$-2.11$&$-0.48\pm0.10$\\
ET069&$ 23.0\pm 4.1$&$-0.06$&$-1.82$&$ 0.27\pm0.10$&$ 19.0\pm 5.4$&$-0.05$&$-2.20$&$-0.11\pm0.15$\\
ET071&$ 85.0\pm 5.5$&$-0.41$&$-1.47$&$-0.14\pm0.06$&$ 97.0\pm 5.2$&$-0.55$&$-1.68$&$-0.35\pm0.06$\\
ET073&$ 44.0\pm 7.8$&$-0.13$&$-1.67$&$-0.16\pm0.10$&$ 51.0\pm12.3$&$-0.17$&$-1.88$&$-0.37\pm0.14$\\
ET083&$ 	   $&$     $&$     $&$	      	  $&$ 26.0\pm 2.7$&$-0.06$&$-2.17$&$-0.22\pm0.06$\\
ET094&$ 43.0\pm 5.0$&$-0.09$&$-1.96$&$-0.12\pm0.08$&$ 56.0\pm 4.7$&$-0.13$&$-2.10$&$-0.26\pm0.07$\\
ET097&$ 21.0\pm 3.7$&$-0.04$&$-2.03$&$ 0.11\pm0.08$&$ 35.0\pm 5.1$&$-0.07$&$-2.14$&$-0.25\pm0.08$\\
ET103&$104.0\pm 4.3$&$-0.46$&$-1.34$&$-0.15\pm0.06$&$113.0\pm 6.9$&$-0.55$&$-1.58$&$-0.39\pm0.07$\\
ET104&$ 31.0\pm 4.4$&$-0.08$&$-1.89$&$-0.29\pm0.08$&$ 38.0\pm 4.2$&$-0.11$&$-2.08$&$-0.48\pm0.07$\\
ET109&$ 68.0\pm 3.4$&$-0.14$&$-1.93$&$-0.10\pm0.05$&$ 72.0\pm 2.2$&$-0.15$&$-2.18$&$-0.35\pm0.05$\\
ET121&$ 19.0\pm 2.8$&$-0.05$&$-1.81$&$ 0.52\pm0.09$&$		 $&$	 $&$	 $&$	    	$\\
ET126&$111.0\pm 4.1$&$-0.72$&$-1.41$&$-0.32\pm0.06$&$129.0\pm 4.9$&$-0.98$&$-1.60$&$-0.51\pm0.06$\\
ET132&$ 50.0\pm 4.3$&$-0.17$&$-1.70$&$-0.22\pm0.06$&$ 47.0\pm 4.4$&$-0.15$&$-2.02$&$-0.54\pm0.06$\\
ET133&$127.0\pm 4.7$&$-0.93$&$-1.29$&$-0.24\pm0.06$&$128.0\pm 4.6$&$-0.97$&$-1.60$&$-0.55\pm0.06$\\
ET137&$187.0\pm 6.3$&$-1.60$&$-1.30$&$-0.43\pm0.09$&$185.0\pm 9.0$&$-1.60$&$-1.62$&$-0.75\pm0.10$\\
ET138&$ 28.0\pm 4.3$&$-0.06$&$-2.13$&$-0.45\pm0.09$&$ 42.0\pm 6.0$&$-0.09$&$-2.20$&$-0.52\pm0.09$\\
ET139&$115.0\pm 8.3$&$-0.44$&$-1.54$&$-0.15\pm0.08$&$143.0\pm 5.3$&$-0.68$&$-1.65$&$-0.26\pm0.07$\\
ET141&$ 42.0\pm 3.4$&$-0.10$&$-1.92$&$-0.26\pm0.07$&$ 70.0\pm 4.3$&$-0.24$&$-1.93$&$-0.27\pm0.06$\\
ET147&$106.0\pm11.8$&$-0.56$&$-1.24$&$-0.11\pm0.11$&$105.0\pm 9.9$&$-0.55$&$-1.54$&$-0.41\pm0.10$\\
ET150&$119.0\pm 6.6$&$-0.85$&$-1.48$&$-0.57\pm0.08$&$ 82.0\pm50.2$&$-0.39$&$-2.01$&$-1.10\pm0.43$\\
ET151&$ 50.0\pm 4.3$&$-0.17$&$-1.72$&$ 0.03\pm0.06$&$ 46.0\pm 3.3$&$-0.15$&$-2.05$&$-0.30\pm0.06$\\
ET160&$ 90.0\pm 5.1$&$-0.47$&$-1.44$&$-0.29\pm0.05$&$111.0\pm 5.8$&$-0.72$&$-1.59$&$-0.44\pm0.05$\\
ET164&$ 30.0\pm 5.3$&$-0.07$&$-1.68$&$ 0.19\pm0.10$&$		 $&$	 $&$	 $&$	    	$\\
ET165&$118.0\pm 4.2$&$-0.83$&$-1.31$&$-0.23\pm0.07$&$129.0\pm 6.6$&$-0.98$&$-1.53$&$-0.45\pm0.08$\\
ET166&$ 50.0\pm 4.5$&$-0.16$&$-1.63$&$-0.16\pm0.06$&$ 72.0\pm 5.3$&$-0.31$&$-1.71$&$-0.24\pm0.06$\\
ET168&$110.0\pm 6.8$&$-0.69$&$-1.24$&$-0.16\pm0.08$&$113.0\pm11.9$&$-0.72$&$-1.50$&$-0.42\pm0.10$\\
ET173&$141.0\pm 6.7$&$-0.61$&$-1.51$&$-0.06\pm0.07$&$123.0\pm20.0$&$-0.47$&$-1.92$&$-0.47\pm0.14$\\
ET198&$ 50.0\pm 4.7$&$-0.16$&$-1.55$&$-0.41\pm0.07$&$ 63.0\pm 7.6$&$-0.23$&$-1.70$&$-0.56\pm0.08$\\
ET200&$ 35.0\pm 6.0$&$-0.08$&$-1.80$&$-0.33\pm0.10$&$ 49.0\pm 5.2$&$-0.13$&$-1.89$&$-0.42\pm0.08$\\
ET202&$ 58.0\pm 7.6$&$-0.20$&$-1.48$&$-0.18\pm0.10$&$ 72.0\pm 7.6$&$-0.29$&$-1.63$&$-0.33\pm0.09$\\
ET206&$ 66.0\pm 4.9$&$-0.25$&$-1.40$&$-0.09\pm0.06$&$ 68.0\pm 3.6$&$-0.27$&$-1.67$&$-0.36\pm0.06$\\
ET232&$ 51.0\pm 4.8$&$-0.22$&$-1.30$&$-0.32\pm0.06$&$ 53.0\pm 4.5$&$-0.23$&$-1.56$&$-0.58\pm0.06$\\
ET237&$ 32.0\pm10.5$&$-0.09$&$-1.72$&$-0.13\pm0.18$&$ 57.0\pm 6.5$&$-0.20$&$-1.68$&$-0.09\pm0.10$\\
ET238&$ 	   $&$     $&$     $&$	      	  $&$ 31.0\pm 4.1$&$-0.08$&$-2.07$&$-0.52\pm0.08$\\
ET240&$ 56.0\pm 5.0$&$-0.23$&$-1.60$&$-0.47\pm0.08$&$ 96.0\pm 5.1$&$-0.65$&$-1.56$&$-0.43\pm0.07$\\
ET241&$ 25.0\pm 5.8$&$-0.06$&$-1.88$&$-0.49\pm0.12$&$ 56.0\pm 4.4$&$-0.20$&$-1.75$&$-0.36\pm0.06$\\
ET242&$ 61.0\pm 6.1$&$-0.22$&$-1.49$&$-0.19\pm0.06$&$ 81.0\pm 7.5$&$-0.37$&$-1.59$&$-0.29\pm0.06$\\
ET244&$ 65.0\pm 4.8$&$-0.24$&$-1.36$&$-0.14\pm0.06$&$ 55.0\pm 4.5$&$-0.19$&$-1.76$&$-0.54\pm0.06$\\
ET270&$ 32.0\pm 3.1$&$-0.08$&$-1.84$&$-0.30\pm0.06$&$ 46.0\pm 5.9$&$-0.14$&$-1.94$&$-0.40\pm0.08$\\
ET275&$ 46.0\pm 7.6$&$-0.12$&$-1.50$&$-0.30\pm0.10$&$ 59.0\pm 6.3$&$-0.17$&$-1.62$&$-0.42\pm0.08$\\
ET300&$ 49.0\pm 6.4$&$-0.16$&$-1.54$&$-0.17\pm0.10$&$ 69.0\pm 6.2$&$-0.29$&$-1.63$&$-0.26\pm0.09$\\
ET317&$ 27.0\pm 3.4$&$-0.07$&$-1.81$&$-0.14\pm0.08$&$ 32.0\pm 4.2$&$-0.08$&$-2.00$&$-0.33\pm0.08$\\
ET320&$ 34.0\pm 3.7$&$-0.10$&$-1.58$&$ 0.11\pm0.07$&$ 20.0\pm 6.5$&$-0.04$&$-2.11$&$-0.42\pm0.17$\\
ET321&$ 	   $&$     $&$     $&$	      	  $&$ 17.0\pm 4.6$&$-0.04$&$-2.35$&$-0.44\pm0.14$\\
ET327&$ 66.0\pm 5.2$&$-0.26$&$-1.49$&$-0.18\pm0.07$&$ 75.0\pm 4.3$&$-0.33$&$-1.70$&$-0.39\pm0.06$\\
ET339&$106.0\pm 5.5$&$-0.65$&$-1.22$&$-0.15\pm0.06$&$117.0\pm 3.7$&$-0.80$&$-1.44$&$-0.37\pm0.06$\\
ET342&$ 55.0\pm 6.8$&$-0.22$&$-1.37$&$-0.04\pm0.09$&$ 41.0\pm10.2$&$-0.13$&$-1.81$&$-0.48\pm0.14$\\
ET354&$ 48.0\pm 5.1$&$-0.20$&$-1.31$&$-0.26\pm0.07$&$ 49.0\pm 7.9$&$-0.20$&$-1.58$&$-0.53\pm0.10$\\
ET363&$ 54.0\pm 5.3$&$-0.14$&$-1.20$&$ 0.06\pm0.07$&$ 47.0\pm 9.9$&$-0.12$&$-1.60$&$-0.34\pm0.12$\\
ET369&$ 	   $&$     $&$     $&$	      	  $&$ 16.0\pm 3.8$&$-0.04$&$-2.14$&$ 0.19\pm0.13$\\
ET376&$ 81.0\pm 5.4$&$-0.29$&$-1.34$&$-0.19\pm0.08$&$ 94.0\pm 6.1$&$-0.38$&$-1.48$&$-0.33\pm0.08$\\
ET378&$ 76.0\pm 4.1$&$-0.36$&$-1.46$&$-0.30\pm0.08$&$ 92.0\pm 4.4$&$-0.55$&$-1.64$&$-0.48\pm0.08$\\
ET379&$ 29.0\pm 3.2$&$-0.08$&$-1.70$&$-0.07\pm0.09$&$ 29.0\pm 6.6$&$-0.08$&$-1.98$&$-0.35\pm0.13$\\
ET382&$ 36.0\pm 4.7$&$-0.14$&$-1.57$&$ 0.15\pm0.09$&$		 $&$	 $&$	 $&$	    	$\\
ET384&$ 28.0\pm 5.0$&$-0.09$&$-1.73$&$-0.29\pm0.11$&$ 32.0\pm 8.4$&$-0.11$&$-1.96$&$-0.52\pm0.15$\\
ET389&$ 	   $&$     $&$     $&$	      	  $&$ 33.0\pm 6.6$&$-0.11$&$-2.07$&$-0.49\pm0.13$\\
ET392&$ 53.0\pm11.0$&$-0.19$&$-1.40$&$ 0.06\pm0.13$&$ 24.0\pm 8.3$&$-0.06$&$-2.09$&$-0.63\pm0.18$\\
\hline
\end{tabular}
\end{table*}

\begin{table*}
\caption{Equivalent widths, HFS corrections (labeled ``HFS'' on top of the respective
columns), and Mn abundance for the Mn\,\textsc{i} $\lambda 5432$ and $\lambda 5516$
lines observed for the stars of the Sculptor dSph galaxy.}
\label{table:abundSclb}
\centering
\scriptsize
\begin{tabular}{l|rrrr|rrrr}
    &\multicolumn{4}{c|}{$\lambda 5432$}&\multicolumn{4}{c}{$\lambda 5516$} \\
Star& EW & HFS & [Mn/H] & [Mn/Fe]& EW & HFS  & [Mn/H] & [Mn/Fe] \\ \hline

ET009&$131.0\pm 8.0$&$-0.34$&$-2.25$&$-0.59\pm0.09$&$ 29.0\pm 4.8$&$-0.04$&$-1.99$&$-0.33\pm0.10$\\
ET013&$ 52.0\pm 9.4$&$-0.12$&$-2.60$&$-0.94\pm0.13$&$ 36.0\pm 4.3$&$-0.11$&$-1.67$&$-0.01\pm0.09$\\
ET024&$250.0\pm 8.8$&$-0.79$&$-1.26$&$-0.03\pm0.15$&$129.0\pm 3.5$&$-0.51$&$-1.41$&$-0.18\pm0.07$\\
ET026&$ 67.0\pm 3.9$&$-0.14$&$-2.55$&$-0.77\pm0.06$&$		 $&$	 $&$	 $&$	    	$\\
ET027&$148.0\pm13.0$&$-0.44$&$-2.08$&$-0.60\pm0.13$&$ 36.0\pm 2.0$&$-0.06$&$-1.89$&$-0.41\pm0.05$\\
ET028&$201.0\pm15.5$&$-0.92$&$-1.58$&$-0.39\pm0.23$&$106.0\pm 3.6$&$-0.42$&$-1.37$&$-0.18\pm0.04$\\
ET031&$ 87.0\pm 3.0$&$-0.14$&$-2.19$&$-0.54\pm0.06$&$ 28.0\pm 3.9$&$-0.05$&$-1.78$&$-0.13\pm0.09$\\
ET033&$ 51.0\pm 5.0$&$-0.09$&$-2.59$&$-0.84\pm0.07$&$		 $&$	 $&$	 $&$	    	$\\
ET039&$ 28.0\pm 6.2$&$-0.05$&$-2.41$&$-0.32\pm0.15$&$		 $&$	 $&$	 $&$	    	$\\
ET043&$130.0\pm12.1$&$-0.63$&$-1.96$&$-0.74\pm0.14$&$ 59.0\pm 7.7$&$-0.18$&$-1.49$&$-0.27\pm0.09$\\
ET048&$            $&$     $&$     $&$	      	  $&$		 $&$	 $&$	 $&$	    	$\\
ET051&$240.0\pm16.4$&$-0.73$&$-0.84$&$ 0.06\pm0.26$&$134.0\pm 7.0$&$-0.92$&$-1.28$&$-0.38\pm0.09$\\
ET054&$ 63.0\pm 4.8$&$-0.12$&$-2.45$&$-0.66\pm0.06$&$		 $&$	 $&$	 $&$	    	$\\
ET057&$180.0\pm14.4$&$-0.89$&$-1.64$&$-0.33\pm0.20$&$ 69.0\pm 5.5$&$-0.19$&$-1.55$&$-0.24\pm0.06$\\
ET059&$118.0\pm 5.4$&$-0.33$&$-2.15$&$-0.64\pm0.06$&$ 30.0\pm 3.0$&$-0.05$&$-1.84$&$-0.33\pm0.06$\\
ET060&$107.0\pm 8.9$&$-0.40$&$-2.24$&$-0.70\pm0.08$&$ 37.0\pm 5.4$&$-0.08$&$-1.74$&$-0.20\pm0.08$\\
ET063&$188.0\pm15.4$&$-1.07$&$-1.51$&$-0.35\pm0.26$&$ 86.0\pm 5.1$&$-0.36$&$-1.43$&$-0.27\pm0.08$\\
ET064&$149.0\pm13.9$&$-0.64$&$-1.92$&$-0.56\pm0.16$&$ 58.0\pm 4.3$&$-0.14$&$-1.58$&$-0.22\pm0.06$\\
ET066&$136.0\pm12.0$&$-0.53$&$-1.96$&$-0.68\pm0.12$&$ 48.0\pm 4.4$&$-0.11$&$-1.61$&$-0.33\pm0.06$\\
ET067&$ 91.0\pm10.4$&$-0.27$&$-2.26$&$-0.63\pm0.10$&$ 26.0\pm 3.3$&$-0.05$&$-1.89$&$-0.26\pm0.08$\\
ET069&$            $&$     $&$     $&$	      	  $&$		 $&$	 $&$	 $&$	    	$\\
ET071&$133.0\pm 5.0$&$-0.65$&$-1.99$&$-0.66\pm0.06$&$ 50.0\pm 5.7$&$-0.13$&$-1.62$&$-0.29\pm0.08$\\
ET073&$ 69.0\pm 4.3$&$-0.16$&$-2.26$&$-0.75\pm0.06$&$ 28.0\pm 3.5$&$-0.07$&$-1.76$&$-0.25\pm0.07$\\
ET083&$ 37.0\pm 5.3$&$-0.05$&$-2.61$&$-0.66\pm0.09$&$		 $&$	 $&$	 $&$	    	$\\
ET094&$111.0\pm 5.1$&$-0.21$&$-2.46$&$-0.62\pm0.06$&$		 $&$	 $&$	 $&$	    	$\\
ET095&$ 33.0\pm 4.5$&$-0.04$&$-2.80$&$-0.66\pm0.07$&$ 17.0\pm 2.4$&$-0.02$&$-1.98$&$ 0.16\pm0.07$\\
ET097&$ 33.0\pm 4.5$&$-0.04$&$-2.80$&$-0.49\pm0.05$&$		 $&$	 $&$	 $&$	    	$\\
ET103&$ 73.0\pm 3.3$&$-0.12$&$-2.38$&$-0.43\pm0.19$&$ 64.0\pm 7.2$&$-0.15$&$-1.48$&$-0.29\pm0.09$\\
ET104&$174.0\pm15.0$&$-0.78$&$-1.62$&$-0.51\pm0.06$&$ 29.0\pm 2.9$&$-0.06$&$-1.76$&$-0.16\pm0.06$\\
ET109&$ 92.0\pm 5.5$&$-0.29$&$-2.11$&$-0.57\pm0.10$&$ 48.0\pm 2.2$&$-0.06$&$-1.96$&$-0.13\pm0.05$\\
ET121&$            $&$     $&$     $&$	   	  $&$		 $&$	 $&$	 $&$	    	$\\
ET126&$168.0\pm 3.9$&$-0.98$&$-1.68$&$-0.59\pm0.07$&$ 83.0\pm 5.5$&$-0.35$&$-1.42$&$-0.33\pm0.07$\\
ET132&$ 94.0\pm 4.5$&$-0.32$&$-2.16$&$-0.68\pm0.06$&$ 32.0\pm 2.2$&$-0.07$&$-1.76$&$-0.28\pm0.05$\\
ET133&$190.0\pm18.0$&$-1.04$&$-1.30$&$-0.25\pm0.30$&$ 79.0\pm 5.6$&$-0.31$&$-1.43$&$-0.38\pm0.07$\\
ET137&$253.0\pm17.4$&$-0.64$&$-0.88$&$-0.01\pm0.26$&$164.0\pm10.3$&$-1.16$&$-1.13$&$-0.26\pm0.13$\\
ET138&$127.0\pm11.8$&$-0.36$&$-2.19$&$-0.51\pm0.11$&$		 $&$	 $&$	 $&$	    	$\\
ET139&$213.0\pm15.8$&$-0.79$&$-1.63$&$-0.24\pm0.22$&$ 88.0\pm 7.4$&$-0.21$&$-1.54$&$-0.15\pm0.09$\\
ET141&$114.0\pm 4.0$&$-0.35$&$-2.31$&$-0.65\pm0.06$&$ 24.0\pm 5.0$&$-0.04$&$-2.06$&$-0.40\pm0.11$\\
ET147&$154.0\pm21.8$&$-0.79$&$-1.79$&$-0.66\pm0.27$&$ 67.0\pm 8.2$&$-0.20$&$-1.38$&$-0.25\pm0.11$\\
ET150&$212.0\pm16.4$&$-0.94$&$-1.07$&$-0.16\pm0.28$&$108.0\pm 4.7$&$-0.61$&$-1.33$&$-0.42\pm0.07$\\
ET151&$ 74.0\pm 4.1$&$-0.17$&$-2.40$&$-0.65\pm0.05$&$		 $&$	 $&$	 $&$	    	$\\
ET160&$153.0\pm14.4$&$-0.86$&$-1.73$&$-0.58\pm0.18$&$ 64.0\pm 5.6$&$-0.21$&$-1.48$&$-0.33\pm0.06$\\
ET164&$ 41.0\pm 4.8$&$-0.06$&$-2.30$&$-0.43\pm0.08$&$		 $&$	 $&$	 $&$	    	$\\
ET165&$160.0\pm 6.1$&$-0.91$&$-1.66$&$-0.58\pm0.10$&$ 90.0\pm 5.0$&$-0.41$&$-1.28$&$-0.20\pm0.07$\\
ET166&$116.0\pm14.6$&$-0.49$&$-1.89$&$-0.42\pm0.15$&$ 34.0\pm 5.1$&$-0.08$&$-1.68$&$-0.21\pm0.09$\\
ET168&$151.0\pm33.9$&$-0.84$&$-1.70$&$-0.62\pm0.41$&$ 70.0\pm 6.7$&$-0.23$&$-1.35$&$-0.27\pm0.09$\\
ET173&$212.0\pm 7.3$&$-0.74$&$-1.92$&$-0.47\pm0.10$&$ 87.0\pm 7.9$&$-0.19$&$-1.71$&$-0.26\pm0.08$\\
ET198&$110.0\pm12.6$&$-0.43$&$-1.81$&$-0.67\pm0.13$&$ 45.0\pm 6.1$&$-0.11$&$-1.44$&$-0.30\pm0.09$\\
ET200&$ 74.0\pm 9.5$&$-0.14$&$-2.17$&$-0.70\pm0.10$&$		 $&$	 $&$	 $&$	    	$\\
ET202&$114.0\pm 8.2$&$-0.46$&$-1.84$&$-0.54\pm0.10$&$ 35.0\pm 3.1$&$-0.08$&$-1.60$&$-0.30\pm0.08$\\
ET206&$101.0\pm10.3$&$-0.36$&$-1.93$&$-0.62\pm0.11$&$ 27.0\pm 5.0$&$-0.06$&$-1.73$&$-0.42\pm0.10$\\
ET232&$ 83.0\pm 3.1$&$-0.36$&$-1.62$&$-0.64\pm0.05$&$ 36.0\pm 3.7$&$-0.11$&$-1.33$&$-0.35\pm0.07$\\
ET237&$            $&$     $&$     $&$	      	  $&$ 28.0\pm 5.5$&$-0.06$&$-1.62$&$-0.03\pm0.12$\\
ET238&$ 69.0\pm 8.5$&$-0.16$&$-2.20$&$-0.65\pm0.09$&$ 38.0\pm 4.1$&$-0.09$&$-1.52$&$ 0.03\pm0.07$\\
ET240&$146.0\pm12.8$&$-0.98$&$-1.57$&$-0.44\pm0.19$&$ 59.0\pm 4.6$&$-0.22$&$-1.40$&$-0.27\pm0.08$\\
ET241&$ 69.0\pm 6.3$&$-0.17$&$-2.12$&$-0.73\pm0.08$&$		 $&$	 $&$	 $&$	    	$\\
ET242&$125.0\pm12.1$&$-0.57$&$-1.77$&$-0.47\pm0.12$&$ 31.0\pm 5.7$&$-0.07$&$-1.70$&$-0.40\pm0.10$\\
ET244&$ 88.0\pm 7.1$&$-0.26$&$-1.95$&$-0.73\pm0.08$&$ 34.0\pm 4.2$&$-0.08$&$-1.57$&$-0.35\pm0.08$\\
ET270&$ 77.0\pm 4.1$&$-0.20$&$-2.19$&$-0.65\pm0.06$&$ 37.0\pm 5.4$&$-0.09$&$-1.61$&$-0.07\pm0.09$\\
ET275&$ 90.0\pm11.8$&$-0.19$&$-1.74$&$-0.54\pm0.12$&$ 41.0\pm 5.7$&$-0.08$&$-1.40$&$-0.20\pm0.09$\\
ET299&$ 26.0\pm 5.6$&$-0.04$&$-2.02$&$-0.20\pm0.12$&$		 $&$	 $&$	 $&$	    	$\\
ET300&$ 62.0\pm10.4$&$-0.14$&$-2.18$&$-0.81\pm0.13$&$ 25.0\pm 3.5$&$-0.06$&$-1.74$&$-0.37\pm0.10$\\
ET317&$ 58.0\pm 8.3$&$-0.12$&$-2.19$&$-0.52\pm0.10$&$		 $&$	 $&$	 $&$	    	$\\
ET320&$ 18.0\pm 4.2$&$-0.02$&$-2.60$&$-0.91\pm0.12$&$		 $&$	 $&$	 $&$	    	$\\
ET321&$ 51.0\pm 7.0$&$-0.08$&$-2.41$&$-0.50\pm0.09$&$ 20.0\pm 4.2$&$-0.04$&$-1.84$&$ 0.07\pm0.11$\\
ET327&$113.0\pm 5.4$&$-0.46$&$-1.93$&$-0.62\pm0.07$&$ 50.0\pm 5.2$&$-0.13$&$-1.48$&$-0.17\pm0.07$\\
ET339&$134.0\pm 6.4$&$-0.66$&$-1.76$&$-0.69\pm0.09$&$ 62.0\pm 5.4$&$-0.19$&$-1.38$&$-0.31\pm0.07$\\
ET342&$ 43.0\pm 7.8$&$-0.08$&$-2.19$&$-0.86\pm0.12$&$ 42.0\pm 7.4$&$-0.12$&$-1.34$&$-0.01\pm0.11$\\
ET354&$            $&$     $&$     $&$	      	  $&$ 31.0\pm 6.8$&$-0.09$&$-1.39$&$-0.34\pm0.13$\\
ET363&$ 56.0\pm 8.1$&$-0.08$&$-1.87$&$-0.61\pm0.10$&$ 23.0\pm 7.3$&$-0.04$&$-1.58$&$-0.32\pm0.17$\\
ET369&$            $&$     $&$     $&$	      	  $&$ 16.0\pm 3.1$&$-0.03$&$-1.70$&$ 0.63\pm0.10$\\
ET376&$125.0\pm 8.4$&$-0.40$&$-1.70$&$-0.55\pm0.10$&$ 67.0\pm 6.1$&$-0.17$&$-1.31$&$-0.16\pm0.09$\\
ET378&$126.0\pm 4.5$&$-0.66$&$-1.90$&$-0.74\pm0.09$&$ 30.0\pm 4.7$&$-0.07$&$-1.82$&$-0.66\pm0.11$\\
ET379&$            $&$     $&$     $&$	      	  $&$		 $&$	 $&$	 $&$	    	$\\
ET382&$ 26.0\pm 6.8$&$-0.06$&$-2.51$&$-0.79\pm0.15$&$ 27.0\pm 6.6$&$-0.08$&$-1.56$&$ 0.16\pm0.14$\\
ET384&$ 56.0\pm 9.2$&$-0.16$&$-2.07$&$-0.63\pm0.12$&$		 $&$	 $&$	 $&$	    	$\\
ET389&$ 58.0\pm 7.0$&$-0.15$&$-2.30$&$-0.72\pm0.11$&$		 $&$	 $&$	 $&$	    	$\\
ET392&$ 87.0\pm 8.7$&$-0.27$&$-1.79$&$-0.33\pm0.10$&$ 28.0\pm 7.4$&$-0.06$&$-1.57$&$-0.11\pm0.15$\\
\hline
\end{tabular}
\end{table*}

\begin{table*}
\caption{Equivalent widths, HFS corrections (labeled ``HFS'' on top of the respective
columns), and Mn abundance for the Mn\,\textsc{i} $\lambda 5407$, $\lambda 5420$, $\lambda 5432$,
and $\lambda 5516$ lines observed for the stars of the Sextans dSph galaxy.}
\label{table:abundSex}
\centering
\scriptsize
\begin{tabular}{l|rrrr|rrrr|rrrr|rrrr}
    &\multicolumn{4}{c|}{$\lambda 5407$}&\multicolumn{4}{c}{$\lambda 5420$}
    &\multicolumn{4}{c|}{$\lambda 5432$}&\multicolumn{4}{c}{$\lambda 5516$} \\
Star& EW & HFS & [Mn/H] & [Mn/Fe]& EW & HFS  & [Mn/H] & [Mn/Fe]
    & EW & HFS & [Mn/H] & [Mn/Fe]& EW & HFS  & [Mn/H] & [Mn/Fe] \\ \hline
 S05\_010&$       $&$	  $&$	  $&$	    $&$       $&$     $&$     $&$	 $&$ 36.0  $&$-0.05$&$-2.59$&$-0.74  $&$       $&$     $&$     $&$	 $\\
         &$       $&$	  $&$	  $&$	    $&$       $&$     $&$     $&$	 $&$\pm 3.3$&$     $&$     $&$\pm0.06$&$       $&$     $&$     $&$	 $\\
 S05\_047&$ 36.0  $&$-0.12$&$-1.40$&$-0.06  $&$       $&$     $&$     $&$	 $&$	   $&$     $&$     $&$       $&$       $&$     $&$     $&$	 $\\
         &$\pm 5.8$&$	  $&$	  $&$\pm0.14$&$       $&$     $&$     $&$	 $&$	   $&$     $&$     $&$       $&$       $&$     $&$     $&$	 $\\
%S07\_083&$       $&$	  $&$	  $&$	    $&$ 54.0  $&$-0.24$&$-1.07$&$ 0.09   $&$	   $&$     $&$     $&$       $&$       $&$     $&$     $&$	 $\\
%        &$       $&$	  $&$	  $&$	    $&$\pm 5.6$&$     $&$     $&$\pm0.13 $&$	   $&$     $&$     $&$       $&$       $&$     $&$     $&$	 $\\
 S08\_003&$       $&$	  $&$	  $&$	    $&$ 29.0  $&$-0.06$&$-2.39$&$-0.50   $&$ 75.0  $&$-0.13$&$-2.58$&$-0.69  $&$       $&$     $&$     $&$	 $\\
         &$       $&$	  $&$	  $&$	    $&$\pm 1.3$&$     $&$     $&$\pm0.03 $&$\pm 1.5$&$     $&$     $&$\pm0.02$&$       $&$     $&$     $&$	 $\\
 S08\_006&$ 54.0  $&$-0.17$&$-1.78$&$-0.38  $&$ 67.0  $&$-0.24$&$-1.93$&$-0.53   $&$106.0  $&$-0.35$&$-2.20$&$-0.80  $&$ 36.0  $&$-0.08$&$-1.82$&$-0.42  $\\
         &$\pm 1.3$&$	  $&$	  $&$\pm0.04$&$\pm 2.0$&$     $&$     $&$\pm0.04 $&$\pm 3.5$&$     $&$     $&$\pm0.07$&$\pm 3.9$&$     $&$     $&$\pm0.08$\\
 S08\_038&$       $&$	  $&$	  $&$	    $&$       $&$     $&$     $&$	 $&$ 33.0  $&$-0.06$&$-2.53$&$-0.54  $&$       $&$     $&$     $&$	 $\\
         &$       $&$	  $&$	  $&$	    $&$       $&$     $&$     $&$	 $&$\pm 2.3$&$     $&$     $&$\pm0.06$&$       $&$     $&$     $&$	 $\\
%S11\_097&$       $&$	  $&$	  $&$	    $&$ 27.0  $&$-0.08$&$-1.87$&$ 0.54   $&$ 48.0  $&$-0.12$&$-1.90$&$ 0.51  $&$ 38.0  $&$-0.12$&$-1.28$&$1.13   $\\
%        &$       $&$	  $&$	  $&$	    $&$\pm 3.5$&$     $&$     $&$\pm0.12 $&$\pm 4.3$&$     $&$     $&$\pm0.12$&$\pm 4.1$&$     $&$     $&$\pm0.12$\\
%S11\_111&$170.0  $&$-1.38$&$-0.63$&$ 1.47  $&$ 96.0  $&$-0.71$&$-1.50$&$ 0.60   $&$ 76.0  $&$-0.27$&$-1.86$&$ 0.24  $&$ 40.0  $&$-0.12$&$-1.38$&$0.72   $\\
%        &$\pm10.9$&$	  $&$	  $&$\pm0.18$&$\pm 8.9$&$     $&$     $&$\pm0.18 $&$\pm 8.7$&$     $&$     $&$\pm0.16$&$\pm18.6$&$     $&$     $&$\pm0.38$\\
\hline
\end{tabular}
\end{table*}

\begin{table*}
\caption{Equivalent widths, HFS corrections (labeled ``HFS'' on top of the respective
columns), and Mn abundance for the Mn\,\textsc{i} $\lambda 5407$, $\lambda 5420$, $\lambda 5432$,
and $\lambda 5516$ lines observed for the stars of the Carina dSph galaxy.}
\label{table:abundCar}
\centering
\scriptsize
\begin{tabular}{l|rrrr|rrrr|rrrr|rrrr}
    &\multicolumn{4}{c|}{$\lambda 5407$}&\multicolumn{4}{c}{$\lambda 5420$}
    &\multicolumn{4}{c|}{$\lambda 5432$}&\multicolumn{4}{c}{$\lambda 5516$} \\
Star& EW & HFS & [Mn/H] & [Mn/Fe]& EW & HFS  & [Mn/H] & [Mn/Fe]
    & EW & HFS & [Mn/H] & [Mn/Fe]& EW & HFS  & [Mn/H] & [Mn/Fe] \\ \hline
MKV\_0458&$       $&$	  $&$	  $&$	    $&$       $&$     $&$     $&$	$&$ 59.0  $&$-0.12$&$-2.17$&$-0.57   $&$      $&$     $&$     $&$       $\\
         &$       $&$	  $&$	  $&$	    $&$       $&$     $&$     $&$	$&$\pm 4.1$&$	  $&$	  $&$\pm 0.09$&$      $&$     $&$     $&$       $\\
MKV\_0484$^1$&$53.3$&$-0.11$&$-1.70$&$-0.17 $&$  79.1 $&$-0.22$&$-1.77$&$-0.24	$&$141.4  $&$-0.35$&$-1.86$&$-0.33   $&$ 30.5 $&$-0.04$&$-1.83$&$-0.30  $\\
         &$\pm 4.5$&$	  $&$	  $&$\pm0.05$&$\pm 2.9$&$     $&$     $&$\pm0.03$&$\pm 5.9$&$	  $&$	  $&$\pm 0.05$&$\pm9.3$&$     $&$     $&$\pm0.16$\\
MKV\_0514&$       $&$     $&$     $&$       $&$       $&$     $&$     $&$	$&$ 22.0  $&$-0.01$&$-2.84$&$-0.52   $&$      $&$     $&$     $&$       $\\
         &$       $&$	  $&$	  $&$       $&$       $&$     $&$     $&$	$&$\pm 2.3$&$  	  $&$	  $&$\pm 0.07$&$      $&$     $&$     $&$       $\\
MKV\_0524$^1$&$38.9$&$-0.07$&$-1.68$&$ 0.07$&$  37.5  $&$-0.06$&$-1.98$&$-0.23	$&$ 86.3  $&$-0.11$&$-2.05$&$-0.30   $&$      $&$     $&$     $&$       $\\
         &$\pm 1.8$&$	  $&$	  $&$\pm0.03$&$\pm 3.5$&$     $&$     $&$\pm0.05$&$\pm 4.5$&$  	  $&$	  $&$\pm 0.05$&$      $&$     $&$     $&$       $\\
MKV\_0596&$ 21.0  $&$-0.05$&$-1.76$&$ -0.22 $&$  31.0 $&$-0.09$&$-1.86$&$ -0.32 $&$ 53.0  $&$-0.12$&$-1.90$&$-0.36   $&$      $&$     $&$     $&$       $\\
         &$\pm 2.6$&$	  $&$	  $&$\pm0.08$&$\pm 3.1$&$     $&$     $&$\pm0.08$&$\pm 3.5$&$	  $&$	  $&$\pm 0.07$&$      $&$     $&$     $&$       $\\
MKV\_0612$^1$&$57.2$&$-0.16$&$-1.48$&$-0.18 $&$  65.6 $&$-0.19$&$-1.68$&$-0.38  $&$134.5  $&$-0.45$&$-1.45$&$-0.15   $&$ 32.4 $&$-0.06$&$-1.61$&$-0.31  $\\
         &$\pm 3.8$&$	  $&$	  $&$\pm0.04$&$\pm 4.3$&$     $&$     $&$\pm0.04$&$\pm 5.2$&$	  $&$	  $&$\pm 0.05$&$\pm4.4$&$     $&$     $&$\pm0.07$\\
MKV\_0640&$ 33.0  $&$-0.08$&$-1.88$&$ -0.15 $&$  26.0 $&$-0.06$&$-2.28$&$ -0.55 $&$ 65.0  $&$-0.13$&$-2.34$&$-0.61   $&$      $&$     $&$     $&$       $\\
         &$\pm 2.9$&$	  $&$	  $&$\pm0.08$&$\pm 3.2$&$     $&$     $&$\pm0.09$&$\pm 4.5$&$	  $&$	  $&$\pm 0.08$&$      $&$     $&$     $&$       $\\
MKV\_0677&$ 30.0  $&$-0.08$&$-2.12$&$ -0.37 $&$  34.0 $&$-0.09$&$-2.34$&$ -0.59 $&$ 90.0  $&$-0.26$&$-2.44$&$-0.69   $&$      $&$     $&$     $&$       $\\
         &$\pm 3.1$&$	  $&$	  $&$\pm0.08$&$\pm 2.9$&$     $&$     $&$\pm0.08$&$\pm 4.6$&$	  $&$	  $&$\pm 0.07$&$      $&$     $&$     $&$       $\\
MKV\_0698&$ 91.0  $&$-0.35$&$-1.75$&$ -0.27 $&$ 115.0 $&$-0.58$&$-1.88$&$ -0.40 $&$	  $&$	  $&$	  $&$	     $&$62.0  $&$-0.15$&$-1.81$&$-0.33  $\\
         &$\pm 2.8$&$	  $&$	  $&$\pm0.05$&$\pm 3.3$&$     $&$     $&$\pm0.05$&$	  $&$     $&$	  $&$	     $&$\pm2.1$&$     $&$     $&$\pm0.05$\\
MKV\_0705$^1$&$74.5$&$-0.24$&$-1.32$&$ 0.03 $&$  72.5 $&$-0.23$&$-1.63$&$-0.28	$&$128.3  $&$-0.42$&$-1.51$&$-0.16   $&$ 38.0 $&$-0.07$&$-1.53$&$-0.18  $\\
         &$\pm 3.9$&$	  $&$	  $&$\pm0.04$&$\pm10.6$&$     $&$     $&$\pm0.09$&$\pm 6.2$&$	  $&$	  $&$\pm 0.06$&$\pm5.9$&$     $&$     $&$\pm0.08$\\
MKV\_0729&$ 53.0  $&$-0.13$&$-1.47$&$ -0.08 $&$       $&$     $&$     $&$	$&$	  $&$	  $&$	  $&$	     $&$27.0  $&$-0.05$&$-1.66$&$-0.27  $\\
         &$\pm 5.6$&$	  $&$	  $&$\pm0.11$&$       $&$     $&$     $&$	$&$	  $&$	  $&$	  $&$	     $&$\pm3.4$&$     $&$     $&$\pm0.11$\\
MKV\_0733&$ 48.0  $&$-0.13$&$-1.59$&$  0.05 $&$       $&$     $&$     $&$	$&$	  $&$	  $&$	  $&$	     $&$      $&$     $&$     $&$       $\\
         &$\pm 5.4$&$	  $&$	  $&$\pm0.10$&$       $&$     $&$     $&$	$&$	  $&$	  $&$	  $&$	     $&$      $&$     $&$     $&$       $\\
MKV\_0740&$       $&$	  $&$	  $&$	    $&$  83.0 $&$-0.26$&$-1.45$&$ -0.25 $&$	  $&$	  $&$	  $&$	     $&$      $&$     $&$     $&$       $\\
         &$       $&$	  $&$	  $&$	    $&$\pm 7.7$&$     $&$     $&$\pm0.13$&$	  $&$	  $&$	  $&$	     $&$      $&$     $&$     $&$       $\\
MKV\_0769$^1$&$   $&$	  $&$	  $&$	    $&$  74.8 $&$-0.21$&$-1.47$&$  0.21 $&$ 76.3  $&$-0.11$&$-1.86$&$-0.18   $&$ 34.8 $&$-0.06$&$-1.44$&$ 0.24  $\\
         &$       $&$	  $&$	  $&$	    $&$\pm 6.7$&$     $&$     $&$\pm0.06$&$\pm 5.0$&$	  $&$	  $&$\pm 0.05$&$\pm5.9$&$     $&$     $&$\pm0.12$\\
MKV\_0780&$       $&$	  $&$	  $&$	    $&$       $&$     $&$     $&$	$&$ 65.0  $&$-0.08$&$-2.03$&$-0.25   $&$      $&$     $&$     $&$       $\\
         &$       $&$	  $&$	  $&$	    $&$       $&$     $&$     $&$	$&$\pm 6.1$&$	  $&$	  $&$\pm 0.10$&$      $&$     $&$     $&$       $\\
MKV\_0825&$ 62.0  $&$-0.24$&$-1.70$&$ -0.27 $&$       $&$     $&$     $&$	$&$123.0  $&$-0.55$&$-2.01$&$-0.58   $&$      $&$     $&$     $&$       $\\
         &$\pm 4.8$&$	  $&$	  $&$\pm0.08$&$       $&$     $&$     $&$	$&$\pm 5.5$&$	  $&$	  $&$\pm 0.09$&$      $&$     $&$     $&$       $\\
MKV\_0840&$       $&$	  $&$	  $&$	    $&$  53.0 $&$-0.21$&$-2.01$&$ -0.83 $&$ 89.0  $&$-0.36$&$-2.24$&$-1.06   $&$      $&$     $&$     $&$       $\\
         &$       $&$	  $&$	  $&$	    $&$\pm 4.8$&$     $&$     $&$\pm0.10$&$\pm 3.0$&$	  $&$	  $&$\pm 0.10$&$      $&$     $&$     $&$       $\\
MKV\_0880&$ 50.0  $&$-0.13$&$-2.04$&$ -0.46 $&$  67.0 $&$-0.23$&$-2.17$&$ -0.59	$&$116.0  $&$-0.37$&$-2.48$&$-0.90   $&$32.0  $&$-0.06$&$-2.11$&$-0.53  $\\
         &$\pm 3.9$&$	  $&$	  $&$\pm0.06$&$\pm 3.0$&$     $&$     $&$\pm0.06$&$\pm 4.9$&$	  $&$	  $&$\pm 0.06$&$\pm2.7$&$     $&$     $&$\pm0.07$\\
MKV\_0900&$ 36.0  $&$-0.10$&$-2.21$&$ -0.49 $&$       $&$     $&$     $&$	$&$111.0  $&$-0.43$&$-2.53$&$-0.81   $&$29.0  $&$-0.06$&$-2.16$&$-0.44  $\\
         &$\pm 2.8$&$	  $&$	  $&$\pm0.06$&$       $&$     $&$     $&$	$&$\pm 4.1$&$	  $&$	  $&$\pm 0.06$&$\pm3.6$&$     $&$     $&$\pm0.08$\\
MKV\_0902&$       $&$	  $&$	  $&$	    $&$       $&$     $&$     $&$	$&$ 54.0  $&$-0.08$&$-2.61$&$-0.62   $&$      $&$     $&$     $&$       $\\
         &$       $&$	  $&$	  $&$	    $&$       $&$     $&$     $&$	$&$\pm 6.7$&$	  $&$	  $&$\pm 0.08$&$      $&$     $&$     $&$       $\\
MKV\_1007&$ 58.0  $&$-0.28$&$-1.76$&$ -0.37 $&$  53.0 $&$-0.24$&$-2.10$&$ -0.71 $&$102.0  $&$-0.55$&$-2.23$&$-0.84   $&$37.0  $&$-0.12$&$-1.84$&$-0.45  $\\
         &$\pm 4.3$&$	  $&$	  $&$\pm0.09$&$\pm 4.0$&$     $&$     $&$\pm0.09$&$\pm 2.8$&$	  $&$	  $&$\pm 0.09$&$\pm3.1$&$     $&$     $&$\pm0.09$\\
MKV\_1009&$       $&$	  $&$	  $&$	    $&$       $&$     $&$     $&$	$&$ 71.0  $&$-0.13$&$-2.11$&$-0.36   $&$      $&$     $&$     $&$       $\\
         &$       $&$	  $&$	  $&$	    $&$       $&$     $&$     $&$	$&$\pm 7.9$&$	  $&$	  $&$\pm 0.12$&$      $&$     $&$     $&$       $\\
MKV\_1013$^1$&$32.9$&$-0.06$&$-1.66$&$-0.36 $&$  89.3 $&$-0.29$&$-1.40$&$ -0.10	$&$ 92.0  $&$-0.17$&$-1.79$&$-0.49   $&$      $&$     $&$     $&$       $\\
         &$\pm 4.6$&$	  $&$	  $&$\pm0.07$&$\pm 4.2$&$     $&$     $&$\pm0.04$&$\pm 4.9$&$	  $&$	  $&$\pm 0.05$&$      $&$     $&$     $&$       $\\
\hline
\end{tabular}
\begin{list}{}{}
\item[$^1$] From FLAMES-UVES spectra \citep{VSI12}
\end{list}
\end{table*}
\end{document}